\documentclass{aa}

\usepackage[varg]{txfonts}

\usepackage{natbib}

\bibpunct{(}{)}{;}{a}{}{,} 

\usepackage{graphicx}

\usepackage{longtable}

\usepackage{color}

\usepackage{lscape}

\begin{document}

\title{Wind properties of variable B supergiants}
\subtitle{Evidence of pulsations connected with mass-loss episodes\thanks{Based on observations taken with the J. Sahade Telescope at Complejo Astron\'omico El Leoncito (CASLEO), operated under an agreement 
between the Consejo Nacional de Investigaciones Cient\'{\i}ficas y T\'{e}cnicas de la Rep\'{u}blica 
Argentina, the Secretar\'{\i}a de Ciencia y Tecnolog\'{\i}a de la Naci\'on, and the National 
Universities of La Plata, C\'ordoba, and San Juan.}}


\author{M. Haucke\inst{1,2}
\and L. S. Cidale\inst{1,2,3}
\and R. O. J. Venero\inst{1,2}
\and M. Cur\'e\inst{3}
\and M. Kraus\inst{4,5}
\and S. Kanaan\inst{3}
\and C. Arcos\inst{3}
}

\institute{Departamento de Espectroscop\'{i}a, Facultad de Ciencias Astron\'{o}micas y Geof\'{i}sicas, Universidad
  Nacional de La Plata, Paseo del Bosque S/N, La Plata, Argentina.\\
  \email{mhaucke@fcaglp.unlp.edu.ar}
\and Instituto de Astrof\'{i}sica de La Plata, CONICET-UNLP, Paseo del Bosque S/N, La Plata, 
Argentina.
\and Instituto de F\'{i}sica y Astronom\'{i}a, Facultad de Ciencias, Universidad de Valpara\'{\i}so,  Av. Gran Breta\~na 1111, Casilla 5030, Valpara\'{\i}so, Chile.
\and Astronomick\'{y} \'{u}stav, Akademie v\v{e}d \v{C}esk\'{e} republiky v.v.i.,
Fri\v{c}ova 298, 251\,65, Ond\v{r}ejov, Czech Republic.
\and Tartu Observatory, T\~{o}ravere, 61602 Tartumaa, Estonia.
}

\date{Received date /  Accepted date }

\authorrunning{Haucke et al.}

\titlerunning{Wind Properties of Variable B Supergiants}

\abstract
    {Variable B supergiants (BSGs) constitute a heterogeneous group of stars with complex photometric and spectroscopic behaviours. They exhibit mass-loss variations and experience different types of oscillation modes, and there is growing evidence that variable stellar winds and photospheric pulsations are closely related.}
    {To discuss the wind properties and variability of evolved B-type stars, we derive new stellar and wind parameters for a sample of 19 Galactic BSGs by fitting theoretical line profiles of H, He, and Si to the observed  ones and compare them with previous determinations.}
{The synthetic line profiles are computed with the non-local thermodynamic equilibrium (NLTE) atmosphere code FASTWIND, with a $\beta$-law for hydrodynamics.}
{The mass-loss rate of three stars has been obtained for the first time. The global properties of stellar winds of mid/late B supergiants are well represented by a $\beta$-law with $\beta > 2$. All stars follow the known empirical wind momentum--luminosity relationships, and the late BSGs show the trend of the mid BSGs. \object{HD\,75149} and \object{HD\,99953} display significant changes in the  shape and intensity of the H$\alpha$ line (from a pure absorption to a P Cygni profile, and vice versa). These  stars  have mass-loss variations of almost a factor of 2.8. A comparison among mass-loss rates from the literature reveals discrepancies of a factor of 1 to 7.  This large variation is a consequence of the uncertainties in the determination of the stellar radius. Therefore, for a reliable comparison of these values we used the invariant parameter $Q_r$. Based on this parameter, we find an empirical relationship that associates the amplitude of mass-loss variations with  photometric/spectroscopic variability on timescales  of tens of days. We find that stars located on the cool side of the bi-stability jump show a decrease in the ratio $V_\infty/V_{\rm{esc}}$, while their corresponding mass-loss rates are similar to or lower than the values found for stars on the hot side. Particularly, for those variable stars a decrease in $V_\infty/V_{\rm{esc}}$ is accompanied by a decrease in $\dot{M}$.}
{Our results also suggest that radial pulsation modes with periods longer than 6 days might be responsible  for the wind variability in the  mid/late-type BSGs. These radial modes might be identified with strange modes, which are known to facilitate (enhanced) mass loss.  On the other hand, we propose that the wind behaviour of stars on the cool side of the bi-stability jump could fit with predictions of the $\delta-slow$ hydrodynamics solution for radiation-driven winds with highly variable ionization.}


\maketitle





\section{Introduction}

Massive stars have a significant impact on the ionization, structure, and evolution of the interstellar medium (ISM). 
Via their stellar winds they  release energy and momentum to their surroundings leading to the formation of stellar wind-blown bubbles,  bow-shocks, 
and circumstellar shells. On the other hand, the huge amount of mass lost from the stars modifies the later stages of their  lives (evolutionary timescales and 
final core masses)  as well as the type of SN remnants left \citep{Meynet1994,Woosley2002}.

 Stellar winds of hot massive stars are mainly driven by line scattering
of UV photons coming from the stars' continuum radiation. The 
standard stationary radiation-driven  wind theory  \citep{CAK1975,
 Pauldrach1986, Friend1986} predicts mass-loss rates and terminal velocities as
a function of stellar parameters \citep{Abbott1982, Vink2000, Vink2001},
and also predicts a wind momentum--luminosity relationship (WLR) \citep{Kudritzki1995}. 
This theory has long described  the global properties of stellar winds of OBA stars. However,  advances in the observational techniques (high spatial 
and spectral resolution) together with   progress in more realistic atmosphere models 
have revealed discrepancies in the mass-loss rates not only between theory and observations, but also amongst different observational tracers themselves 
\citep[cf.][]{Puls2008}.

As \citet{Owocki1984, Owocki1988}, and later \citet{Feldmeier1998}  demonstrated, strong hydrodynamic instabilities occur in line-driven winds of massive O-type stars, leading to the formation of wind-clumping (small-scale density inhomogeneities distributed across the wind). Wind-clumping has traditionally been used to solve the issue of the mass-loss discrepancy found in O supergiants (OSGs) when different diagnostic methods are employed  \citep{Bouret2005,Fullerton2006}. Mass-loss rates derived from unsaturated UV resonance lines give lower values than those obtained from the H$\alpha$ line \citep[cf.][]{Puls2008}.  The   macroclumping approach (with optically thick clumps  at  certain  frequencies)  has provided a fully self-consistent and simultaneous fit to both UV and optical lines  \citep{Oskinova2007, Sundqvist2010, Sundqvist2011, Surlan2013}.
In previous works an artificial reduction of the stellar mass-loss rate, an extremely high clumping factor, or an anomalous chemical abundance for specific elements were invoked; however, the macroclumping approach means that none of these is necessary.

The situation is different for B supergiants (BSGs). Stellar winds of BSGs are fairly well represented by the analytical velocity $\beta$-law, with $\beta$ typically in the range $1-3$ \citep{Crowther2006,MarkovaPuls2008}. Moreover, these stars show a drop in the ratio $V_{\infty}/V_{\rm{esc}}$ of a factor of two at  around   21\,000 K, from 2.6 for stars on the hot side to 1.3 for stars on the cool side  \citep{Pauldrach1990}, known as the bi-stability jump \citep[][]{Lamers1995}. This jump was interpreted as a consequence of the recombination of Fe IV to Fe III in the wind  \citep{Vink1999}. These authors    predicted that this drop should be accompanied by a steep increase in the mass-loss rate of about a factor of 3 to 5, 
 whilst some observations indicate a decrease \citep{Crowther2006,MarkovaPuls2008}. A recent study of the effects of  micro- and macroclumping in the opacity behaviour of the H$\alpha$ line  at the bi-stability jump  suggests that the macroclumping should also play an important role at temperatures on the cool side of the jump \citep{Petrov2014}.

It is also well known that stellar winds of BSGs are highly variable. To date, most of the studies  related to  $\alpha$ Cyg variables and hypergiants have focused on their photometric and/or optical spectroscopic  
(mainly H$\alpha$) variability, and many surveys have been carried out to search for variability 
periods  \citep[e.g.][]{Kaufer2006,Lefever2007,Lefevre2009}.  
Long-term space-based photometry and spectroscopy have linked this variability  to  opacity-driven
radial and non-radial oscillations \citep{Glatzel1999,Lefever2007,Kraus2015}, and new 
instability domains have been established in the Hertzsprung--Russell (HR) diagram covering the region of 
BSGs \citep{Saio2006, Godart2017}. As massive stars can cross the BSGs' domain more than once, the pulsation 
activity of these stars can drastically change between their red- and blue-ward evolution 
\citep{Saio2013}. BSGs on a blue-ward evolution (i.e. a post-red supergiant)  tend to undergo a 
significantly larger number of pulsations, even including radial strange-mode pulsations with 
periods between 10  and 100 days (or more). It has been suggested that strange-mode pulsations  
cause time-variable mass-loss rates in very luminous evolved massive stars (Glatzel et al. 1999). 
The first observational evidence of the presence of strange modes have been found in two BSGs: \object{HD\,50064} \citep{Aerts2010} and 
\object{55\,Cyg}  \citep{Kraus2015}. In the latter, the authors found line variations with periods in the range of 2.7 hrs to 22.5 days. 
They interpreted these variations in terms of oscillations in p-, g-, and strange modes.  The last could lead to phases of enhanced mass loss. 
Finally, the connection between pulsation and mass loss in \object{55\,Cyg} was confirmed by \citet{Yadav2016}, based on a  linear non-adiabatic stability 
analysis with respect to radial perturbations.  These authors demonstrated that, as a consequence of the instabilities, the non-linear simulations revealed 
finite amplitude pulsations consistent with the observations.

To obtain further insights on the wind structure and wind variability of evolved stars of 
intermediate mass, we studied a sample of Galactic BSGs (classified either as irregular or $\alpha$ 
Cyg-type variables)  in both  the blue and the H$\alpha$ spectral region. The H$\alpha$
emission is quite sensitive to the  wind properties \citep{Puls2008}, and we used the Fast Analysis of 
STellar atmospheres with WINDs   \citep[FASTWIND,][]{fastwind1997, FWmio, Rivero2012} to obtain terminal velocities and mass-loss rates by 
fitting synthetic to observed line profiles. In addition, to derive proper
stellar parameters (effective temperature and surface gravity) we consistently modelled the Si ionization balance together with the photospheric lines of 
H and He. Determining  new, more  accurate 
values of mass-loss rates and stellar parameters of massive stars is crucial for our understanding of stellar wind properties, wind 
interactions with its surroundings, and the plausible mechanisms related with wind variability and its evolution.

The paper is structured as follows: our observations and  modelling are described in Sects. 
 \ref{obs} and  \ref{mod}, respectively. The results of our line profile fittings are shown in  Sect. \ref{res} and the new  wind parameters are compared with 
previous determinations. For three objects (\object{HD\,74371}, \object{HD\,99953} and \object{HD\,111973}) the  wind parameters are 
reported for the first time. As most of the stars in our sample exhibit photometric and spectroscopic variability, on timescales from one to tens 
of days,
and have been modelled previously by several authors,  in  Sect. \ref{dis} we discuss changes in the wind parameters.  We present  for mid and 
late BSGs an empirical  relationship between the amplitude of mass-loss variations and light/spectroscopic  periods, indicating that a 
significant percentage of the mass-loss  could be  triggered  by pulsation modes. 
Our conclusions are given in  Sect. \ref{con}. Finally,  Appendix \ref{ape1} shows model fittings to the photospheric lines of 
each star and summarizes the stellar and wind parameters found in the literature and in this work.


\section{Observations}\label{obs}

We took high-quality  optical spectra for 19 Galactic BSGs of spectral types between B0 and B9. These stars 
were selected from the \emph{Bright Star Catalog} \citep{Hoffleit1991}.
The observations were performed in January 2006, February 2013, April 2014, and  February and March 2015. 
We used  the REOSC spectrograph (in crossed dispersion mode) attached to the \emph{Jorge Sahade} 
2.15 m telescope at the Complejo Astron\'omico El Leoncito (CASLEO), San Juan, Argentina. 
The adopted instrumental configuration was a 400 l/mm grating ($\#580$), a single slit of width 250 $\mu$, and a 1024x1024 TEK CCD detector with a 
gain of 1.98 e$^-$/\,adu. This 
configuration produces spectral resolutions of $R\, \sim{12\,600}$  at 4500 \AA\, and  $R\, \sim{13\,900}$ at 6500 \AA.
Spectra were reduced and wavelength calibrated following the standard procedures using the corresponding IRAF\footnote{IRAF is distributed by the National Optical Astronomy 
Observatory, which is operated by the 
Association of Universities for Research in Astronomy (AURA) under cooperative agreement with 
the National Science Foundation.} routines. The resulting spectra have an average signal-to-noise ratio (S/N)  of $\sim$ 300.

Table \ref{table:1} lists the stars in our programme, indicating name and HD number, spectral 
type, and variable type designation according to  \citet{Lefevre2009} (from  a study of HIPPARCOS 
light curves) or the VSX database. In addition, binary stars are also listed. In the following columns of the same table, we list the reported light curve or spectroscopic periods, 
the observation dates of our data, and the achieved wavelength coverage.
Here, we present six  stars with more than one
observation (\object{HD\,53138},  \object{HD\,58350}, \object{HD\,75149}, \object{HD\,80077}, \object{HD\,99953}, and  
\object{HD\,111973}). One of these  stars (\object{HD\,111973}) was observed on two consecutive nights.

\begin{table*}

    \tabcolsep 3pt
\caption{Log of Observations}
\label{table:1}
\centering
\small
\begin{tabular}{lclcccc}
\hline\hline
Star Name       &   HD Number     &  Sp. Type$^{\,a}$  & Var. Type$^{\,b}$ & Period  &  Observation Date    & Wavelength Interval\\
                &                 &                   &                  &  days           &\tiny{(YY/MM/DD)} & [\AA:\AA]          \\
\hline \\
$\beta$ ~\,Ori   & 34\,085  &  B8Iae     &   ACYG      & 2.075$^{\,c}$, 1.22-74.74$^{\,d}$, 28$^{\,n}$                           & 2\,006/01/15                       &    [3500:7850]     \\

$\kappa$ ~\,Ori  & 38\,771  &  B0.5Ia    &   IA        & 1.047$^{\,e}$, 1.9$^{\,f}$, 4.76$^{\,e}$, 6.5$^{\,f}$, 9.5$^{\,f}$  & 2\,006/01/15          &    [3500:7850]     \\

$\chi^{2}$ Ori   & 41\,117  &  B2Ia      &   ACYG      & 0.92$^{\,e}$, 0.95$^{\,e}$, 2.869$^{\,e}$, 20$^{\,e}$, 40$^{\,e}$, 200$^{\,e}$        & 2\,006/01/15                       &     [3500:7850]     \\

PU\,Gem          & 42\,087  &  B4Ia$^{\,g}$       &   ACYG      & 6.807$^{\,o}$, 25$^{\,e}$                                         & 2\,006/01/15                       &     [6500:7850]     \\

V731\,Mon        & 47\,240  &  B1Ib      &   ACYG, SB$^{\,q}$      & 1.73$^{\,j}$, 2.742$^{\,c,\,e}$, 133$^{\,e}$                                  & 2\,006/01/15                       &      [3500:7850]    \\

$\epsilon$\,CMa  & 52\,089  &  B1.5   II &   IA        &                                                   & 2\,013/02/05                       &      [4300:6850]    \\

V820\,Cas        & 52\,382  &  B1Ia    &   IA        &                                                   &2\,006/01/15                       &      [3500:7850]     \\

$\sigma^{2}$\,CMa& 53\,138  &  B3Ia      &   ACYG      & 3.69$^{\,j}$, 24.39$^{\,j}$, 24.44$^{\,c}$                                       & 2\,006/01/15                       &      [3500:7850]     \\ 

                 &          &            &             &                                                   & 2\,013/02/05                       &      [4300:6850]    \\

$\eta$\,CMa      & 58\,350  &  B5Ia      &   L, ACYG    & 4.7$^{\,i}$, 6.631$^{\,j}$                                       & 2\,006/01/15                       &      [3500:7850]    \\

                 &          &            &             &                                                   &2\,013/02/05                       &      [4300:6850]     \\

J\,Pup           & 64\,760  & B0.5Ib     &   IA        & 1.2$^{\,m}$, 1.8$^{\,j}$, 2.4$^{\,m}$, 2.8$^{\,j}$, 6.8$^{\,m}$                                &2\,013/02/05                       &      [4300:6850]     \\

LN\,Vel          & 74\,371  & B6Iab/b    &   IA, ACYG  & 8.29$^{\,c,\,i}$, 1, 15-20$^{\,h}$                                 & 2\,006/01/15                       &    [3500:7850]     \\

OP\,Vel          & 75\,149  &  B3Ia      &  SPB?, ACYG & 1.086$^{\,c,\,n}$, 1.215$^{\,j}$, 2.214 $^{\,j}$                               & 2\,006/01/15                       &    [3500:7850]     \\

                 &          &            &             &                                                   & 2\,013/02/05                       &      [4300:6850]    \\

                 &          &            &             &                                                   & 2\,013/02/07                       &     [4300:6850]     \\

                 &          &            &             &                                                   & 2\,014/04/14                       &      [4200:6650]    \\

GX\,Vel          & 79\,186  &  B5Ia      &   IA        &                                                   & 2\,006/01/15                       &     [3500:7850]     \\

PV\,Vel          & 80\,077  & B2Ia+e     &  GCAS?, SDOR & 3.115$^{\,c}$, 21.2$^{\,l}$, 41.5$^{\,h}$, 55.5$^{\,h}$, 66.5$^{\,h}$, 76.0$^{\,h}$        & 2\,006/01/15                       &     [3500:7850]     \\

                 &          &            &             &                                                   & 2\,014/04/12                       &      [4200:6650]    \\

V519\,Car        & 92\,964  &  B2.5Ia    &   ACYG      & 2.119$^{\,j}$, 4.71$^{\,c}$, 14.706$^{\,j}$                              & 2\,013/02/05                       &      [4300:6850]     \\

V808 Cen          & 99\,953  &  B1/2Iab/b &   IA        & 17.7$^{\,k}$                                              & 2\,014/04/14                       &     [4200:6650]     \\

                 &          &            &             &                                                   & 2\,015/02/13                       &     [4200:6650]     \\

                 &          &            &             &                                                   & 2\,015/03/13                       &     [4200:6650]     \\

$\kappa$\,Cru    & 111\,973 &  B2/3Ia    &   ACYG?, IA, SB$^{\,p}$ & 9.536$^{\,i}$, 57.11$^{\,i}$                                      & 2\,014/04/11                       &     [4200:6650]     \\

                 &          &            &             &                                                   & 2\,014/04/12                       &     [4200:6650]     \\

ALS\,3038        & 115\,842 &  B0.5Ia/ab &   ACYG?, IA & 10.309$^{\,c}$, 13.38$^{\,i}$                                     & 2\,014/04/11                       &    [4200:6650]     \\

V1058\,Sco       & 148\,688 &  B1Iaeqp   &   ACYG      & 1.845$^{\,c}$, 6.329$^{\,c}$                                      & 2\,014/04/11                       &    [4200:6650]     \\

\hline

\end{tabular}

\tablefoot{\tablefoottext{a}{From the SIMBAD astronomical database \citep{Wenger2000}}.
\tablefoottext{b}{IA (poorly studied variables of early spectral type), L
  (slow irregular variables), GCAS ($\gamma$ Cassiopeiae type), ACYG ($\alpha$ Cyg type), SPB (slowly pulsating B stars), and  S Dor (S Doradus type)  \citep[classification taken from][]{Lefevre2009}. The question mark means that the classification was based on a new period determination or that the type could not be clearly  identified from the HIPPARCOS light curve.}
\tablefoottext{c}{\citet{Lefevre2009},}
\tablefoottext{d}{\citet{Moravveji2012},} \tablefoottext{e}{\citet{Morel2004},}  \tablefoottext{f}{\citet{Prinja2004},} \tablefoottext{g} {this work}, \tablefoottext{h}{\citet{vanGenderen1989},} \tablefoottext{i}{\citet{Koen2002},}  \tablefoottext{j}{\citet{Lefever2007},} \tablefoottext{k}{\citet{Sterken1977},} \tablefoottext{l}{\citet{Knoechel1982},} \tablefoottext{m}{\citet{Kaufer2006},} \tablefoottext{n}{\citet{Aerts2013},} \tablefoottext{o}{International Variable Star Index (VSX) database, and}
\tablefoottext{p,q}{SB \citep[spectroscopic  binary][]{Chini2012,Prinja2002}}.
}

\end{table*}


\section{Stellar and wind parameters}\label{mod}

To derive stellar and wind parameters we made use of the code FASTWIND (v10.1.7). 
The code computes a spherically expanding line-blanketed 
atmosphere. All background elements are considered with their solar abundances \citep[taken from][]{Grevesse1998}. \ion{H}{}, \ion{He,}{} and \ion{Si}{} atoms are treated 
explicitly with high precision by means of  a complete NLTE approach and an accelerated lambda iteration 
(ALI) scheme is applied to solve the comoving-frame equations of  radiative transfer \citep{Puls1991}. The atmospheric 
stratification is modelled considering a smooth transition between a quasi-hydrostatic photosphere and an 
analytical wind structure described by a velocity $\beta$-law. The temperature structure is
calculated from the electron thermal balance and is consistent with 
the radiative equilibrium condition.

We opted for a wind model without clumping because we want to compare the derived  mass-loss rates with previous determinations and  to  
study the wind variability. In general,  mass-loss rates found in the literature are mainly estimated using unclumped models. A second reason is  that 
we do not have contemporaneous data in the UV and IR spectral regions to evaluate the importance of the clumping factor in the H$\alpha$ line modelling. 
As a consequence, our results will provide upper-limit values for the mass-loss rates. In addition, to obtain an optimal fit to the observed line 
profiles, it was necessary to consider not only a microturbulence velocity, $V_{\rm{micro}}$, and broadening due to the star's rotation, $V\,\sin\,i$, but 
also the effect of a macroturbulence velocity, $V_{\rm{macro}}$. The latter contributes mainly to line broadening and shaping as noted by \citet{SD2007}. 

To obtain the effective temperature  ($T_{\rm{eff}}$) we evaluated the ionization balance (e.g. \ion{Si}{ii}$-$\ion{Si}{iii}, \ion{Si}{iii}$-$\ion{Si}{iv}, if the \ion{Si}{iv}\,$\lambda$\,4089,\,4116 lines are observed, and \ion{He}{i}$-$\ion{He}{ii}).
We used solar abundances for He and Si ($\log\,N_{\rm {He}}/N_{\rm H}$= $-$1.07 and $\log\,N_{\rm {Si}}/N_{\rm H}$= $-$4.45) and found good fits for each modelled spectrum (see details in sections \S \ref{objects} and \S \ref{dis}). To derive an accurate determination of the surface gravity ($\log\,g$), we modelled the H$\gamma$ and H$\delta$ lines. We used a `by eye' fitting procedure to find the best-fitting  synthetic line spectrum to the observed one.

The  H$\alpha$ line was modelled to derive the mass-loss rate ($\dot{M}$) and 
the parameters of the velocity $\beta$-law: the power index $\beta$ and the terminal velocity $V_{\infty}$. It is important to stress  that the major source of uncertainty in the determination of mass-loss  rates is due to ambiguous determinations  of stellar radii and distances to the stars \citep[see][]{Markova2004}. On the other hand,  H$\alpha$ is rather sensitive to $V_{\infty}$ \citep{Garcia2017}. Therefore, the stellar radius and terminal velocities should be derived independently, as  explained below.

Prior to the modelling process, the stellar radius, $R_\star$, was derived using  the measured angular diameter,  the  distance to the star, or  a fit to the  
observed spectral energy distribution (SED). We model the SED with the interactive user interface BeSOS\footnote{The Be Stars Observation Survey (BeSOS; http://besos.ifa.uv.cl) is a database containing reduced spectra acquired using the echelle spectrograph PUCHEROS.} using the best-fitting 
atmospheric (TLUSTY or Kurucz) model \citep{Hubeny1995, Kurucz1979}. BeSOS  reads  stellar photometry data of any star from photometry-enabled 
catalogues  in VizieR and  the parallax distance from the HIPPARCOS catalogue. Sometimes, it was not possible to obtain a good fit to the SED (see details below on the procedures applied to each star). In these cases, a slightly 
modified distance was required to improve the fit.  To find the best model, BeSOS uses the IDL program {\it mpfit} that searches for the best fit 
using the Levenberg--Marquardt (LM) method, which is a standard technique for solving
non-linear least squares problems.

BeSOS  provides, for a given colour  excess, a new set of stellar parameters: $T_{\rm {eff}}$, $\log\,g$, and $R_\star$ (in units of the solar radius).  As initial entries we use the $T_{\rm {eff}}$ and  $\log\,g$ values obtained in this work from the Si and He ionization balances and from H$\gamma$ and H$\delta$ line widths, respectively, and the code searches then for the most optimal values to fit the SED. The colour excess, $E(B-V)$, was calculated 
using the observed $B-V$ and the interpolated $(B-V)_0$ values calculated from the $T_{\rm {eff}}-(B-V)$ scale for supergiants \citep{Flower1996}. 
In a few cases,  the $E(B-V)$ had to be modified to fit  the 2200~\AA\, bump properly. The stellar radius obtained using the SED is averaged with the values derived  via the angular diameter and $M_{\rm{bol}}$ (the bolometric magnitude). Table \ref{table:besos} lists  the intrinsic properties for each 
star: the spectral type found in the literature, the visual apparent magnitude, the observed and  intrinsic colour indexes, 
the calculated colour excess,  the stellar parameters ($T_{\rm {eff}}$ and $\log\,g$, obtained from fittings to the SED), the distance to the star ($d$), the bolometric correction ($BC$), the calculated bolometric magnitude, and the stellar radius. The computed parameters are listed with their corresponding errors.
Details on the best-fitting model obtained with BeSOS are given in  section \S \ref{res}.

Typical error bars derived using the BESOS code for  $T_{\rm {eff}}$,  $\log\,g$, and  $R_\star$ are of about  $1\%$, $2\%$, and $5\%$, respectively. Nevertheless, considering uncertainties in the distance estimates between $5\%$ and $30\%$ (from parallax measurements) and differences less than 1\,500 K  between the $T_{\rm {eff}}$ values derived from BESOS and FASTWIND, the propagation of errors yields  an uncertainty in $M_{\rm{bol}}$ lower than $10\%$ and  a margin of error between $5\%$ and $20\%$ for  $R_\star$.

In relation to the stellar parameters derived from FASTWIND, we can adopt the parameter errors estimated by \citet{Kraus2015}, who found uncertainties in $T_{\rm{eff}}$ of 300 K -- 500 K from the Si ionization balance and 1\,000 K for the He lines. We adopt  $\Delta T_{\rm{eff}}$= 1\,000 K for those stars for which only the He lines were modelled and $\Delta T_{\rm{eff}}$= 500 K if the  \ion{Si}{ii}, \ion{Si}{iii},  and \ion{He}{} lines were used to derive the temperature. From measurements of the wings of H$\gamma$ and H$\delta$ we estimated an error of 0.1 dex in $\log\,g$. The error bars in $V_{\rm{micro}}$ are 2 km s$^{-1}$ for Si and He lines, 5~km~s$^{-1}$ for the H lines, and 10 km s$^{-1}$ for H$\alpha$. We model all the photospheric lines using the same $V_{\rm{micro}}$.

Both $V\,\sin\,i$ and $V_{\rm{macro}}$ affect the line broadening.
  To model the line profiles we adopted the value of  $V\,\sin\,i$ ($\pm10$ km s$^{-1}$) found in the literature \citep[][given in Table \ref{table:A1}]{Kudritzki1999, Lefever2007,Fraser2010} or average values when large scatters are present. We derived the macroturbulent velocity that reproduces the extra broadening seen in the line profile. We found uncertainties in $V_{\rm{macro}}$ from 10 km s$^{-1}$ to  20 km s$^{-1}$.

Regarding the wind parameters, we searched for  $V_{\infty}$  measurements from UV observations  \citep{Prinja2010,Prinja1998,Howarth1997} and used them as initial values to reproduce the H$\alpha$ line. Good fits were obtained using slight variations of  $V_{\infty}$. These values are listed in Table \ref{table:2} with deviations of $10\%$ with respect to the UV measurements. In a few cases, the UV terminal velocities were not able to fit the observed H$\alpha$ line profile and our own $V_{\infty}$ determinations are provided. In this case, the discrepancies obtained between the derived  $V_{\infty}$ and  the UV data can increase up to  30$\%$. Therefore, we consider errors for  $V_{\infty}$ between 10\% and 30\%.

Adopting the values  for $V_{\infty}$ from the UV and errors of $10\%$ for $R_\star$, we observe that late and early B supergiants may show a change of  up to  $20\%$ in the equivalent width (EW) of the emission component of a P Cygni line profile if $\dot{M}$ varies  by an amount of  $10\%$ and  $20\%$, respectively. Larger errors in  $V_{\infty}$ might give uncertainties on $\dot{M}$ of $30\%$. However,  when the H$\alpha$ line  is seen in absorption, the error on $\dot{M}$ can be larger than a factor of 2 \citep{Markova2004}.

\begin{table*}

    \tabcolsep 3pt

\caption{Stellar parameters calculated in the present study or adopted from the literature}.

\label{table:besos}

\centering 

\small

\begin{tabular}{rlcrccccccrr}

\hline\hline  

HD~~~ &      ~~  ST  &          $m_{\rm V}$       &        $B-V$    &            $(B-V)_0^{~a}$      &      $E(B-V)$   &   $T_{\rm {eff}}$   &        $\log\,g$  &   $ d$~~  &  ~$BC^{~a}$ &  $ M_{\rm{bol}}$~~  &   $R_\star$~~~\\

Number                   &              &                                  &                      &                                 &                     &   K                          &         dex               &     pc~~       &                   &           &         R$_\odot$~~       \\       

\hline

\object{34\,085}   & B8\,Iae   &   $0.13$   &   $-0.03$ & $-0.07$ &  $0.04$ & $11\,760\pm120$  & $2.00\pm0.10$  &  $264\pm23$   & $-0.83$  & $-7.9\pm0.2$ & $72\pm10$\\

\object{38\,771} & B0.5\,Ia  &  $2.06$   &    $-0.18$ & $-0.23$ &  $0.05$ & $25\,700\pm260$  &  $2.70\pm0.05$  &  $198\pm~8$     & $-2.41$    &$-7.0\pm0.2$ & $13\pm~~ 1$\\

\object{41\,117}   &   B2\,Ia   &  $4.63$   &    $0.28$  & $-0.18$ &  $0.42$ & $17\,940\pm180$  & $2.20\pm0.05$  & $~552\pm85$      & $-1.84$    &$-7.3\pm0.4$ & $23\pm~~ 3$ \\

\object{42\,087}  &   B4\,Ia     & $5.78$    &   $0.20$  & $-0.14$ &  $0.34$ & $15\,000\pm150$  &  $2.39\pm0.05$  &  $~2\,075\pm1\,000^{s}$ & $-1.46$ &  $-8.3\pm1.1$ & $55\pm11$\\

\object{47\,240}  &   B1\,Ib     & $6.15$   &    $0.15$  & $-0.18$ &  $0.33$  &  $17\,500\pm180$  & $2.40\pm0.05$ &  $1\,598\pm479$  & $-1.78$ & $-7.7\pm0.7$  &  $30\pm~ ~4$\\

\object{52\,089}  & B1.5\,II   & $1.50$   &    $-0.21$ & $-0.20$ &  $0.00$  & $21\,000\pm210$   &  $3.00\pm0.05$  &  $124\pm~2$   &  $-2.20$ &  $-6.2\pm0.1$  &  $11\pm~~ 1$\\

\object{52\,382}  &   B1\,Ia & $6.50$   &   $0.19$   &$-0.20$  & $0.44$ & $23\,140\pm230$  &  $2.47\pm0.05$ &  $1\,301\pm430^{s}$ &  $-2.05$ & $ -7.5\pm0.8$ &  $21\pm~~ 2$\\

\object{53\,138}  &   B3\,Ia      & $3.02$   &   $-0.08$  &  $-0.21$ & $0.10$ & $18\,000\pm180$ &  $2.20\pm0.05$ &   $~847\pm287$  &  $-1.66$ & $-8.5\pm0.8$  &  $46\pm~~ 4$\\

\object{58\,350}  &   B5\,Ia       & $2.45$ &    $-0.09$ &  $-0.17$ & $0.08$  &  $15\,000\pm500^f$ & $2.00\pm0.10^f$   & $~609\pm148$    & $-1.24$ &  $-8.0\pm0.6$   & $54\pm~~6$\\

\object{64\,760}  &   B0.5\,Ib  & $4.24$ &    $-0.14$  & $-0.21$ & $0.07$  &  $22\,370\pm220$ &  $2.50\pm0.05$ &   $507\pm25$ &  $-2.21$ &  $-6.7\pm0.2$   &    $12\pm~~ 1$ \\

\object{74\,371}  &   B6\,Iab/b  & $5.23$   &  $0.20$ &  $-0.15$  &  $0.35$  & $13\,800\pm140$ & $2.00\pm0.05$ & $1\,800\pm360^{s}$ & $-1.02$& $-8.7\pm0.5$ &  $73\pm10$\\

\object{75\,149}  &  B3\,Ia        &    $5.46$    & $0.27$ & $-0.19$  &   $0.46$ & $15\,000\pm150$ & $2.12\pm0.05$   &  $1\,642\pm330^{s}$ & $-1.39$ & $-8.4\pm0.7$ &  $61\pm 13$ \\

\object{79\,186} &  B5\,Ia  &    $5.00$ & $0.22 $ & $-0.13$  &  $0.35$  &    $15\,000\pm150$          &   $2.12\pm0.05$       &      $1\,449\pm609$ & $-1.36$ &$-8.3\pm0.9$ &    $61\pm~~7$\\

\object{80\,077} &    B2Ia+e  &   $7.56$   &   $1.34$  & $-0.16$    & $1.50$ &    $18\,000\pm180$   &    $2.17\pm0.05$   &  $3\,600\pm600^{s}$  & $-1.62$ &$-11.6\pm0.4$ &  $195\pm 47$ \\    

\object{92\,964}   &   B2.5Ia  &  $5.38$  & $0.27$    & $-0.16$     &  $0.48$ &   $18\, 000\pm180$  &  $2.19\pm0.05$ &  $1851\pm994$ & $-1.66$& $-9.1\pm1.2$ & $70\pm14$ \\      

\object{99\,953}  &   B1/2\,Iab/b       & $6.5$    & $0.38$  &  $-0.18$  &  $0.56$ & $18\,830\pm190$ &   $2.30\pm0.05$ &    $1\,075\pm427$ & $-1.78$ &  $-7.2\pm0.9$ &  $25\pm~~ 3$\\

\object{111\,973} & B2/3\,Ia     & $5.94$  &  $0.24$ & $-0.14$    & $0.38$  & $17\,180\pm170$ &   $2.18\pm0.05$ & $1\,660\pm350^{s}$   & $-1.46$ &  $-7.8\pm0.5$ &  $46\pm~~ 9$\\

\object{115\,842}&   B0.5\,Ia/ab                           & $6.03$   &  $0.30$ & $-0.23$   &  $0.60$ & $25\,830\pm260$  &  $ 2.75\pm0.05$ & $1\,538\pm750$  & $-2.46$ & $-9.2\pm1.2$  &  $35\pm~ ~4$ \\                 

\object{148\,688}&   B1\,Iaeqp       &  $5.31$  &  $0.35$ & $- 0.20$  &  $0.54$ & $20\,650\pm210$ &   $2.20\pm0.05$  &     $~~833\pm229$ & $-2.00$  & $-8.0\pm0.7$  &  $31\pm~~4$ \\

\hline

\end{tabular}

\tablefoot{

  {Values of $T_{\rm{eff}}$ and $\log\,g$ were derived by fitting the SED, with the exception of those flagged with $f$.}

  {$R_\star$ is an average of the values obtained from three different approaches (see details in section \S \ref{mod}).}

  {d is the distance to the star in parsecs derived from HIPPARCOS, with the exception of those data flagged with $s$.}

  \tablefoottext{a}{BC: bolometric correction taken from \citet{Flower1996},}

  \tablefoottext{f}{parameters derived with the code FASTWIND (see details in section \S \ref{objects}), and }

  \tablefoottext{s}{distances derived from other methods, see details in the text.}

  }

\end{table*}


\section{Results}\label{res}

For each star in our programme we modelled the line profiles of H, He, and Si. Figure  \ref{halpha} shows  the observed 
H$\alpha$ line and the best-fitting synthetic model. The rest of the lines and their corresponding fits are shown 
in the Appendix (Figs. \ref{fig:7}-\ref{fig:13}).  In Table  \ref{table:2} we list the obtained stellar and 
wind parameters with their errors: $T_{\rm{eff}}$, $\log\,g$, $\beta$, $\dot{M}$, $V_{\infty}$, $V_{\rm{micro}}$, $V_{\rm{macro}}$, $V\,sin\,i$, 
$R_\star$ (stellar radius in solar radius units), $\log L/L_\odot$  (stellar luminosity referenced to the solar value), $\log\,D_{\rm {mom}}$ (the modified wind momentum rate, see section \S \ref{dis}),  $\log L/M$ (both L and M  in solar units, where M is the stellar mass), and $Q$ (optical-depth invariant, discussed in section \S \ref{dis}). The parameters  $V_{\rm{micro}}$ and $V_{\rm{macro}}$ are   related to the  photospheric lines.  Initially $V_{\rm{micro}}$ was fixed at 10 km\,s$^{-1}$  and then varied by $\pm 5$ km\,s$^{-1}$  to achieve the best fit to the intensities of He and Si lines.

The photospheric lines of each star (at a given epoch) were modelled with the same set of  $V_{\rm{micro}}$ and $V_{\rm{macro}}$ values. 
However, to  properly match  the H$\alpha$ and \ion{He}{i}\,$\lambda$\,6678 line widths and shapes we often needed a different turbulent velocity since 
these lines are affected by the velocity dispersion of the flow.

Regarding the determination of the wind parameters, we were able to reproduce the H$\alpha$ line of many BSGs. However, in a few cases, even when the 
emission component of the P\,Cygni profile looked well fitted, it was impossible to reproduce the intensity of the absorption component.  Similar 
problems were found when reproducing the H$\alpha$ absorption profile of \object{HD\,38771}, \object{HD\,75149,} and 
\object{HD\,111973} because of the presence  of an incipient emission in the line core. Contrary to the H$\alpha$ line,  
synthetic line profiles of the photospheric lines  match the observations very well.

\subsection{Comments on individual objects}

\label{objects}

In the following we summarize the previous and current data of the stellar and 
wind parameters for each star. In addition, we complement our data with images taken with the Wide-field Infrared Survey Explorer (WISE)  
\citep{Wright2010}, covering the range 3.5 $\mu$ - 22 $\mu$ (bands W1 and W4). These images are used to identify 
former phases  of strong stellar winds.

\begin{figure*}[t]

\includegraphics[width=1.\textwidth, angle=0]{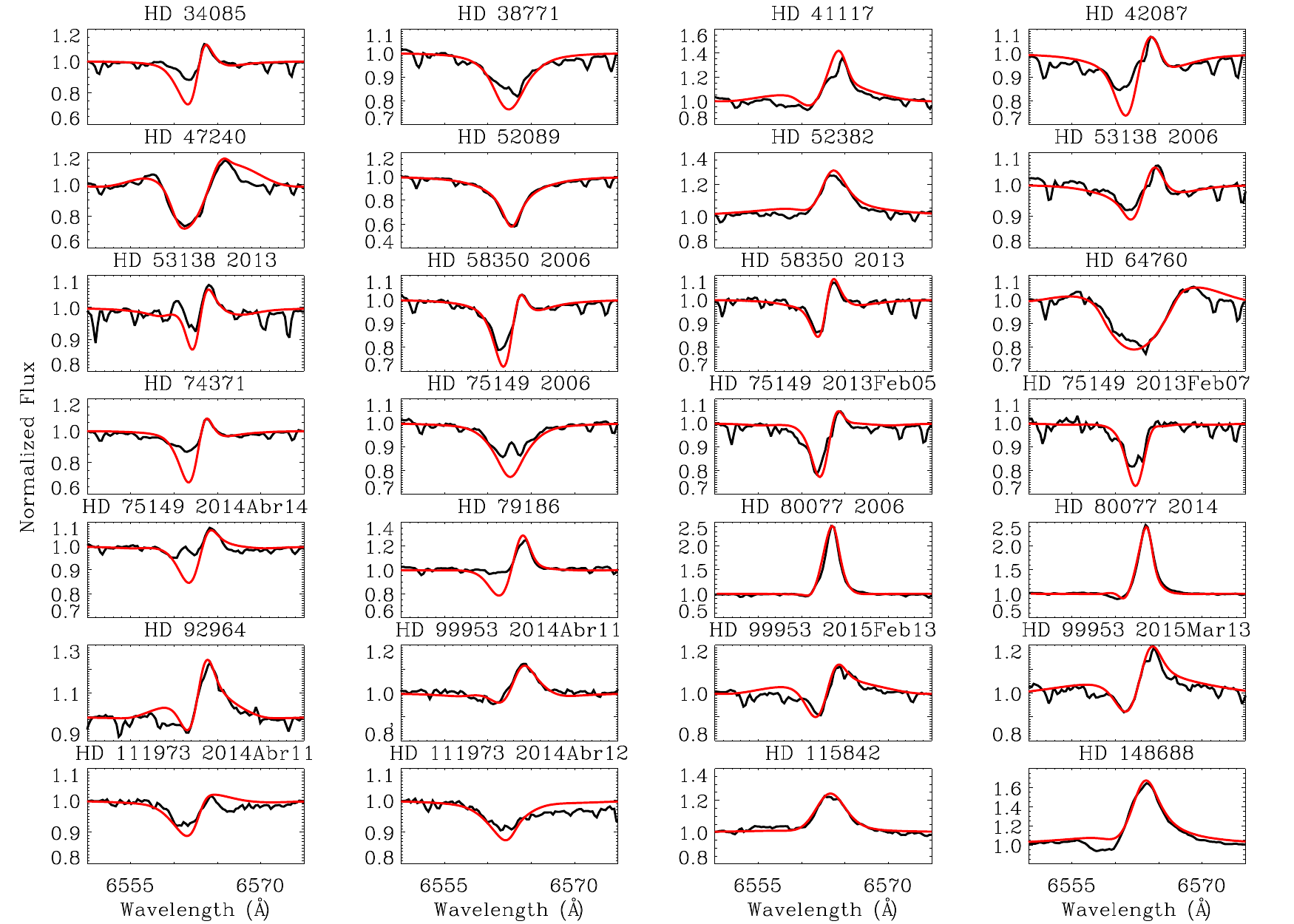}

\caption{ H$\alpha$ line and the {best-fitting} synthetic model.\label{halpha} For \object{HD\,53138}, \object{HD\,58350}, \object{HD\,75149}, \object{HD\,80077}, \object{HD\,99953,} and \object{HD\,111973} more than one plot are shown due to multi-epoch observations.}

\end{figure*}

\begin{table*}

  \tabcolsep 1.8pt

\caption{Stellar and wind parameters derived from line fitting procedures.}

\label{table:2}

\centering 

\small

\begin{tabular}{llcccccrcccrcl}

\hline\hline

STAR       & ~~~$T_{\rm {eff}}$ & $\log\,g$ & $\beta$  &           $\dot{M}$           & ~~~$V_{\infty}$& $V_{\rm{macro}}$ &  $V_{\rm{micro}}$ & $V\,\sin\,i$ & $R_{\star}$  & $\log\,(L/L_{\odot})$ & $\log\,D_{\rm mom}$ &  ~~~$\log\,(L/M)$& ~~~~~ $\log\,Q$ \\

                   &    ~~~~  K       &     dex      &          & $10^{-6} \rm M_{\odot}\,yr^{-1}$ &     ~~~$\rm km\,s^{-1}$ &     $\rm km\,s^{-1}$   &       $\rm km\,s^{-1}$       &       $\rm km\,s^{-1}$      & $\rm R_{\odot}$ & dex & dex &dex & ~~~~~~dex\\

\hline

\object{HD\,34085}   &  $12\,700\pm~~500$ &   $1.70\pm0.1$  & $2.6$    & 0.23$\pm$0.02 & 155$\pm$46        &   52$\pm$3~~      &   10 $\pm$ 2  &   $30$ &    $72$  &  $5.09\pm0.09$ & $27.22\pm0.19$  & $4.13$ & $-12.71\pm0.32$   \\

\object{HD\,38771}   &        $25\,000\pm1\,000$ &    $2.70\pm0.1$  & $1.5$    & 0.14$\pm$0.04&        1500$\pm$150        &   60$\pm$10       &   13 $\pm$ 2  &   $80$ &    $13$  & $4.78\pm0.07$  & $27.69\pm0.19$ & $4.27$ & $-13.29\pm0.24$ \\

\object{HD\,41117}   &        $19\,000\pm1\,000$ &    $2.30\pm0.1$  & $2.0$        & 0.17$\pm$0.03& 510$\pm$51         &    65$\pm$20     &   10 $\pm$ 5  &   40 &    23  & $4.84\pm0.16$  & $27.38\pm0.15$  & $4.27$ & $-12.78\pm0.23$   \\

\object{HD\,42087}   &        $16\,500\pm1000$ &      $2.45\pm0.1$  & $2.0$    & 0.57$\pm$0.05  &      700$\pm$70          &   80$\pm$15~      &   15 $\pm$ 5  &   $80$ &    $55$  & $5.31\pm0.43$  & $28.27\pm0.13$  & $3.83$ & $-13.12\pm0.23$ \\

\object{HD\,47240}   &  $19\,000\pm1000$ &         $2.40\pm0.1$  &    $1.0$    & 0.24$\pm$0.02 &       450$\pm$90          &   60$\pm$10       &   10 $\pm$ 3  &   $95$&    $30$  & $5.03\pm0.30$ & $27.57\pm0.15$ & $4.08$ & $-12.82\pm0.26$\\

\object{HD\,52089}   &  $23\,000\pm1\,000$ &  $3.00\pm0.1$  & $1.0$    & 0.02$\pm$0.006        &       900$\pm$270         &   65$\pm$5~~      &   8 $\pm$ 2  &   $10$ &    $11$  & $4.49\pm0.05$ & $26.58\pm0.28$ & $3.88$ & $-13.69\pm0.38$\\

\object{HD\,52382}   &        $21\,500\pm1\,000$ &    $2.45\pm0.1$  & $2.2$    & 0.24$\pm$0.04 &        1\,000$\pm$100      &  65$\pm$5~~      &   10 $\pm$ 2  &   $55$ &    $21$  &  $4.94\pm0.32$ & $27.84\pm0.14$ & $4.29$ & $-13.10\pm0.21$\\

\object{HD\,53138}   &        $18\,000\pm1000$ &      $2.25\pm0.2$  & $2.0$        & 0.24$\pm$0.02 &        600$\pm$120         &   50$\pm$10       &   10 $\pm$ 4   &   $40$ &    $46$  & $5.31\pm0.32$ & $27.79\pm0.14$ & $4.19$ & $-13.09\pm0.22$\\

                                    &  $18\,000\pm1000$ &       $2.25\pm0.2$  &      $2.0$    & 0.20$\pm$0.01  &     450$\pm$135  &  60$\pm$5~~      &   9 $\pm$ 3  &   $40$ &    $46$  & $5.31\pm0.32$ & $27.59\pm0.14$ & $4.19$ & $-13.36\pm0.22$\\

\object{HD\,58350}   &  $15\,000\pm~~500$ &   $2.00\pm0.1$  &     $3.0$        & 0.15$\pm$0.02 &       233     $\pm$23    &   70$\pm$10        &   12 $\pm$ 2  &   $40$ &    $54$  & $5.13\pm0.23$ & $27.21\pm0.13$ & $4.12$ & $-12.97\pm0.20$\\    

                                    &  $16\,000\pm~~500$ &      $2.00\pm0.1$  &      $3.0$        & 0.12$\pm$0.01 &  175$\pm$18         & 50$\pm$10  &   10 $\pm$ 5  &   $40$ &    $54$  & $5.24\pm0.23$ & $26.98\pm0.13$ & $4.23$ & $-12.90\pm0.20$\\  

\object{HD\,64760}   &        $23\,000\pm1\,000$ &    $2.90\pm0.1$   & $0.5$      & 0.42$\pm$0.06 &    $1\,500\pm150$      &   $100\pm50$~~    &   $15\pm5$  &   $230$ &    $12$  & $4.57\pm0.08$ & $28.14\pm0.12$ & $3.73$ &  $-12.76\pm0.18$\\

\object{HD\,74371}   &        $13\,700\pm~~500$ &     $1.80\pm0.1$   &        $2.0$        & 0.28$\pm$0.03 &        200$\pm$60  &   60$\pm$10       &   10 $\pm$ 3  &   $30$ &    $73$  & $5.23\pm0.19$ & $27.48\pm0.21$ & $4.17$ & $-12.80\pm0.33$\\  

\object{HD\,75149}   &        $16\,000\pm1\,000$ &    $2.10\pm0.1$   & $2.5$      & 0.09$\pm$0.03 &    400$\pm$40         &    62$\pm$12     &   9 $\pm$ 3  &   $40$ &    $61$  & $5.35\pm0.29$ & $27.25\pm0.11$ & $4.14$ & $-13.63\pm0.22$\\     

&        $16\,000\pm1\,000$ &   $2.10\pm0.1$  & $2.5$      & 0.20$\pm$0.01  &      350$\pm$35         &    52$\pm$12     &   11 $\pm$ 1  &   $40$ &    $61$  & $5.35\pm0.29$ & $27.54\pm0.11$ & $4.14$& $-13.19\pm0.22$\\

                       &        $16\,000\pm1\,000$ &    $2.10\pm0.1$   &        $2.5 $     & 0.16$\pm$0.05 & 400$\pm$40         & 57$\pm$7~~    &   17 $\pm$ 1  &   $40$ &    $61$  & $5.35\pm0.29$ & $27.50\pm0.11$ & $4.14$ & $-13.37\pm0.22$\\

                       &        $16\,000\pm1\,000$ &    $2.10\pm0.1$   &        $2.5$      & 0.25$\pm$0.01 &  350$\pm$35         & 55$\pm$10     &   15 $\pm$ 1  &   $40$ &    $61$  & $5.35\pm0.29$ & $27.64\pm0.11$ & $4.14$& $-13.10\pm0.22$\\

\object{HD\,79186}   &        $15\,800\pm~~500$ & $2.00\pm0.1$     &  $3.3$    & 0.40$\pm$0.02  &      400$\pm$40         &    53$\pm$7~~     &   11 $\pm$ 1  &   $40$ &    $61$  & $5.33\pm0.38$ & $27.90\pm0.11$ & $4.21$ & $-12.98\pm0.20$\\

\object{HD\,80077}   &  $17\,700\pm1\,000$ & $2.20\pm0.1$  &  $3.2$      & 5.4$\pm$0.50  &     200$\pm$20        &     $\cdots$      &   $\cdots$          &   $10$ &    $195$ & $6.53\pm0.17$  & $28.86\pm0.14$ &  $4.23$ &  $-12.15\pm0.26$ \\

                       &        $17\,700\pm1\,000$ & $2.20\pm0.1$   &   $3.0$        & 5.4$\pm$0.50  &        150$\pm$15          &   60$\pm$5~~      &   10 $\pm$ 5  &   $10$ & $195$ & $6.53\pm0.17$ & $28.98\pm0.14$ & $4.23$ & $-11.97\pm0.26$\\

\object{HD\,92964}   &        $18\,000\pm1\,000$ &    $2.20\pm0.1$   &        $2.0$    & 0.49$\pm$0.03   &     370$\pm$55         &    40$\pm$ 5~~      &   11 $\pm$ 1  &   $45$ &   $70$  & $5.67\pm0.49$ & $27.98\pm0.13$ & $4.25$ & $-12.93\pm0.25$\\

\object{HD\,99953}   &        $19\,000\pm1\,000$ &    $2.30\pm0.1$   &        $2.0$ & 0.08$\pm$0.01   &     250$\pm$50         &    50$\pm$ 5~~     &   18 $\pm$ 2  &   $50$ &    $25$  & $4.87\pm0.37$ & $26.80\pm0.14$ & $4.22$ & $-12.79\pm0.24$\\

                                    &   $19\,000\pm1\,000$ &    $2.30\pm0.1$   &     $2.0$    & $0.13\pm0.01$  &     500$\pm$100         &   50 $\pm$ 5~~      &   18 $\pm$ 2  &   $50$ &    $25$  & $4.87\pm0.37$ & $27.33\pm0.14$ & $4.22$& $-13.03\pm0.24$\\

                                    &   $19\,000\pm1\,000$ &    $2.30\pm0.1$   &     $2.0$    & $0.22\pm0.01$   &    700$\pm$140         &   50 $\pm$ 5~~      &   18 $\pm$ 2  &    $50$ &    $25$ & $4.87\pm0.37$ & $27.69\pm0.14$ & $4.22$ & $-13.02\pm0.24$\\

\object{HD\,111973}  &        $16\,500\pm1\,000$ &    $2.10\pm0.1$   &        $2.0$    & $0.21\pm0.01$ &350$\pm$70         &   57$\pm$8~~     &   12 $\pm$ 2 &   $35$ &    $46$  & $5.16\pm0.20$ & $27.51\pm0.15$ & $4.19$ & $-13.02\pm0.27$\\

                       &        $16\,500\pm1\,000$ &    $2.10\pm0.1$   &        $2.0$    & $0.14\pm0.004$ &      $350\pm105$         &   63$\pm$ 3~~     &   12 $\pm$ 2  &   $35$ &    $46$  & $5.16\pm0.20$ & $27.33\pm0.15$ & $4.19$ & $-13.19\pm0.27$\\

\object{HD\,115842}  &        $25\,500\pm1\,000$ &    $2.75\pm0.1$  & $2.5$    & 1.80$\pm$0.30  &      1\,700$\pm$340        & 63$\pm$3~~     &   23 $\pm$ 7  &   $50$ &    $35$  & $5.67\pm0.62$ & $29.06\pm0.18$ & $4.30$ & $-12.91\pm0.28$\\

\object{HD\,148688}  &        $21\,000\pm1\,000$ &    $2.45\pm0.1$  & $2.5$    & 1.15$\pm$0.20  &      1\,200$\pm$360         &        65$\pm$5~~     &   11 $\pm$ 1  &   $50$ &   $31$  & $5.23\pm0.28$ & $28.69\pm0.23$ & $4.26$ & $-12.80\pm0.36$\\

\hline

\end{tabular}

\end{table*}

\vskip 0.3cm




{\bf \object{HD\,34085}} is a B8\,Iae star that has been studied by many authors.
Using FASTWIND we estimated the stellar fundamental parameters:  $T_{\rm {eff}}$ = 12\,700 K and $\log\,g$ = 1.7. The best-fitting model to the SED was 
obtained with the BeSOS interactive interface using a Kurucz model with $T_{\rm {eff}}$ = 11\,760 K, $\log\,g$ = 2.0,  $R_\star$ = 71 R$_\odot$,  
$E(B-V)$ = 0.044 (calculated as explained in section \S \ref{mod}), and  $d$ = 259 pc (which is close to the distance of 264 pc given by HIPPARCOS). Using the HIPPARCOS distance and the mentioned value of $E(B-V)$,  we estimated $M_{\rm{bol}}$=$-7.95$ mag and  $R_\star$ = 70 R$_\odot$; instead, from the angular diameter \citep[2.713 
mas,][]{Zorec2009} we calculated $R_\star$ = 76 R$_\odot$. We assumed  a mean value for
this star of $R_\star$ = 72 R$_\odot$.

Our estimate of $T_{\rm {eff}}$  (12\,700 K) agrees with previous determinations (see Table \ref{table:A1}). This star shows a photometric variation 
of 2.075 days  \citep{Lefevre2009}. Moreover, \citet{Moravveji2012} found 19 significant pulsation modes from radial velocities with variability 
timescales ranging from 1.22 days  to 74.74 days.
It also presents a variable stellar wind with  changes in the  mass-loss rate of at least 20\% on a timescale of one year \citep{Chesneau2014}. Using 
spectro-interferometric monitoring, these authors found  time variations 
in the differential visibilities and phases. For some epochs, the temporal evolution of the signal suggests the rotation
of circumstellar structures. However, at some periods, no phase signal was observed at all. This result was interpreted in 
the context of second-order perturbations of an underlying spherical wind.

The  H$\alpha$ line is highly variable \citep[with profiles in absorption, filled in by emission, double-peaked, 
or inversed P Cygni;][]{Chesneau2014}. Our observation resembles the one reported by 
\citet{Przybilla2006}. Our estimate for the mass-loss rate is $\sim 1.5$ times lower  than the value obtained by \citet{Markova2008} ,
and at least one-third of the value reported by \citet{Chesneau2014}. We also derive a low terminal velocity. This value
is quite 
uncertain because we were not able to fit the absorption component of the  H$\alpha$ P Cygni profile.  Moreover, a high 
$V_{\rm {macro}}$ (85 km s$^{-1}$) was needed to reproduce the emission component of the line profile. The photospheric lines were modelled 
very well, 
with the exception of the \ion{He}{i} $\lambda$ 4471 line, which does not exhibit the forbidden component. 

\vskip 0.3cm




{\bf \object{HD\,38771}} (B0.5Ia): The stellar and wind parameters of this star have been derived by many authors: 
\citet{Nerney1980, Garmany1981, Lamers1982, Kudritzki1999, Crowther2006, Searle2008,Zorec2009}. Their values range from 26\,000 K to 
27\,500 K in $T_{\rm{eff}}$, from 2.9 dex to 3.07 dex in $\log\,g$, from 13.0 $R_{\odot}$ to 28 $R_{\odot}$ in stellar radius, 
from 0.27\,$\times\,10^{-6}$ $\rm M_{\odot}\,yr^{-1}$ to 1.20\,$\times\,10^{-6}$ $\rm M_{\odot}\,yr^{-1}$ in $\dot{\rm M}$, and from 
1\,350 $\rm km\,s^{-1}$ to 1\,870 $\rm km\,s^{-1}$ in $V_{\infty}$ (see Table \ref{table:A1}). The star presents a variable magnetic 
field \citep{Nerney1980} and exhibits spectral variations. The observed H$\alpha$ line is quite variable,  
showing a pure absorption profile \citep{Kudritzki1999}, a double-peaked absorption profile 
with a central emission \citep{Rusconi1980}, and an absorption profile with a strong central  emission \citep{Crowther2006}. 
\citet{Morel2004} also described changes in morphology and profile amplitude of 32.6\%, which clearly suggest 
variations in the wind conditions. These authors also found two photometric periods of 4.76 days and 1.047 days and \citet{Prinja2004} 
reported additional spectroscopic periods of 1.9 days, 6.5 days,  and 9.5 days.

At the time of our observation the H$\alpha$ line displayed an asymmetric absorption profile with a weak  emission in the core 
(Fig. \ref{halpha}). We also note that the core of the  H$\beta$ line might be filled in by an incipient emission 
(Fig. \ref{fig:7}). Extra broadening effects are present in all the photospheric lines  leading to large $V_{\rm{macro}}$ estimates 
(of the order of  the projected rotation velocity).

We derived $T_{\rm {eff}}$ = 25\,000 K  using   \ion{Si}{ii} and \ion{Si}{iii,} and also using  the \ion{He}{i} and \ion{He}{ii} ionization balance. 
The best-fitting SED model was obtained with a TLUSTY model  for $T_{\rm {eff}} = 25\,700$ K, $\log\,g = 2.70$, $R_{\star}$ = 14 R$_\odot$, and 
$E(B-V) = 0.05$ mag for a  distance of 191 pc \citep[close to the HIPPARCOS distance of $d_{\rm H}$= 198 pc; ][]{VanLeeuwen2007}.  Using  $d_{\rm H}$ 
in  Pogson's formula we derived a  $M_{\rm{bol}}$ = -6.99 mag and  $R_{\star} \sim 12$ R$_\odot$. The adopted stellar radius,  
$R_{\star} \sim 13$ R$_\odot$, which is consistent with the value derived from  the angular diameter calculated by \citet[][0.62 mas]{Zorec2009}.

 Our $T_{\rm {eff}}$ value is 1\,000 K lower than the values obtained 
by \citet{Searle2008} and \citet{Gathier1981}, while the stellar radius is similar to that  obtained by \citet{Kudritzki1999}. 
Compared with previous determinations, our mass-loss rate has the lowest value. As H$\alpha$ is in pure absorption  $\dot{M}$ might 
have a large uncertainty.


Considering the diversity of the  stellar and wind  parameters found in the literature, and  that 
the H$\alpha$ line profile shows important variations, this star is a good candidate to search for pulsating or magnetic activity connected to  
cyclic wind variability.

\vskip 0.3cm




{\bf \object{HD\,41117}} (B2Ia): This is another deeply studied star  \citep{Zorec2009,Crowther2006,Morel2004,Kudritzki1999,Scuderi1998,Nerney1980}. \citet{Morel2004} found H${\alpha}$ variations, both in shape and intensity, and reported light variability on a 
period of 2.869 days. Additional sets of spectroscopic periods were obtained from H$\alpha$ equivalent width time series \citep[][Table 1]{Morel2004}.

From Table \ref{table:A1} we see that the fundamental stellar parameters and luminosity are  consistent with those already published, in 
contrast with the  wind parameters (mainly $\dot{M}$) that are quite different (by a factor of up to 3.6).

The best-fitting SED model corresponds to  $T_{\rm {eff}} = 17\,940$ K, $\log\,g = 2.2$, $R_{\star}$ = 26 R$_\odot$, $E(B-V) = 0.42$ mag 
\citep[instead we derived  $E(B-V) = 0.46$ mag  from][]{Flower1996} and $d$= 477 pc (while the distance measured by HIPPARCOS is $552\pm85$ pc). On the  other hand, using   $d=447$ pc and $E(B-V)$,   the angular diameter \citep[0.371 mas,][]{Zorec2009}, or the computed bolometric magnitude ($-7.35$ 
mag)  the stellar radius is  $R_{\star} \sim 23$ R$_\odot$. We  therefore adopted $R_{\star}$ = 23 R$_\odot$, which  is also consistent with the stellar 
radius expected from  the HIPPARCOS distance. The stellar radius is lower than the value derived by other authors (see Table \ref{table:A1}); however, if 
we use the distance derived by \citet[][1.62 kpc]{Megier2009}  we obtain $R_{\star}$ = 62 R$_\odot$.

From the line modelling we obtained higher $T_{\rm {eff}} = 19\,500$ K and $\log\,g = 2.3$ values.
Our spectrum reveals a H${\alpha}$ emission line with a complex blue-ward absorption component (see Fig. \ref{halpha}). 
The mass-loss rate derived here  is lower (by a factor of about 5) than the  value reported by \citet{Kudritzki1999}. We also obtained a lower terminal 
velocity. The rest of the lines are 
in pure absorption and are excellently modelled (see Fig. \ref{fig:7}). 
Values of  $V_{\rm{micro}}$ and $V_{\rm{macro}}$ as high as the rotational velocity of the star  ($V\,\sin\,i$ = 40 km\,s$^{-1}$) are needed to 
reproduce the extra broadening of the photospheric lines. A larger macroturbulent velocity (65 km\,s$^{-1}$) was also obtained by \citet{IACOB}. 


\vskip 0.3cm




{\bf \object{HD\,42087}} (B4Ia):
The SED was fitted with a TLUSTY model using $T_{\rm {eff}}$ = 15\,000 K, $\log\,g = 2.39$, $R_{\star}$ = 60 R$_\odot$, $E(B-V) = 0.34$ mag,  and $d$= 2\,021 pc 
\citep[which is similar to the distance of 2\,075 pc derived by] [from the Ca II H+K lines]{Megier2009}. 
On the other hand, a distance $d$ = 2\,075 pc gives  $M_{\rm{bol}}=-8.33$ mag and $R_{\star}$ = 50 R$_\odot$, while the angular distance  
calculated by \citet{Zorec2009} leads to  $R_{\star}$ = 57 R$_\odot$. We adopted a mean value of $R_{\star}$ = 55 R$_\odot$.

Our optical spectrum only covers  the H$\alpha$ region; therefore, we were not able to determine the effective temperature using the silicon lines. Thus, we adopted  the value calculated by \citet{Zorec2009} 
($T_{\rm {eff}}$ = 16\,500 K) and derived from the H lines a $\log\,g=$ 2.45. This  $T_{\rm {eff}}$ is lower than the value reported by \citet{Benaglia2007} and \citet{Searle2008}, and greater than the value we derived from the fitting of the SED.

This star presents a significant H$\alpha$ variability of 91.2\% and both the H$\alpha$ and \ion{He}{i}\,$\lambda$6678 lines show a cyclic behaviour on 
a  periodicity  of $\sim$ 25 days \citep{Morel2004}. This period is larger than the 6.807-day period found from the HIPPARCOS light curve \citep{Morel2004, Lefevre2009}.

Our H$\alpha$ line shows a P Cygni feature with a weak emission and a broad absorption component with a superimposed incipient emission structure. 
We fitted the H$\alpha$ line profile using a wind model with $\dot{M} = 0.57\,10^{-6}$ M$_\sun$ yr$^{-1}$ and  $V_\infty$= 700 km s$^{-1}$. The obtained $V_\infty$ 
is similar to the one measured in the  UV spectral range \citep[735 km s$^{-1}$,][]{Howarth1997}. 
\vskip 0.3cm




{\bf \object{HD\,47240}} is a fast rotating B1Ib star \citep[$V\,\sin\,i = 114$ km\,s$^{-1}$,][]{IACOB} that lies behind the Monoceros Loop supernova 
remnant (SNR). It presents very broad photospheric absorption 
lines, periodic light variations of  2.742 days \citep{Lefevre2009, Morel2004}, periodic 
motions due to binarity, and the presence of discrete absorption components  
\citep[DACs,][]{Prinja2002} that appear double.  We were able to get a good fit to the SED for the following parameters: 
$T_{\rm {eff}}$= 17\,500 K, $\log\,g$= 2.4, $E(B-V)$= 0.33 mag, $d$= 1\,515 pc, and $R_\star$ = 35 R$_\odot$. The derived distance is consistent with 
the distance estimate of 1\,598 pc by \citet{Megier2009} and the colour excess also matches the 2\,200 \AA\, bump. We also calculated a 
$M_{\rm {bol}}$= $-7.66$ mag and $R_\star$ = 28 R$_\odot$ (for $d$ = 1\,598 pc). This stellar radius agrees with the computed value from the angular 
diameter (0.157 mas) by \citet{Zorec2009}. We adopted a mean value of $R_\star$ = 30 R$_\odot$.

The stellar parameters  derived with FASTWIND agree with the values given in one
of the models calculated by \citet{Lefever2007} 
(who  found two models with very different $T_{\rm {eff}}$ and $\log\,g$ and both fit very well the \ion{Si}{iii} line 
profiles). The $T_{\rm{eff}}$ obtained here is lower than the 21\,670 K derived with the BCD spectrophotometric method \citep{Zorec2009}.

To model the photospheric lines we needed a high $V_{\rm {macro}}$ value (60 km\,$\rm s^{-1}$) that is much higher 
than those used to model the H lines formed in 
the wind (3 km\,$\rm s^{-1}$). As this object is a pulsating variable star of $\alpha$\,Cyg type,  a  high $V_{\rm{macro}}$ value is expected.

Our spectrum shows H$\alpha$ as a double-peaked emission line with an intense central absorption. 
This profile was considered as an indicator of the presence of a disk-like structure
\citep{Lefever2007}. We achieved a good fit to the observed H$\alpha$ line 
 for  the intensity of the absorption and for the two emission components. We derived a
 mass-loss rate of 2.4\,$\times\,10^{-7} \rm M_{\odot}\,yr^{-1}$. This value  agrees with the upper limit of 
the range reported by \citet[][1.7\,$\times\,10^{-7} \rm M_{\odot}\,yr^{-1}$ -- 2.4\,$\times\,10^{-7} \rm M_{\odot}\,yr^{-1}$]{Lefever2007} and is lower than the 
value calculated by \citet[][0.31\,$\times\,10^{-7} \rm M_{\odot}\,yr^{-1}$]{Morel2004} from the theoretical WLR for OBA supergiants. For the terminal velocity, we found 
almost half of the value obtained by \citet{Lefever2007,Prinja2013}. 

\vskip 0.3cm




{\bf \object{HD\,52089}}  (B1.5 II) is the brightest EUV source in the sky. It shows a UV emission line spectrum and X-ray emission consistent 
with a wind-shock model. Using the observed SED and TLUSTY models we derived $T_{\rm {eff}}$= 21\,000 K, $\log\,g$ = 3.0, and 
$R_\star$ = 11 R$_\odot$ for a distance of 124 pc \citep{VanLeeuwen2007} and  $E(B-V)$ = 0.00 mag.  This stellar radius is also consistent with the 
bolometric magnitude ($M_{\rm {bol}}$= -6.18 mag) and the  measurement of the angular diameter \citep[0.801 mas,][]{Zorec2009} that lead to 10 R$_\odot$ 
and 10.6 R$_\odot$, respectively. From FASTWIND we derived a higher  $T_{\rm{eff}}$ (23\,000 K) that agrees with the values given in 
\citet{Lefever2010, Zorec2009}, and \citet{Morel2008}, but it is higher than the value of 20\,100 K obtained by \citet{Fraser2010}. Our surface gravity 
value (3 dex) agrees more closely with the \citet{Fraser2010} estimate and it is 0.2 dex lower than those reported by \citet{Lefever2010} and  \citet{Morel2008}. 
The star has a longitudinal magnetic field of 149\,G \citep[][]{Morel2008}.
Our spectrum displays all the H lines in absorption. Using optical data we derive a mass-loss rate of $2\,\times\,10^{-8}$ M$_\odot$yr$^{-1}$, which  is 
similar to  the  value obtained phenomenologically  by \citet[][$3\,\times\,10^{-8}$ M$_\sun$yr$^{-1}$ -- $8\,\times\,10^{-8}$ M$_\sun$yr$^{-1}$]{Cohen1996}.

\vskip 0.3cm




{\bf \object{HD\,52382}} (B1Ia) is an O-B2 runaway candidate \citep{Peri2012}.
Using  FASTWIND to fit the  ratios of  Si\,III/Si\,II and He\,II/He\,I we obtain   $T_{\rm {eff}}$ = 21\,500 K, and from the H$\gamma$ and H$\delta$ lines a 
$\log\,g$ = 2.45. From  \citet{Flower1996} we have $(B-V)_0 $= -0.20 mag and derive a colour excess of 0.39 mag. The best fit to the observed SED corresponds to a TLUSTY model 
with $T_{\rm {eff}}$= 23\,140 K,  $\log\,g$ = 2.47, $R_\star$ = 21.6 R$_\odot$, and $d$ = 1.23 kpc, but with a larger $E(B-V)$ = 0.44 mag.  
This distance agrees with  the 1.1 kpc found by \citet{Megier2009} and  
with the parallactic distance of 1.3 kpc measured by Gaia  \citep[0.768 mas;][]{Gaia2016}. Using the distance of 1.3 kpc in  Pogson's formula and the 
expression for the angular radius \citep[with $0.17$ mas;][]{Pasinetti2001}, we obtain a stellar radius of 18.6 R$_\odot$ and 23.5 R$_\odot$, respectively. Therefore, 
we adopted  R$_\odot$= 21 R$_\odot$ as the mean value.  We rejected the HIPPARCOS distance of 471 pc because it leads  to a value of $M_{\rm {bol}}$  that is too low for a B supergiant.

\begin{figure}[t]
\includegraphics[width=0.48\textwidth, angle=0]{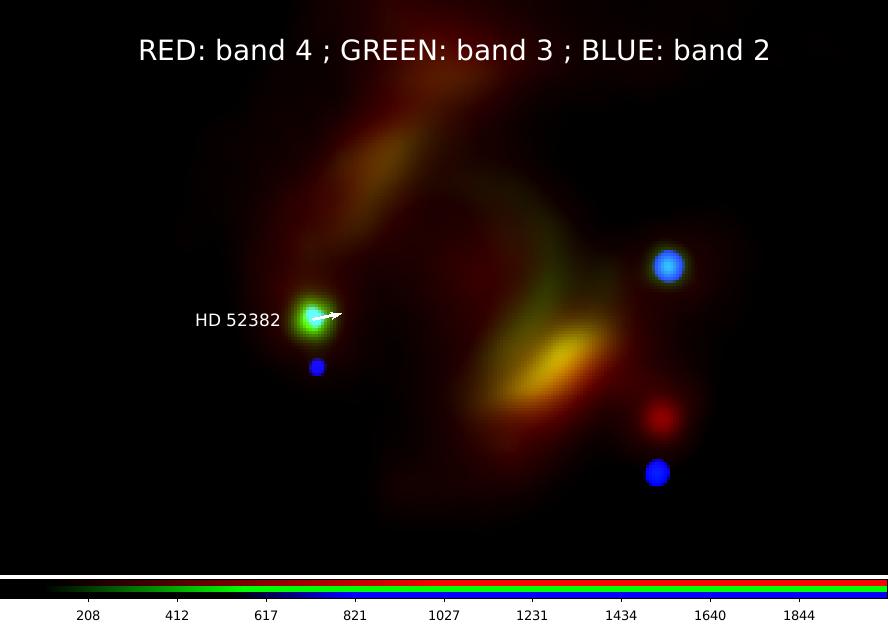}
 \caption{WISE images (W2 = 4.6 $\mu$m in blue, W3 = 12 $\mu$m in green, and W4 = 22 $\mu$m in red) showing a bow-shock structure near \object{HD\,52382}.
   The white arrow indicates the direction of the proper motion of the star. \label{WISE-HD52382}}

\end{figure}

 The derived $T_{\rm{eff}}$ = 21\,500 K is between  the values calculated by \citet{Lefever2010} and \citet{Kri2001}.
The star presents a variable H$\alpha$ line profile displaying a P\,Cygni or a pure emission feature \citep{Morel2004}. Our spectrum shows a pure emission 
line. 
We were able to reproduce all the spectral lines very well. We derived a terminal
velocity of 1\,000 km\,$\rm s^{-1}$, which is of the same order as that obtained from  the UV region \citep[900 km\,$\rm s^{-1}$,][]{Prinja1990},  but 
200 km\,$\rm s^{-1}$ lower than the value found by \citet{Kri2001} for a similar mass-loss rate.
The combined WISE W2, W3, and W4 images (Fig. \ref{WISE-HD52382}, at 4.6 $\mu$m, 12 $\mu$m, and 22 $\mu$m, respectively) reveal a complex structure 
close to the star and in the direction of its proper motion (indicated in Fig. \ref{WISE-HD52382} with a white arrow) that could be related to wind  interactions with the ISM, due to high mass-loss 
episodes in the past.

We want to stress that for an optimal fitting of the photospheric lines we have to introduce 
high values of $V_{\rm{macro}}$,  of around 65 km $\rm s^{-1}$ \citep[a similar result was reported by][]{Lefever2010}, 
which could be an indication of pulsation activity. 
The \ion{He}{i} $\lambda$ 6678 line shows a little emission in the core.
\vskip 0.3cm

{\bf \object{HD\,53138}} (B3Ia) has been studied by several authors (see summary in Table \ref{table:A1}).
The derived $T_{\rm {eff}}$ ranges from 15\,400 K to 18\,500 K, and $\log\,g$ from 2.05 dex to 2.35 dex. 
The star exhibits irregular variations in the H$\alpha$ line \citep{Morel2004}. Our spectra also show 
variations: in 2006 the star displayed a typical P\,Cygni profile, while in 2013 a double-peaked emission was observed. We also note differences in the 
intensities and line widths of the H$\gamma$ and the \ion{Si}{iii} lines for the two epochs that lead to slightly different values of 
$T_{\rm{eff}}$ and $\log\,g$.  We obtained for both epochs the following fundamental parameters: 18\,000 K for $T_{\rm{eff}}$ and  2.25 dex for $\log\,g$ with uncertainties of $1\,000$ K and $0.2$ dex. The best-fitting SED model suggests  
$T_{\rm{eff}}$ =  $18\,000$ K, $\log\,g$= $2.2$,  $R_\star$= $46$ R$_\odot$, $d$ = 822 pc (close to 847 pc the HIPPARCOS estimate), and  $E(B-V)$= 0.10 \citep[lower than the colour excess of  0.131 mag 
we obtained from][]{Flower1996}.  The calculation of   $R_\star$ by means of  the angular diameter gives 51 R$_\odot$, while the computed $M_{\rm {bol}}$ (-8.53 mag) suggests  $R_\star$ = 46  R$_\odot$.  We adopted  $R_\star$= 46  R$_\odot$

The photospheric lines are broadened by macroturbulent motion ($V_{\rm{macro}}$ = 60 km 
s$^{-1}$). This high value of  $V_{\rm{macro}}, $ which is almost twice the measured value for the 
projected rotation velocity, is consistent with the suggestion of a pulsating variable star of 
$\alpha$\,Cyg type \citep{Lefevre2009}.

Our model fits  the observed \ion{He}{i}\,$\lambda$\,4471\,\AA\, line and the emission component of H$\alpha$ fairly well, 
although we were not able to reproduce  the absorption component of the profile in 2006 or the emission component 
observed in 2013. Our mass-loss  estimates are in the range of previously determined values of other authors, being higher in 2006. 
The terminal velocity measured in the spectrum of 2013 is lower than in 2006 (see Table \ref{table:A1}). \citet{BC77} measured a 
significant infrared excess at 10 $\mu$m requiring a mass-loss rate  a factor of 20 times higher than ours to account for it.
\vskip 0.3cm




{\bf \object{HD\,58350}} (B5Ia)  is an MK standard star. The set of stellar parameters derived with FASTWIND ($\log\,g$ and $T_{\rm{eff}}$) agree very well with those derived by \citet{McErlean1999,Searle2008,Fraser2010}. 
We were not been able to obtain a good fit to the SED using simultaneously the UV, visual, and IR photometry. However, based on the distance given by  HIPPARCOS (609 pc) and 
the calculated angular diameter \citep[0.882 mas,][]{Zorec2009}, we obtained $R{_\star}$ = 57 R$_{\odot}$. Based on HIPPARCOS parallax measurement and the derived $E(B-V)$ = 0.083 
mag from \citet{Flower1996}, we obtained $M_{\rm {bol}} = -7.97$ mag and $R{_\star}$ = 51 R$_{\odot}$. We adopted $R_\star$= 54 R$_\odot$ as the mean value.

The H$\alpha$ line shows a variable P\,Cygni profile with a tiny emission component. We were able to match 
the observations taken in 2006 and to get a fairly good fit to the one acquired in 2013. In the former, we failed to fit the absorption component. Our mass-loss rate is similar to that given by \citet{Lefever2007} and lower than the value given by 
\citet{Searle2008}.
As these last authors do not show the stellar spectrum in the H$\alpha$ region, we cannot discuss the origin of the 
discrepancies.
\citet{Morel2004} show time-series of the H$\alpha$ line where the profile is seen in absorption, while  the observations of \citet{Ebbets1982} display a P Cygni profile.  The light curve  presents one period of 4.70 days \citep{Koen2002} and another one of 6.631 days \citep{Lefever2007}.

\vskip 0.3cm




{\bf \object{HD\,64760}} (B0.5Ib) is a  rapid rotator. The star was extensively studied as part of the IUE `MEGA Campaign' by \citet{massa1995} and  
\citet{Prinja1995}. The line variability observed in the UV argues in favour of rotationally modulated wind variations.
It displays a double-peaked emission line profile in H$\alpha$ and very  
broad absorption lines of H and He. Direct observations reveal  a connection between multi-periodic non-radial  pulsations  
(NRPs)  in  the  photosphere  and  spatially structured winds \citep{Kaufer2006}. These observations also seem to be compatible with the presence of  
co-rotating interaction regions.

The fitting of the SED provides the following parameters: $T_{\rm {eff}}$ = 22\,370 K, $\log\,g$= 2.50, $R_\star$= 15 R$_\odot$,  $E(B-V)$= 0.07 mag, and $d$= 486 pc 
(while HIPPARCOS parallax gives $d$=507 pc). The derived  $M_{\rm {bol}}$ = -6.72 mag gives $R_\star$= 12 R$_\odot$, while the angular diameter leads to $R_\star$= 10 
R$_\odot$. We adopted   $R_\star$= 12 R$_\odot$ as the mean value.

Our stellar parameters derived with FASTWIND ($T_{\rm {eff}}$ = 23\,000 K, $\log\,g$= 2.90) agree fairly well with 
those of \citet{Lefever2007}, but show large discrepancies with the values reported by  \citet{Searle2008}. Although 
\citet{Searle2008} do not show the spectrum, they  mentioned some difficulties in deriving accurate values from the models due to the large 
width of the spectral lines.

\begin{figure}[t]

\includegraphics[width=0.48\textwidth, angle=0]{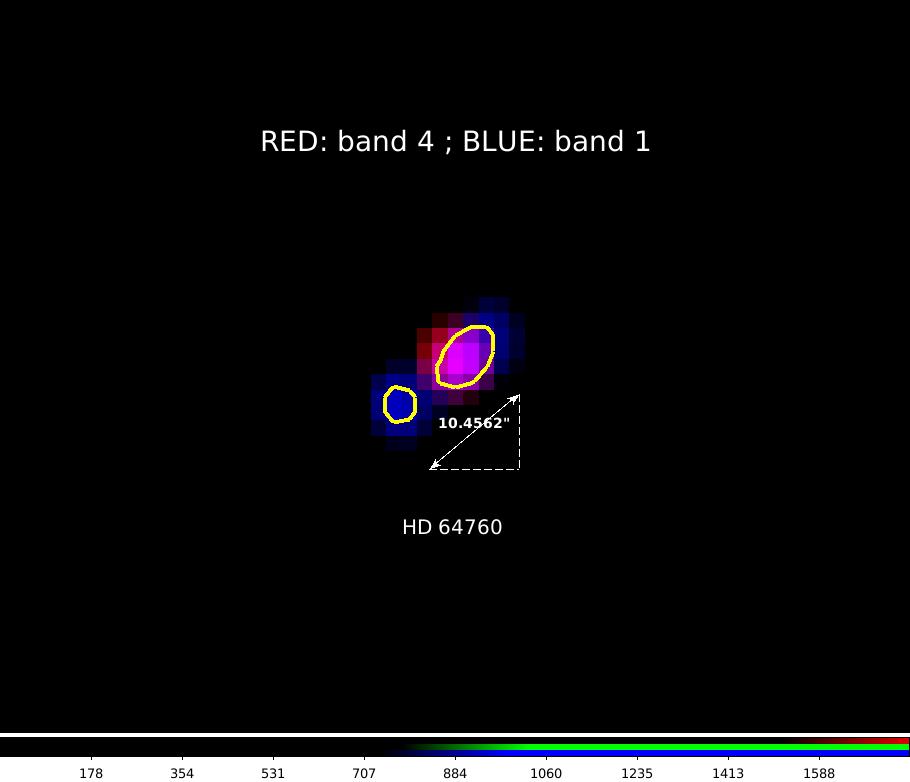}

\caption{WISE W1 and W4 images (3.4 $\mu$m and 22 $\mu$m, respectively) showing a double component or a wind-lobe structure.
Yellow curves represent regions with the same brightness. \label{WISE-HD64760}}

\end{figure}

In general, we obtained very good fits for all the lines, even for the double-peaked H$\alpha$ profile. Our model gives the same set of wind parameters as 
\citet{Lefever2007}, but for the stellar radius a half of the value published by these authors was obtained.
The combined WISE W1 and W4 images (3.4 $\mu$m and 22 $\mu$m, respectively) show the presence of either a double visual component or a wind-lobe structure 
(see Fig. \ref{WISE-HD64760}).  
\vskip 0.3cm

{\bf \object{HD\,74371}} (B6Iab/b) displays light variations with periods of $5-20$ days \citep{vanGenderen1989} and 8.291 days   
\citep{Koen2002,Lefevre2009}. The stellar parameters $T_{\rm{eff}}$ and $\log\,g$ were determined by \citet{Fraser2010} and agree very well with the values that we derived with FASTWIND ($T_{\rm {eff}}$= 13\,700 K, $\log\,g$= 1.8).  From the SED we were not able to get a good fit using  the HIPPARCOS parallax \citep[0.43 mas, d = 2.3 kpc][]{VanLeeuwen2007}. The best fit was achieved 
using a distance of 1.8 kpc,  $E(B-V)=0.35$, $T_{\rm {eff}}$= 13\,800 K, $\log\,g$= 2.0, and  $R_\star$= 73 R$_\odot$. The found distance of 1.8 kpc agrees with the 
value reported by  \citet[][1.9 kpc]{Humphreys1978}.

Our model fits  all the lines very well, with the exception of the absorption component of the H$\alpha$ P\,Cygni profile and 
the core of the H$\beta$ line, which seems to be filled in by an incipient emission (Fig. \ref{fig:10}). This is the first time that the wind parameters of 
\object{HD\,74371} have been determined (see Table \ref{table:2}).

\vskip 0.3cm




{\bf \object{HD\,75149}} (B3Ia):  The stellar parameters obtained in this work with FASTWIND agree very well with those given by \citet{Lefever2007}, \citet{Fraser2010}, 
and our SED model with $T_{\rm {eff}}$ = 15\,000 K, $\log\,g$ = 2.12, $R_\star$ = 71 R$\odot$, $E(B-V)=$ 0.46 mag, and a distance of 1\,642 pc. We were not able to 
fit the SED with the parallax measurement of 0.37 mas (2.7 kpc) given by HIPPARCOS, but our value of $d$ is within the corresponding error bars (1.4 kpc). The derived 
distance of 1.64 kpc leads to $M_{\rm V} = -7.05$, which is in agreement with the value calculated by \citet[][-7.0]{Humphreys1978},  $M_{\rm{bol}} = -8.45$ and $R_\star$ = 56 R$\odot$.  The angular diameter also gives $R_\star$ = 58 R$\odot$. We adopted a mean value of $R_\star$ = 61 R$\odot$.
The star displays small amplitude light variations with a period of 1.086 days
\citep{Koen2002,Lefevre2009}, in addition to the  variability of 1.2151 days and 2.2143 days reported by   \citet{Lefever2007}.

From our spectra,
we found that the star shows  important variations in the H$\alpha$ line: an absorption line profile with a weak emission at 
the core is seen in 2006. A P\,Cygni feature  is seen on 2013 February 5; it turns into
an absorption profile two days later. In 2014 a P Cygni profile with a complex absorption component is seen again (see Fig. \ref{halpha}); this profile also
presents two weak emission components that resemble the one published by \citet{Lefever2007}. Using the spectrum taken in 2006 the derived  $\dot{M}$ is similar to 
the value reported by these authors, although our value of $V_{\infty}$ is a bit lower. To explain the line profile changes we have to assume a 
variable wind structure with a mass-loss enhancement of a factor of about 1.8 and 2.2, between 
our observation in 2\,013 and in 2\,014, and of a factor of about 2.8, between data taken in 2\,014 and 2\,006.  We want to stress that in order to model the 
H$\alpha$ line width at various epochs, it was necessary  to consider different and also large values for $V_{\rm{macro}}$. The terminal velocity is very similar 
in all the models. Nevertheless, we have to keep in mind that 
the values derived from a pure absorption profile are more uncertain since the line is less sensitive to  the wind conditions. 

We were able to model  all the photospheric lines quite well with the exception of the H$\beta$ line core observed in 2006, which is weaker. 

\vskip 0.3cm




{\bf \object{HD\,79186}} (B5Ia ): From the SED we were able to match a TLUSTY model using the following parameters: $T_{\rm{eff}}$ = 15\,000 K, $\log\,g$ = 2.12, $R_\star$ = 67  R$_\odot$, and $d$= 1.42 kpc, adopting $E(B-V)$= 0.35 mag. The distance is very similar to  that obtained from the HIPPARCOS parallax 
(1.45 kpc), which yields $M_{{bol}}$= $-8.26$ mag and  $R_\star$ = 53 R$_\odot$, while the angular diameter (0.4 mas) gives  61 R$_\odot$.  We used a mean stellar 
radius of 61 R$_\odot$.

Our estimation of $T_{\rm {eff}}$ is slightly higher than the values given by \citet{Fraser2010} and \citet{Prinja2010}, respectively  $\Delta\,T\,\sim\,700\,K$ and 
$\Delta\,T\,\sim\,800\,K$,  and even higher than those obtained by \citet{Kri2001} and \citet{Under1984}. 

The logarithm of the surface gravity agrees with the unique
value (2.0 dex) available in the literature \citep{Fraser2010}. The stellar radius of our model is similar to that estimated by \citet{Under1984}.

Although we obtained a very good fit for all the photospheric lines, we were not able to reproduce the shallow absorption component of the H$\alpha$ 
P Cygni profile. Nevertheless, the terminal velocity agrees with the model parameters given by \citet{Kri2001} and with measurements  from UV 
observations \citep[435 km s$^{-1}$,][]{Prinja2010}. Our $\dot{M}$ is slightly lower than the value given by  \citet{Kri2001}.

\vskip 0.3cm



{\bf \object{HD\,80077}} (B2Ia+e) might be a member of the open cluster \object{Pismis 11}, located at a distance of 3.6 kpc. 
With an absolute bolometric magnitude of -10.4, this star is among the brightest of the known B-type supergiants in our Galaxy \citep{Knoechel1982, Marco2009}.  

\citet{Carpay1989,Carpay1991} detected light variations with an amplitude $\sim$ 0.2 mag and suggested 
that the star could be a luminous blue variable (LBV).
Using HIPPARCOS and V photometric data, \citet{vanLeeuwen1998} obtained a periodogram that reveals that the most significant peaks are those near 66.5 days and 55.5 days, and peaks with much lower significance near 76.0 days  and 41.4 days. 

Light variations of 0.151 mag over a period of 3.115 days  were found by \citet{Lefevre2009} and  another period of 21.2 days   was determined from the polarimetry \citep{Knoechel1982}. 
From the SED we obtained  $T_{{\rm eff}}$ = 18\,000 K,  $\log\,g$ =  2.17,  $E(B-V)= 1.5$ mag,  $R_\star$ = 200 R$_\odot$, and  $d$ = 3\,600 pc (while the HIPPARCOS 
distance is 877 pc). Using the distance to the Pismis 11 cluster and the obtained $E(B-V),$ we calculated  $M_{\rm{bol}}$ = $-11.49$ mag and $R_\star$ = 187 R$_\odot$. We 
adopted a stellar radius of  195 R$_\odot$.

\citet{Carpay1989} derived $T_{\rm{eff}}$ = 17\,700 K,  $\log\,g$ = 2, and a mass-loss rate of 
5.11\,$\times\,10^{-6}$ M$_{\sun}$\, yr$^{-1}$. Based on radio data, \citet{Benaglia2007} estimated a significantly lower mass-loss rate 
(1.7 \,$\times\,10^{-6}$ M$_{\sun}$\, yr$^{-1}$)  that agrees with our result (see Table \ref{table:A1}). We observed a P Cygni  H$\alpha$ profile in the spectra 
taken in 2006 and 2014. Small changes in the emission and absorption components can be seen  in Fig. \ref{halpha}.

Finally, we  want to stress that \object{HD\,80077} is a very massive post-main sequence star and, hence, an enhancement of the He abundance is expected. However, we do not observe any  noticeable  contribution of  forbidden components and the \ion{He}{i}\,$\lambda$\,4471 line was well-matched with a solar He abundance model.

\vskip 0.3cm




{\bf \object{HD\,92964}} (B2.5Ia) displays photometric variations with  periods of 2.119 days and 14.706 days  \citep{Lefevre2009}. The H$\beta$ and H$\gamma$ 
line profiles  show asymmetries.  \citet{Lefever2007} ascribed this behaviour 
to the strong stellar wind which  affects the photospheric lines. The same  authors also noted  that the  \ion{He}{i} $\lambda$ 6678 line requires 
a $V_{\rm {macro}}$ that is two times higher than the value derived for the Si lines.

The stellar parameters derived from the SED are $T_{{\rm eff}}$ = 18\,000 K, $\log\,g$ =  2.19, $E(B-V)= 0.481$,  $R_\star$ = 76 R$_\odot$, and  $d = 1\,806$~pc 
(close to the HIPPARCOS distance,  $d$ = 1\,851 pc). We used the extinction curve from \citet{vanBreda1974}, who found $R=3.5$ for this star.  From the HIPPARCOS 
distance and the calculated $A_{\rm V}$= 1.68 mag  we obtained  $M_{\rm{bol}}~=~-9.14$ mag and $R_\star$ = 62 R$_\odot$, and from the angular diameter 
\citep[0.37 mas][]{Pasinetti2001} the derived stellar radius is 73 R$_\odot$. As a mean value we used  $R_\star$ = 70 R$_\odot$.

From our  modelling of the photospheric lines, done with FASTWIND, we obtained $T_{{\rm eff}} = 18\,000$ K and $\log\,g = 2.2$. 
The derived $T_{{\rm eff}}$  agrees with the values  adopted by  \citet{Lefever2007} and \citet{Kri2001}, but present large departures from the atmospheric 
parameters derived by \citet{Fraser2010}. To model the spectrum we used a projected rotational velocity (40 km s$^{-1}$) that is slightly higher than 
  that given in the literature (28 km s$^{-1}$ and 31 km s$^{-1}$). As did \citet{Lefever2007},  we found a very high value for V$_{\rm macro}$.

Previous observations reveal important intensity variations in both  absorption and emission  components of the P\,Cygni 
profile of the  H$\alpha$ line \citep[see Fig. A.1 given in][]{Lefever2007}. Comparing our spectra with that observation, 
the H$\alpha$ line observed in 2\,013 shows a wider absorption and a higher  emission component. We were able  to reproduce only the emission component 
of the H$\alpha$ line profile. Nevertheless, we derived a value for  V$_{\infty}$= 370  km s$^{-1}$  that is close to that measured using the UV lines  \citep[435 km s$^{-1}$,][]{Prinja1990} and lower than the values quoted in Table \ref{table:A1}. However, the mass-loss rate obtained in this work is consistent with the value measured by \citet{Lefever2007}.

\vskip 0.3cm




{\bf \object{HD\,99953}} (B1/2\,Iab/b) is a poorly studied star. \citet{Fraser2010} derived  the following stellar parameters: $T_{{\rm eff}}$ = 16\,800 K and $\log\,g$ = 2.15. Based on our modelling of the H$\gamma$, \ion{Si}{iii}, and 
\ion{He}{i} lines, we obtained a new  set of values: $T_{{\rm eff}}$ =19\,000 K and $\log\,g$ = 2.30. The latter are compatible with the assigned spectral 
classification and the observed SED ($T_{{\rm eff}}$ = 18\,830 K, $\log\,g$ =  2.30, $E(B-V)= 0.56$,  $R_\star$ = 28 R$_\odot$) for  d = 1\,077 pc (which is 
similar to the HIPPARCOS distance, 1\,075 pc). This distance and $E(B-V)$ predict  $M_{\rm{bol}}$ = $-7.16$ mag and $R_\star$ = 22 R$_\odot$. We adopted
$R_\star$ = 25 R$_\odot$ as the mean value.

The H$\alpha$ line shows a P\,Cygni profile and its intensity varies with time. The mass-loss and terminal velocity change 
by a factor of $\sim$ 2.8. Different values of $V_{\rm{macro}}$ were needed to model the line at different epochs.

\begin{figure}[t]

\includegraphics[width=0.48\textwidth, angle=0]{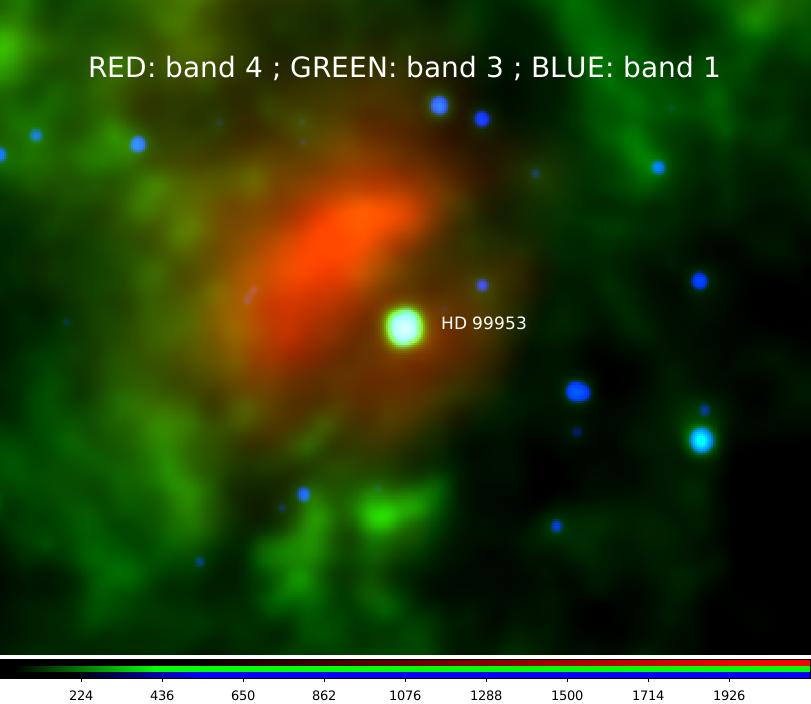}

\caption{WISE images showing wind-blown and warm dust structures around \object{HD\,99953}.\label{WISE-HD99953}}

\end{figure}

This work reports the first determinations of the mass-loss rate of the star ($\dot{M}$= $0.08\, 10^{-6}$ M$_\odot$ yr$^{-1}$, $0.13\, 10^{-6}$ M$_\odot$ yr$^{-1}$, 
and $0.22\, 10^{-6}$ M$_\odot$ yr$^{-1}$). The terminal velocity  derived from the H$\alpha$ 
line ranges from 250 km s$^{-1}$ to 700  km s$^{-1}$. This wide range includes the value measured from the UV spectral 
region \citep[510 km s$^{-1}$,][]{Prinja1990}. The combined WISE W3 and W4 images show warm dust heated by radiation coming from the star (see Fig. \ref{WISE-HD99953}).

\vskip 0.3cm

{\bf \object{HD\,111973}} (B2/3Ia): This spectroscopic binary was classified as a B3\,I star \citep{Chini2012}. It presents light variations 
with periods of 57.11 days  and 9.536 days  \citep{Koen2002}.

The SED was  fitted with a TLUSTY model  with $T_{\rm {eff}}$ = 17\,180 K, $\log\,g$ = 2.18, $R_\star$ = 48.6 R$_\odot$, $E(B-V)$ = 0.38 mag, and $d$ = 1\,660 pc. 
The colour excess was obtained with the calibration scale of  \citet[][]{Flower1996}. 
Using this distance and the derived colour excess, we obtained $M_{\rm {bol}}$= $-7.81$ mag and  $R_\star$ = 40 R$_\odot$.
On the other hand, when using the angular distance (0.26 mas) we have  $R_\star$ = 46 R$_\odot$, a value  that is close to that derived by 
\citet[][47 R$_\odot$]{Pasinetti2001}. We adopted  $R_\star$ = 46 R$_\odot$.

We were able to match all the photospheric lines.  The  derived $T_{{\rm eff}}$ and $\log\,g$ (16\,500 K and 2.1 dex)  are similar to those reported by 
\citet[][16\,000 K and 2.3 dex]{Fraser2010} and  \citet[][16\,500 K]{Prinja2010}. 

The H$\alpha$ line shows short-term variations: we observed a P\,Cygni profile with a weak emission component that turned into an absorption profile
the following night  (see Fig. \ref{halpha}). This variation is similar to the expected dynamical timescale of a prototypical BSG wind ($\sim 1.3$ days).
We modelled the  H$\alpha$ line and derived a lower terminal velocity than the value measured from the UV lines \citep[520 km s$^{-1}$][]{Prinja2010}. Furthermore, 
a high $V_{{\rm macro}}$ = 160-190 km/s$^{-1}$ was needed to model this line.

This work reports the mass-loss rate of the star for the first time. It ranges from  $0.14\, 10^{-6}$ M$_\sun$ yr$^{-1}$ to $0.21\, 10^{-6}$ 
M$_\sun$ yr$^{-1}$.

\vskip 0.3cm




{\bf \object{HD\,115842}} (B0.5Ia/ab) presents  photometric variations on a timescale of 13.38 days     \citep{Koen2002} and variations in the 
H$\alpha$ line profile.  Our spectrum displays  the H$\alpha$ line in pure emission (see Fig. \ref{halpha}),  while the one given 
by \citet{Crowther2006} displayed a P Cygni feature. The H$\beta$ line is seen in absorption and the profile is slightly asymmetric (see Fig. \ref{fig:13}).

From the SED we derived the following parameters: $T_{\rm {eff}}$ = 25\,830 K, $\log\,g$ = 2.73, $R_\star$ = 38 R$_\odot$, $E(B-V)$ = 0.6 mag, and $d$ = 1\,543 pc. 
This colour excess fits the observed 2\,200 \AA\, bump feature and is higher than the $E(B-V)$ = 0.53 mag derived from \citet{Flower1996}.
  The obtained distance is consistent with the HIPPARCOS and Gaia parallaxes (0.65 mas and 0.6105 mas, respectively) and the distance of 1\,583 pc estimated from the Ca II H 
  and K lines \citep{Megier2009}. Using the HIPPARCOS distance (1\,538 pc) and $E(B-V)$ = 0.6 mag, we calculated $M_{\rm {bol}}= -9.22$ mag and  $R_\star$ = 32 R$_\odot$, 
  while from the angular diameter  \citep[0.22 mas,][]{Pasinetti2001} we  derived $R_\star$ = 33 R$_\odot$. We adopted $R_\star$= 35 R$_\odot$ as the mean value.

We achieved very good fits to the photospheric lines for $T_{\rm {eff}}$ = 25\,500 K and $\log\,g$ = 2.75. The stellar parameters derived in this work agree with 
those found by \citet{Crowther2006} and \citet{Fraser2010} (see Table \ref{table:A1}).  We obtained a lower mass-loss rate (1.8\,$\times\,10^{-6}\, \rm M_{\odot}\,yr^{-1}$) and a higher terminal velocity 
(1\,700 km s$^{-1}$) than the \citet{Crowther2006} estimates (2.0\,$\times\,10^{-6}\, \rm M_{\odot}\,yr^{-1}$, 1\,180 km s$^{-1}$). 

Our value of  $V_\infty$ is also higher than the value determined from the IUE data  \citep[1\,125 km s$^{-1}$,][]{Evans2004}. These authors  reported  a very high 
macroturbulence, $V_{\rm{macro}}$ = 225 km s$^{-1}$, which is twice our estimate.

\begin{figure}[t]
 \includegraphics[width=0.48\textwidth, angle=0]{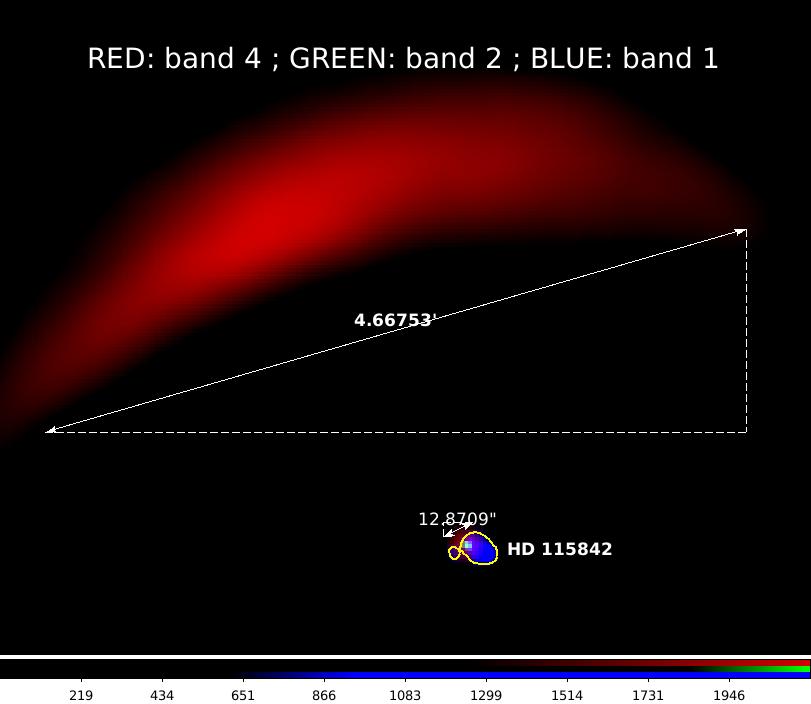}
 \caption{Bow-shock structure and asymmetric density structure associated with \object{HD\,115842}. The yellow curves represent regions with
   the same brightness, each one associated with a different lobe.\label{WISE-HD115842}}

\end{figure}

WISE W3 and W4 images show a large well-defined bow-shock structure (see Fig. \ref{WISE-HD115842}) 
related to a possible strong wind-ISM interaction phase. 
In addition, an asymmetric density structure seems to be present in the  WISE W1 image with an angular 
size that is greater than the WISE resolution.

\vskip 0.3cm

{\bf \object{HD\,148688}} (B1Iaeqp) is a southern oxygen-rich
supergiant that displays line variations \citep{Jaschek1973}. Data from the HIPPARCOS satellite correlate with periods of 1.845 days  and  6.329 days 
\citep{Lefever2007}.

The best-fitting model to the observed SED leads to $T_{{\rm eff}}$ = 20\,650 K, $\log\,g$ = 2.2, $R_\star$ = 34 R$_\odot$, and $d$ = 838 pc (while the HIPPARCOS distance 
is 833 pc). This gives $M_{{\rm bol}}$ = $-7.97$ mag and $R_\star$ = 26 R$_\odot$,  for $E(B-V)$ = 0.54 mag. This  $E(B-V)$ was obtained from \citet{Flower1996} and also fits 
the deep bump at 2\,200 \AA\, present in the SED. We adopted  $R_\star$ = 31 R$_\odot$ as the mean value.

The stellar parameters obtained from FASTWIND ($T_{{\rm eff}}$= 21\,000 K and $\log\,g$ = 2.45) are in better agreement with the works of \citet{Fraser2010} 
and \citet{Lefever2007}. 
The H$\alpha$ line from Fig. \ref{halpha} shows a P\,Cygni feature that looks like the one shown by \citet{Lefever2007},  while the spectrum of \citet{Crowther2006} 
exhibits an emission line.

We were only able to model  the emission component of the P\,Cygni profile. We obtained a mass-loss rate that is a  factor of 1.1 and 1.4 lower than the values found by 
\citet{Lefever2007} and \citet{Crowther2006}, respectively.  The terminal velocity is similar to the value found by \citet[][see Table \ref{table:A1}]{Lefever2007}, 
although it is greater than measurements derived from the UV lines \citep[725 km s$^{-1},$][]{Prinja1990}.


\subsection{Global properties of stellar and wind parameters}

\begin{figure}
  \centering
  \includegraphics[width=0.5\textwidth,angle=0]{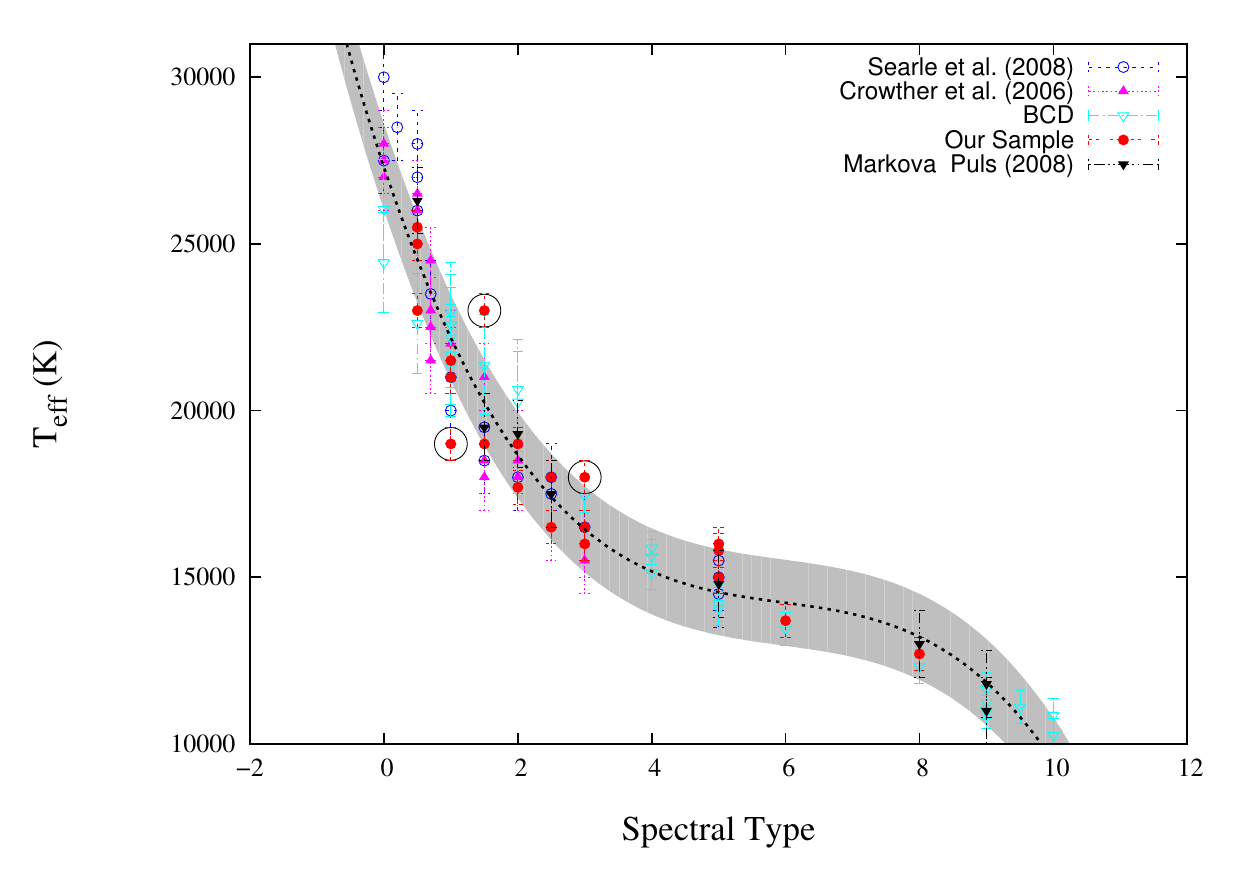}
  \caption{Effective temperature as function of the spectral subtype. The relation was fitted  using a third-order polynomial 
  regression (dotted line)  with a dispersion of 1\,290~K  (grey band, see  text for details). Comparison of  our sample of Galactic B supergiants with a set of BSGs collected from the  literature. Data points of our sample with large open circles are outliers.  
    \label{st-teff}}
\end{figure}

Although the stars present spectral variability, it is still possible to describe their global properties.   Figure \ref{st-teff} shows a comparison of  $T_{\rm {eff}}$ of BSGs  with the spectral subtype. The sample is a collection of data from the literature and this work. 
  With the exception of those stars that have $T_{\rm {eff }}$ based on the BCD spectrophotometric calibration \citep{Zorec2009}, the  effective 
  temperature was obtained via adjustments of  the line profiles of \ion{He}{i}, \ion{He}{ii}, and line intensity ratio of \ion{Si}{iv}/\ion{Si}{iii} or 
  \ion{Si}{iii}/\ion{Si}{ii}, as in this work. 

We performed a fit to the observed  $T_{\rm {eff}}$ versus spectral subtype relation using a third-order polynomial 
  regression ($a=-54.6\pm7.7$, $b=973.4\pm107.5$, $c=-6054.3\pm387.9,$ and $d=27\,316\pm331$). The derived coefficients agree very well with the fit obtained by \citet{MarkovaPuls2008}. We illustrate our fit in Fig. \ref{st-teff} with a grey wide band that has a dispersion of 1\,290~K.  The  supergiants of our sample, with the exception of \object{HD\,52089}, \object{HD\,53138,} and \object{HD\,47240} (indicated in the 
  figure with large open circles), fall inside the traced relation.  Moreover, we can observe that the departures of the effective  temperatures for the  early  BSGs range 
  roughly  from  0~K to 5\,000~K and tend to be lower the later the B spectral subtype. This same tendency is also observed in \citet{Markova2008}.

\begin{figure}
  \centering
  \includegraphics[width=0.5\textwidth,angle=0]{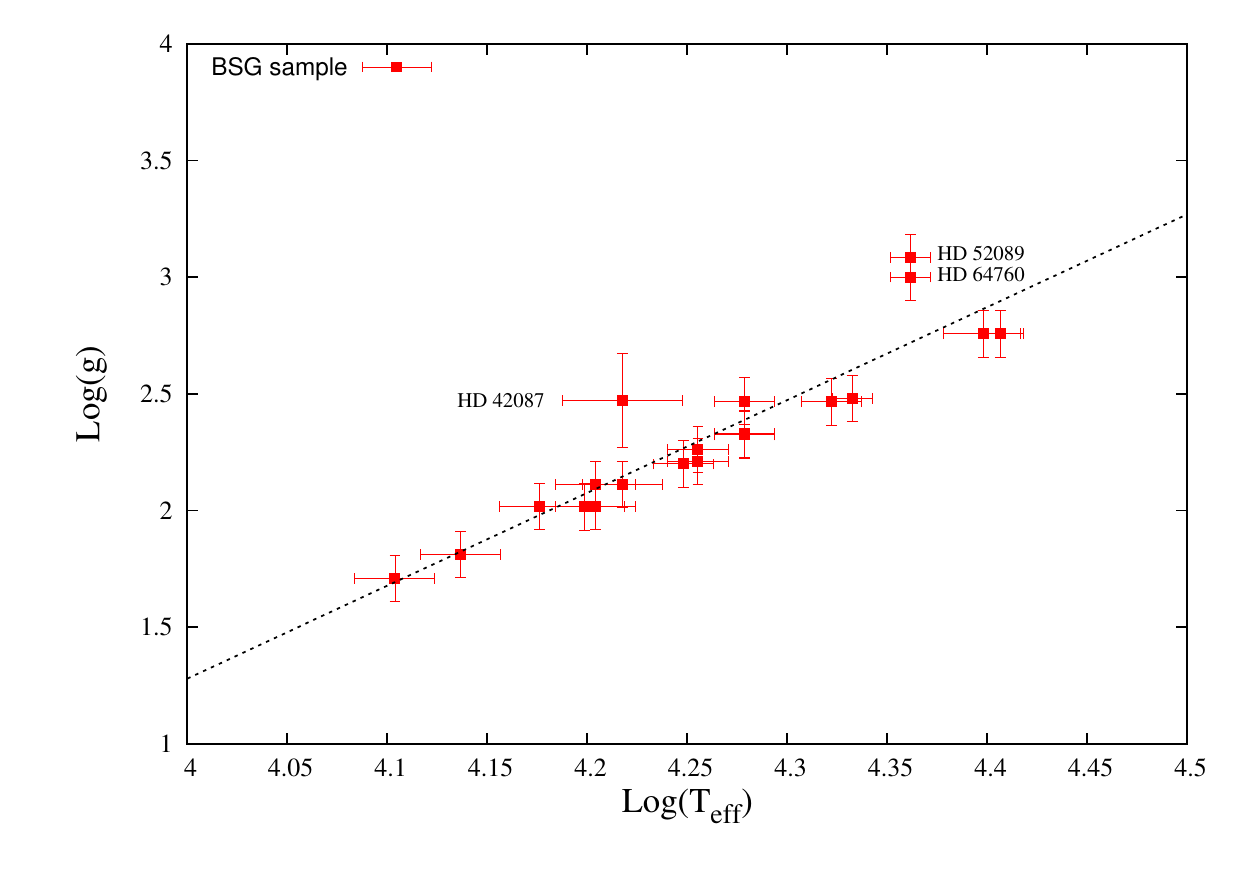}
  \caption{Linear relation ($\log T_{\rm {eff}}$ - $\log\,g_{\rm {cor}}$)  of our sample of B Galactic supergiants (dashed line). The surface gravity was corrected by the centrifugal acceleration. \label{teffvsg}}
\end{figure}

Figure \ref{teffvsg} displays a linear correlation between $\log\,T_{\rm {eff}}$ and $\log\,g_{\rm {cor}}$ (the surface gravity corrected by the centrifugal acceleration, see Table \ref{table:A1}). This kind of correlation was previously reported by \citet{Searle2008}. We performed a linear fit and obtained a slope of $3.98 \pm 0.31$. This indicates that 
most of the  stars in our sample  have a similar luminosity-to-mass ratio 
($\log\,(L/M)\,\sim$\,4), as expected, since $L \propto T_{\rm{eff}}^4$.  

A large scatter is observed around the stars  that have large 
$\log\,g_{\rm{cor}}$ ($\ge 2.5$); in  particular \object{HD\,42087, HD\,52089}, and \object{HD\,64760} do not follow the relation. Two of these stars, \object{HD\,42087} and \object{HD\,64760}, are close to the TAMS so we believe they have just left the main sequence.  
\object{HD\,64760} is a high rotator and, as a consequence,  we should observe higher apparent bolometric luminosities and lower effective temperatures than 
those expected for their non-rotating stellar counterparts \citep{Fremat2005,Zorec2005}, giving a later evolutionary stage of the
object.

From Table \ref{table:2} we can distinguish stars showing two  different wind regimes. One group gathers stars with steep velocity gradients, $\beta\,\leq\,1$, in 
good agreement with the standard wind theory, while the other group shows  values  of $\beta$ between 1.5 and 3.3, similar to the results reported by 
\citet{Crowther2006,Searle2008}, and \citet{Markova2008}. Stars with $\beta$ between 2 and 2.5 are the most common ones. High values of $\beta$ are often found among 
the late B-type stars.

 The terminal velocities range from 155 km\,s$^{-1}$ to 1\,700 km\,s$^{-1}$, while mass-loss rates range generally from 
0.08$\times$10$^{-6}$ M$_\sun$ yr$^{-1}$ to 0.7$\times$10$^{-6}$ M$_\sun$ yr$^{-1}$ (four stars fall outside this mass-loss range).

\begin{figure}

  \centering

  \includegraphics[width=0.51\textwidth,angle=0]{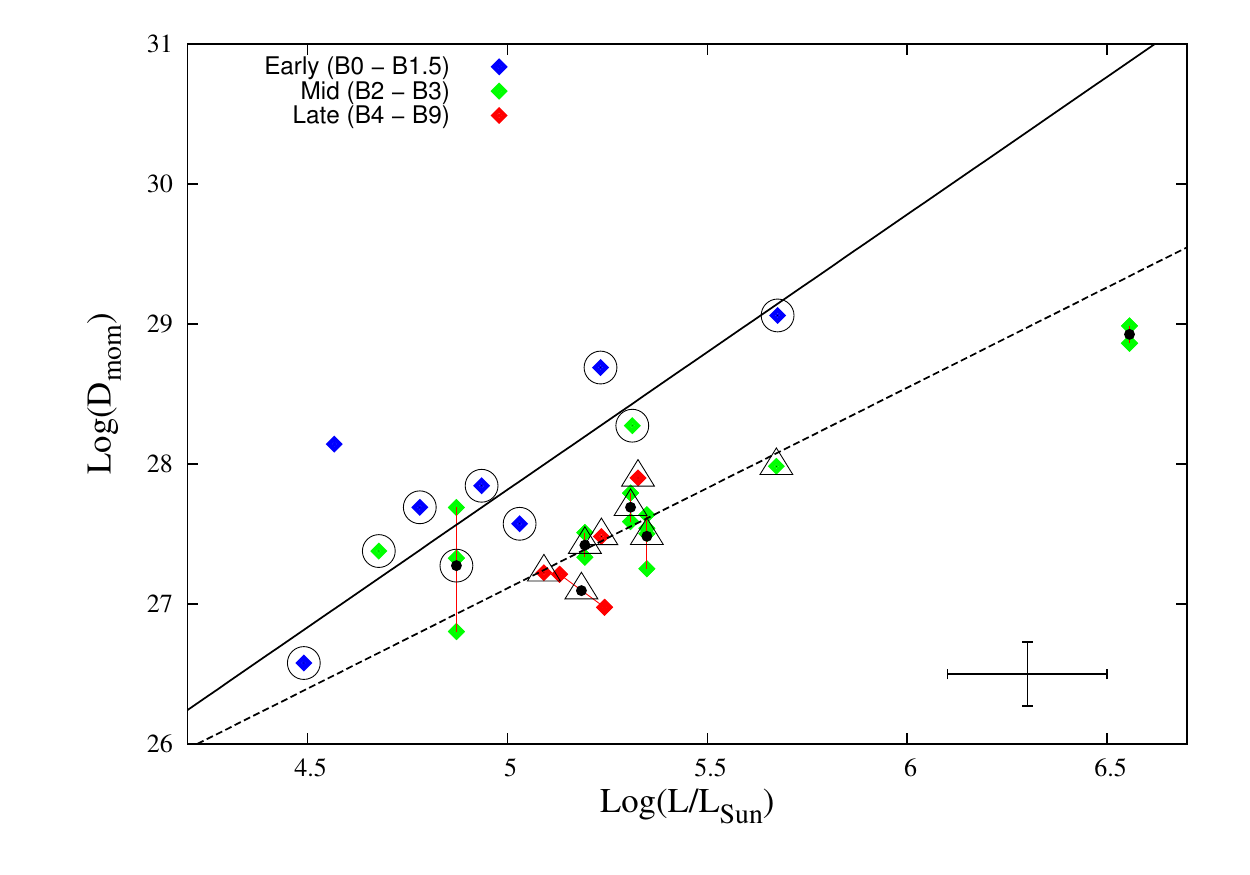}

  \caption{Wind-momentum luminosity relationships (dotted lines). Blue, green, and red diamonds  represent early-, mid-, and late-type BSGs, respectively. The upper  fit line was derived using data points in large circles, while the lower  fit line corresponds to data points with large triangles.  Diamond symbols connected with solid lines correspond to observations of the  same star in different epoch (black dots represent their average values). \label{WMLu}}

\end{figure}

In addition, we calculate the average values for the total mechanical momentum flow of the wind, 
$D_{\rm {mom}}$, as a function of the luminosity of the star. This is known as the modified WLR \citep{Kudritzki1995}, given by
\begin{equation}
D_{\rm mom}=\dot{M}\,v_{\infty}\,{R}^{0.5} \propto L^{1/\alpha_{\rm eff}}, 
\end{equation}
\noindent where $\alpha_{\rm{eff}}$ = $\alpha\,-\delta$ \citep[see details in][]{Kudritzki1999, Puls1996}; $\alpha$ and $\delta$ are the force multipliers related 
to the line opacity and wind ionization, respectively.

Our sample splits  into two different groups (see Fig.\,\ref{WMLu}): the early BSGs with spectral types between B0 and B1.5 (blue diamonds) and mid- to late-type 
BSGs with spectral types from B2 to B9 (green and red diamonds). These two groups seem to be  clearly separated. However,   among  the early B supergiants we found 
three mid B stars (HD\,41117, HD\,42087, and the variable star HD\,99953) with wind momentum rates comparable to the early ones.

From a linear regression, we found that the sample of the early BSGs plus the three mentioned mid B stars (see Fig.\,\ref{WMLu}, large circles) can be fitted by the following relationship:
\begin{equation}
  \log\,D_{\rm{mom}}= 1.96(\pm0.28) \log\,L/L_\sun  + 17.98(\pm1.43),  {\rm \:\:for\; B0-B1.5.}
  \label{WLR1}
\end{equation}

The second WLR was traced using BSG stars with spectral types B2-B9 (see Fig.\,\ref{WMLu},  large triangles). The values of the wind  momentum rates of these stars are clearly lower than those of the early-type objects. Furthermore, the WLR displays a different slope (less steep) with luminosity. For these objects we find a linear regresion of the form
\begin{equation}
\log\,D_{\rm{mom}}= 1.43(\pm0.42) \log\,L/L_\sun  + 19.94(\pm2.23),  {\rm\:\: for\; B2-B9.}
\label{WLR2}
\end{equation}

The observed offset between the WLRs increases from the early- to the mid-type BSGs. Despite the large errors of these quantities, the tendency is opposite to the results observed by \citet{Kudritzki1999} and predicted by \citet{Vink2000}.

In Fig.\,\ref{WMLu}, observations of the same star at different epochs are connected by solid lines. For these stars, average values of $\log\,D_{\rm {mom}}$ (Fig.\,\ref{WMLu},  black dots) were considered to fit the regression line of the WLR for each group, i.e. early B or mid/late B stars.  In the next section, we  return to the issue of mass-loss variations.

It is interesting to stress that the wind
regime of the stars defining each WLR is different. From Table \ref{table:2} we found that the early-type stars have mostly $\beta \lesssim 2$ and terminal velocities greater than 500 km\,s$^{-1}$, 
while the mid- and late-type stars have $\beta\,\geq\,2$ and terminal velocities lower than 500 km\,s$^{-1}$.

On the other hand, the parameter $\alpha_{\rm {eff}}$, which is the inverse of the slope of the WLR, changes from  0.5 for the early BSGs to 0.7 for the mid/late ones. An increase in $\alpha_{\rm {eff}}$ might suggest highly structured ionized outflows that lead to a weakening of the radiation force. Under this condition, different mechanisms (e.g. pulsations, clumping) might also contribute to driving the wind.

\begin{figure}

  \centering

  \includegraphics[width=0.5\textwidth,angle=0]{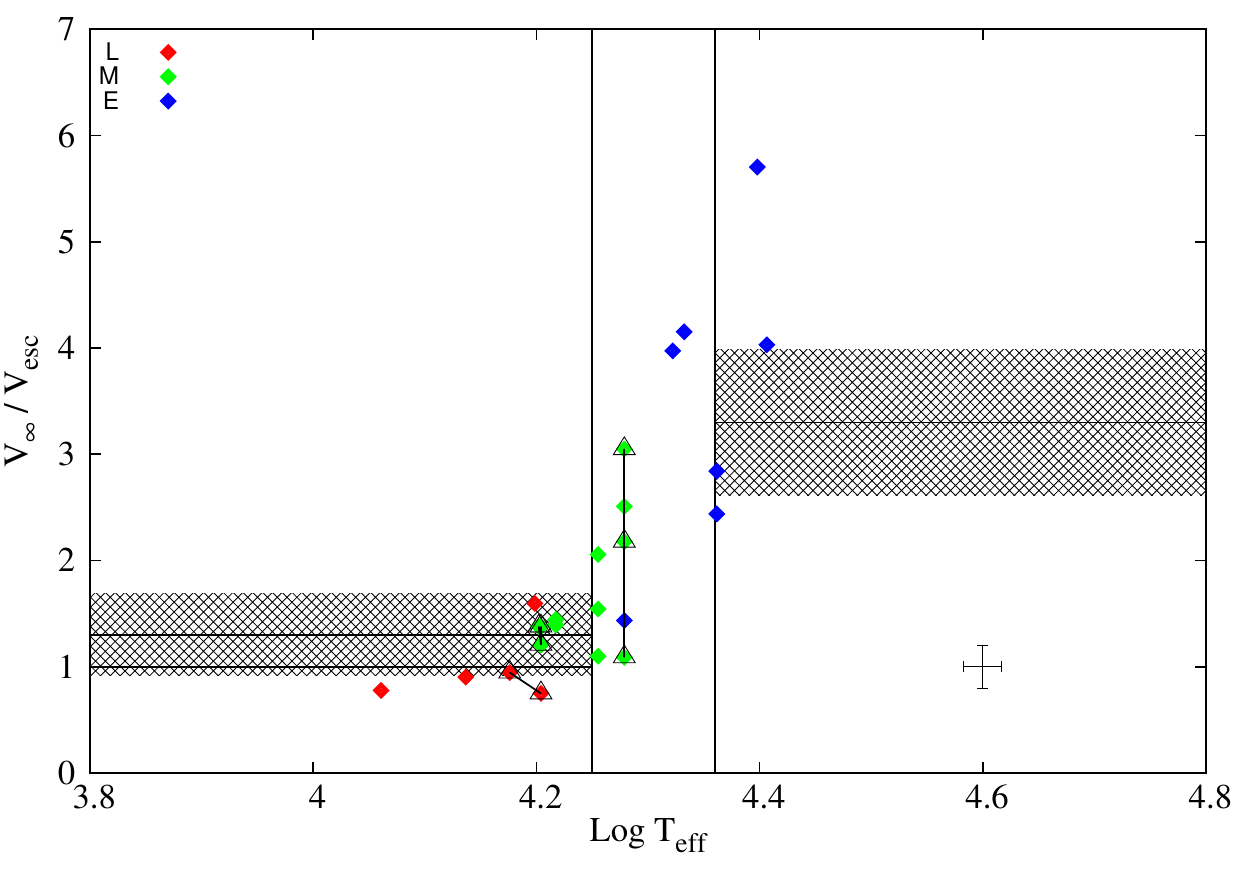}

  \caption{Ratios of $V_\infty/V_{\rm {esc}}$ as a function of $\log\,T_{\rm{eff}}$. Early-type (E), mid-type (M), and late-type (L)  stars are indicated by blue, green, and red diamonds, respectively.  Stars with wind variability are indicated with large open triangles and connected with a solid line. The largest variability is shown by  HD\,99953, which is located in the region limited by vertical black lines (the bi-stability jump). The shaded areas refer to two zones with different $V_\infty/V_{\rm{esc}}$: 3.3 for effective temperatures above 23 000 K, and 1.3 below 18 000 K, with errors of 43\% and 33\%, respectively, which were identified by \citet{MarkovaPuls2008}.\label{velratio}}

\end{figure}

Finally, Fig.\,\ref{velratio} shows the ratios of $V_\infty/V_{\rm {esc}}$ as a function of $\log\,T_{\rm{eff}}$. In the figure we present our results together with the two temperature regions defined by \citet{MarkovaPuls2008}, which show significantly different $V_\infty/V_{\rm {esc}}$ ratios: 3.3 for effective temperatures above 23\,000 K,  and 1.3 below 18\,000 K. Errors of 33\% for the cooler and 43\% for the hotter regions are shown as shaded areas \citep[for details on the error determinations, see][]{MarkovaPuls2008}. 

The region in between, delimited  by vertical black lines, is called the bi-stability jump. \citet{MarkovaPuls2008} showed that the stars located there show a gradual decrease in $V_\infty/V_{\rm {esc}}$. Overall, our results show a similar behaviour to that reported by these authors.
In particular, we found many stars in our sample located  inside the bi-stability region. Moreover, as some stars are variable we also plotted in Fig.\, \ref{velratio} the various positions of those stars observed during different epochs. These positions are connected with a solid black line.

The variation in $V_\infty/V_{\rm {esc}}$  of those stars on the cool side of the bi-stability jump seems to be small. Instead,  HD\,99953 (green diamonds with triangles),   which is located inside the  bi-stability region, presents such a large wind variability that it seems to switch from a slow to a fast regime. The same variation for this star is observed in the WLR. With the exception of \object{HD\,80077}, which  has a huge mass-loss rate, we do not observe that a decrease in   $V_\infty/V_{\rm {esc}}$  is accompanied by an increase in the mass-loss rate, as predicted by \citet{Vink1999}. In general,  mid and late BSGs present values of $\dot{M}$ that are similar to or lower than those of the stars located on the hot side of the bi-stability jump.

\section{Discussion}

\label{dis}

\subsection{Photospheric parameters}

For some stars different estimates of $T_{\rm{eff}}$ and  $\log\,g$ ($\Delta T_{\rm{eff}} > 1\,500$\,K, and  $\Delta \log\,g \sim 0.2$) are found in the literature (see Table \ref{table:A1}).

In relation with these $T_{\rm{eff}}$ estimates, it is important to  point out that this parameter is derived using different methods or  approaches. Quite often the atmospheric parameters are obtained by fitting the observed line profiles with a synthetic spectrum calculated from NLTE line-blanketed plane-parallel hydrostatic model atmospheres, using the codes TLUSTY and SYNSPEC \citep{Hubeny1995,Lanz2007}. This procedure was mainly carried out by \citet{Fraser2010,McErlean1999,Kudritzki1999}, while \citet{Crowther2006, Lefever2007, MarkovaPuls2008, Searle2008} used codes like FASTWIND or CMFGEN \citep{FWmio, Hillier1998} that solve multi-level non-LTE radiative transfer problems in the co-moving frame, based on NLTE line-blanketed spherically symmetric wind models.

  For most of the stars in our sample, we found good agreement between our $T_{\rm{eff}}$ estimates and those obtained by \citet[][six stars in common]{Crowther2006}, \citet[][eight stars in common]{Lefever2007}, and  \citet[][five stars in common]{Searle2008}, within error margins of $\pm 1500\,$ K. Good agreements were also found with the values derived by \citet[][twelve stars in common]{Fraser2010}, who used TLUSTY. Among all the previous mentioned authors, the largest discrepancies in $T_{\rm{eff}}$ are found for six stars,  \object{HD\,42087}, \object{HD\,52089}, \object{HD\,53138}, \object{HD\,64740}, \object{HD\,92964}, and \object{HD\,99953}, which  show appreciable differences ($\Delta T_{\rm{eff}} > 2\,000$\,K).  Two of them (\object{HD\,53\,138} and \object{HD\,64740}) have shown either some degree of ion stratification or  ionization changes in the UV lines  \citep[see details in][]{Prinja2002}. The  star \object{HD\,52089} has a magnetic field  and X-ray emission (see section \ref{objects}) and like  \object{HD\,99953} is located inside the bi-stability region. This might suggest that these stars show intrinsic temperature variations since we do not observe a tendency regarding a particular method used to derive their atmospheric parameters.

We also found that the line spectra do not show He forbidden components suggesting that the He abundance is close to the solar value.  This is in accordance with what \citet{Kraus2015} observed in \object{55\,Cyg}. In addition,  \citet{Crowther2006} did not report  any discrepancies on the Si and He abundances.

We did not find any correlation between $V_{\rm{macro}}$ and $T_{\rm{eff}}$, as was reported by \citet{Fraser2010}. In general, we needed larger values for $V_{\rm{macro}}$ (mostly around $50-60$ km s$^{-1}$ for all the stars) than the ones given in the literature to be able to fit the photospheric line widths. In some cases it was even necessary to slightly increase $V\,sin\,i$ as well.  This could be due to our use of a mean value for $V_{\rm{macro}}$ to model all the lines and,  particularly, the line-forming region for He might   extend into the base of the wind where higher turbulent velocities can be expected.

The major discrepancy found among the stellar parameters with previous works is related to the determination of the stellar radius, due to the different methods used to calculate distances to stars.  We think that the values we estimated in section \S \ref{objects} are quite reliable. Uncertainties in the stellar radius yield an insecure luminosity and a mass-loss discrepancy. This last issue is discussed in the next subsection.

  \subsection{Spectral and wind variability}\label{wind}

Variable B supergiants often show spectroscopic variability in the optical and UV spectral region that can be attributed to the large-scale wind structures \citep{Prinja2002}.   Variations linked to rotational modulation have been reported, for example, in \object{HD\,14134} and \object{HD\,64760} \citep[][respectively]{Morel2004,Prinja2002}, while those  associated with pulsation activity  were found in \object{HD\,50064} and \object{55 Cyg} \citep[][respectively]{Aerts2010,Kraus2015}. Nevertheless, variability related to the presence of weak magnetic fields has also been proposed \citep{Henrichs2003,Morel2004}.

Most of the stars in our sample show H$\alpha$ line variations in both shape and intensity. These variations can take place on a scale of a few days, but daily and even hourly variability has also been reported \citep[e.g.][]{Morel2004,Kraus2012,Kraus2015,Tomic2015}. Among our sample, \object{HD\,75149}, \object{HD\,53138}, and \object{HD\,111973}  display emission-line episodes or dramatic line profile variations with a diversity of alternating shapes (from a pure  absorption to a P Cygni profile and vice versa).

  Regarding the amplitude of the mass-loss variation in individual objects, \citet{Prinja1986}  measured a variability at 10\% level on timescales of a day or longer, while  \citet{Lefever2007} and \citet{Kraus2015} reported on ratios between maximum and  minimum mass-loss rates ranging from 1.05 up to 2.5.

  In this way, based on a carefully study of the variation in the  H$\alpha$ line among the stars we observed at different epochs (modelled with the code FASTWIND), the largest variations in $\dot{M}$ (a factor of  1.5 to 2.7) are detected in \object{HD\,75149}, \object{HD\,99953,} and \object{HD\,111973}.
  As the H$\alpha$ line is very sensitive to changes in the mass density and in the velocity 
gradients at the base of the wind \citep{Cidale1993}, a connection between line variability  and  changes in the wind structure  due to  photospheric instabilities can be expected. 

To gain insight into the origin of the variability of the   BSG winds, we gathered information from the literature for all our objects on periods of light 
and spectroscopic variations (quoted in Table \ref{table:1}), and  on previous determinations of the wind parameters derived from individual modelling attempts. A comparison of the mass-loss estimates listed in Table \ref{table:A1} is  of particular interest as they display ratios between maximum and minimum values that range from 1 to 7. These discrepancies  are not completely true because  these models were computed using different stellar radii, and the differences in  mass-loss rates could be much lower. For instance, for  HD\,34085 we found a discrepancy in $\dot{M}$ of 1.9 when  compared with the result given by \citet{MarkovaPuls2008}, but it could be only 1.3 times higher  ($\dot{M} = 0.44  \times 10^{-6}$ M$_\sun$ yr$^{-1}$)  if a value of  $R_\star$ = 115 R$_\odot$ is adopted (keeping  all the other parameters constant).

  Therefore, to compare the values  obtained by different authors we should discuss the scaling properties of the corresponding models since  line fits are not unique \citep{Puls2008}. Thus we will adopt the optical-depth invariant $Q_{\rm r}= \dot{M}$/$R_\star^{1.5}$, instead of the traditional parameter $Q = \dot{M}$/$(R_\star\,V_{\infty})^{1.5}$ 
\citep{Puls1996} since we assume that $V_\infty$ can change due to the wind variability. The corresponding calculations and errors for $\log\,Q_{\rm r}$ are given in Table \ref{table:A1}. An overview of these values shows that the differences in  $\log\,Q_{\rm r}$ (for the stars we observed at different epochs) are two or three times larger than their corresponding error bars suggesting real variations in the wind parameters. These differences can also be observed  among the $Q_r$ values we calculated from data given by different authors.

It is interesting to note that we found five stars in the domain of the bi-stability region.  \object{HD\,99953} shows pronounced  H$\alpha$ variations from which different estimations of $V_\infty$ ($250$ km s$^{-1}$, $500$ km s$^{-1}$, and  $700$  km s$^{-1}$) and $\dot{M}$ ($0.08\times10^{-6}$ M$_\sun$ yr$^{-1}$, $0.13\times10^{-6}$ M$_\sun$ yr$^{-1}$, and $0.22\times10^{-6}$ M$_\sun$ yr$^{-1}$) were obtained. This implies variations in $V_\infty$ and $\dot{M}$ of a factor of about $2.8$. This result goes against the theoretical prediction made by \citet{Vink1999} who argue that the jump in mass loss is accompanied by a steep decrease in the ratio $V_\infty/V_{\rm{esc}}$ (from $2.6$ to $1.3$), which is close to the observed bi-stability jump in the terminal velocity. Moreover, these authors proposed that if the wind momentum $\dot{M}\,V_\infty$ were about  constant across the bi-stability jump, the mass-loss rate would have increased  by a factor of two from stars with spectral types earlier than B1 to later than B1.

In the particular case of \object{HD\,99953}, our result indicates that  $\dot{M}\,V_\infty$ is not constant (see Figure \ref{WMLu}) nor does the mass loss increases when the  ratio $V_\infty/V_{\rm{esc}}$ decreases, even when the observed changes in the variables are of the order of the expected values ($\sim 2.8$).
 On the other hand, the mass loss derived for \object{HD\,99953} ($\log\,\dot{M}= -6.66$, when $V_\infty$= 700 km s$^{-1}$) agrees with the value computed by  \citet[][$-6.54$ for $V_\infty/V_{\rm{esc}}= 2.6$]{Vink1999} in Table 1; however, the terminal velocity we derived is the half of the theoretical value found by these authors.  We do not believe that the discrepancy could be due to the luminosity of the star ($\log L/L_\sun = 4.87\pm0.37$) since the models were computed for a  $\log L/L_\sun =5$. We also found that this is not related to the velocity law adopted for the model since  \object{HD\,47240} does not fit either.

  Looking for other observed BSGs with  $\log L/L_\sun \sim 5$ (see Table \ref{table:1}), we found that the wind parameters of the stars on the hot side of the bi-stability jump, like \object{HD\,38771} (with $\log\,\dot{M}= -6.85$ and $V_\infty$= 1500 km s$^{-1}$) and \object{HD\,52382} (with $\log\,\dot{M}= -6.62$ and $V_\infty$= 1000 km s$^{-1}$), resemble those listed in Table \ref{table:1} by \citet{Vink1999}. In contrast,  the wind parameters of the stars on the cool side of the  bi-stability jump, like  \object{HD\,34085} (with $\log\,\dot{M}= -6.64$ and $V_\infty/V_{\rm{esc}}= 0.8$), \object{HD\,58350}  (with $\log\,\dot{M}= -6.82$ and $V_\infty/V_{\rm{esc}}= 0.9$),  and \object{HD\,111973} (with $\log\,\dot{M}$ between $-6.7$ and $-6.85$, and $V_\infty/V_{\rm{esc}}= 1.4$) do not fit  any model predictions. They often have lower terminal velocities and mass-loss rates than values expected from the models. This means that a decrease in the ratio $V_\infty/V_{\rm{esc}}$, both inside and on the cool side of the bi-stability jump, is not accompanied by an increase in the mass-loss rate. This result supports the conclusions drawn by \citet{MarkovaPuls2008}  who used the modified optical depth invariant and found that both $\dot{M}$ and  $V_\infty$ are decreasing in parallel.

  We believe that the decrease in both  the mass-loss rate and the terminal velocity, on the cool side of the bi-stability jump, could be explained with the  $\delta-slow$ hydrodynamic solution for radiation line-driven winds \citep{Cure2011}. \citet{Venero2016} found that a change in the ionization of the wind, characterized by the parameter $\delta$ of the line-force multiplier \citep{Abbott1982}, defines two different stationary wind
regimes  (fast and slow).  The fast regime is always present at high effective temperatures ($>$ 25\,000 K), while fast or slow regimes could develop at low effective temperatures. Both wind regimes are separated  by an instability zone where non-stationary flow regimes exist.  \citet{Venero2016} proposed that an initial density perturbation could trigger a switch in the wind regime between fast and slow solutions (and vice versa) in the B mid-type supergiants. In the same work the authors show preliminary computations of  wind hydrodynamics using  time-dependent wind solutions that lead to the formation of kink velocity field structures which might be related to the evolution of absorption components like the ones observed in the UV line spectrum (DACs). In addition, at temperatures $<$ 17\,000 K the models predict a decrease in both terminal velocity and mass-loss rate \citep[see Figure 4 in][]{Venero2016}.  Futher research on the slow hydrodynamics solution is in progress \citep{Venero2017}.

  \subsection{The wind momentum--luminosity relationship}

 The empirical WLR of massive stars is generally represented by a linear regression of the form
\begin{equation}
\log\,D_{\rm{mom}}= X \log\,L/L_\sun + D_0.
\end{equation}

Values for the coefficients $X= 1/\alpha_{\rm {eff}}$ and $D_0$ reported in the literature are given in Table  \ref{table:WLR}, together with the parameter $\alpha_{\rm {eff}}$ and the corresponding errors. It is worth mentioning that previous investigations on the WLR for Galactic BSGs (see Table  \ref{table:WLR}) show that wind momentum rates derived by means of line-blanketed analyses are systematically higher than those reported by \citet{Kudritzki1999}.

  \begin{table*}

    \caption{Coefficient for the WLR derived by different authors.}

      \label{table:WLR}

      \centering

    \begin{tabular}{llccc}

      \hline
      \hline
Reference   & $X$ & $D_0$  &  $\alpha_{\rm{eff}}$   & sample used\\

\hline

\citet{Kudritzki1999} & $1.34\pm0.25$  &  $21.24\pm1.38$      &               & E\\

\citet{Kudritzki1999} & $1.95\pm0.2$  &  $17.07\pm1.05$      &               & M\\

\citet{Vink2000}   &  $1.826\pm0.044$  & $18.68\pm0.26$  & & $T_{\rm{eff}} > 22.5$ kK from theory\\

\citet{Vink2000}   &  $1.914\pm0.043$  & $18.52\pm0.23$  & & $T_{\rm{eff}} < 27.5$ kK from theory\\

\citet{Herrero2002}   & $2.18\pm0.21$ & $16.81\pm1.16$ & $0.46\pm0.04$   &  mostly OSGs\\

\citet{Mokiem2007}& $1.84\pm0.17$ & $18.87\pm0.98$  & $0.54\pm0.05$ & mostly OSGs\\


\citet{Searle2008}& $1.59\pm$ & $19.86\pm0.78$  & $0.63\pm0.06$ & E \& M\\

This work         & $1.96\pm0.28$  & $17.98\pm1.43$  & $0.51\pm0.07$ & E\\

This work         & $1.43\pm0.42$ & $19.94\pm2.23$  & $0.70\pm0.21$ & M \& L\\

\hline

   \end{tabular}

\tablefoot{E, M, and L refer to early-, mid-, and late-type B supergiants.}

\end{table*}

\begin{figure}

  \centering

  \includegraphics[width=0.51\textwidth,angle=0]{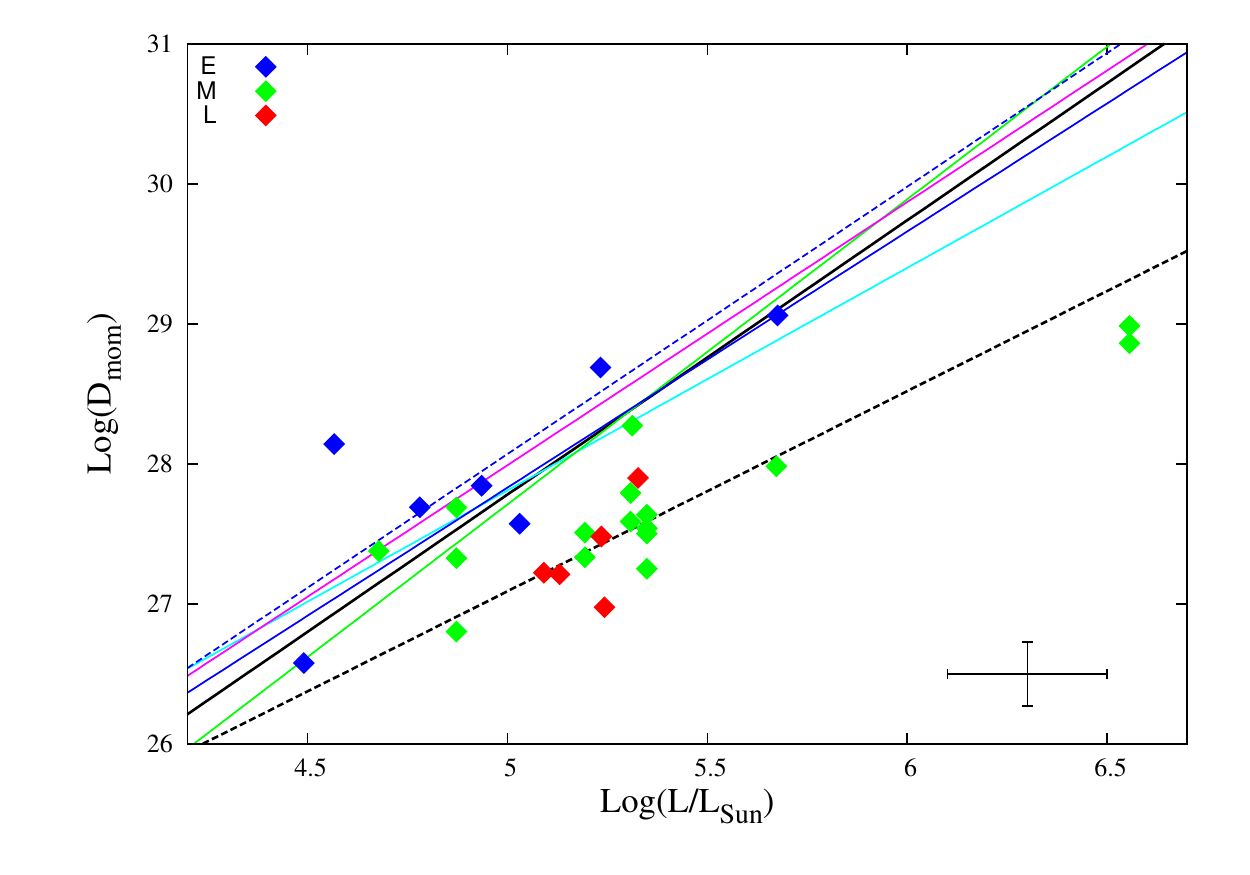}

  \caption{Comparison of various empirical WLR: this work (solid and dashed black lines are for early BSGs and mid- or late-type BSGs, respectively), \citet[][solid green line]{Herrero2002}, \citet[][solid cyan line]{Searle2008}, and  \citet[][magenta solid line]{Mokiem2007}. The theoretical WLR of \citet{Vink2000} is in blue, solid and dashed lines are for $T_{\rm{eff}} > 27\,500$ K and $T_{\rm{eff}} < 22\,500$ K, respectively. The figure includes the sample of stars shown in Fig. \ref{WMLu}. \label{all_rectas}}

\end{figure}

   The coefficients ($X$, $D_o$) derived in this work for the early BSGs  are close to those obtained by \citet{Herrero2002} and \citet{Mokiem2007}, who used a sample of stars consisting mostly of OSGs. Both coefficients also agree with the theoretically predicted values of \citet{Vink2000}, but disagree with the values obtained by \citet{Kudritzki1999} and \citet{Searle2008}. These relations are plotted in Fig. \ref{all_rectas}.

Particularly, we find that the difference between our ($X$, $D_o$) values and those of \citet{Searle2008} could be due to the fact that the sample of stars selected by these authors considered a mix of early and mid BSG types. In fact, if we split their sample,  we obtain  $X = 1.85\pm0.36$ and $D_0$ = $18.47\pm2.00$ for early BSGs,  and  $X = 1.35\pm0.14$ and $D_0$ = $21.08\pm0.73$ for the mid BSGs, which both agree with our values.

\subsection{The pulsation--mass-loss relationship}

\citet{Saio2011} demonstrated that one or more radial strange modes are 
expected to be excited in evolved BSGs. These authors computed the periods of excited modes during the evolution of only two stellar models: 
stars with initial masses  of 20\,M$_{\sun}$ and 25\,M$_{\sun}$. It appears that the periods of these radial strange modes tend to decrease when the effective temperature of the star increases during the blue-ward evolution. 

As radial strange modes are considered suitable to trigger mass loss, we first checked which of the known (photometric and spectroscopic) periods 
(see Table \ref{table:1}) might be considered as radial strange modes, and then opted for a comparison of these periods with the amplitude of the mass-loss 
rates (i.e. ratios between maximum and minimum values) using the $Q_{\rm r}$ parameters given in Table \ref{table:A1}.

\begin{figure}[h]
  \includegraphics[width=0.35\textwidth, angle=270]{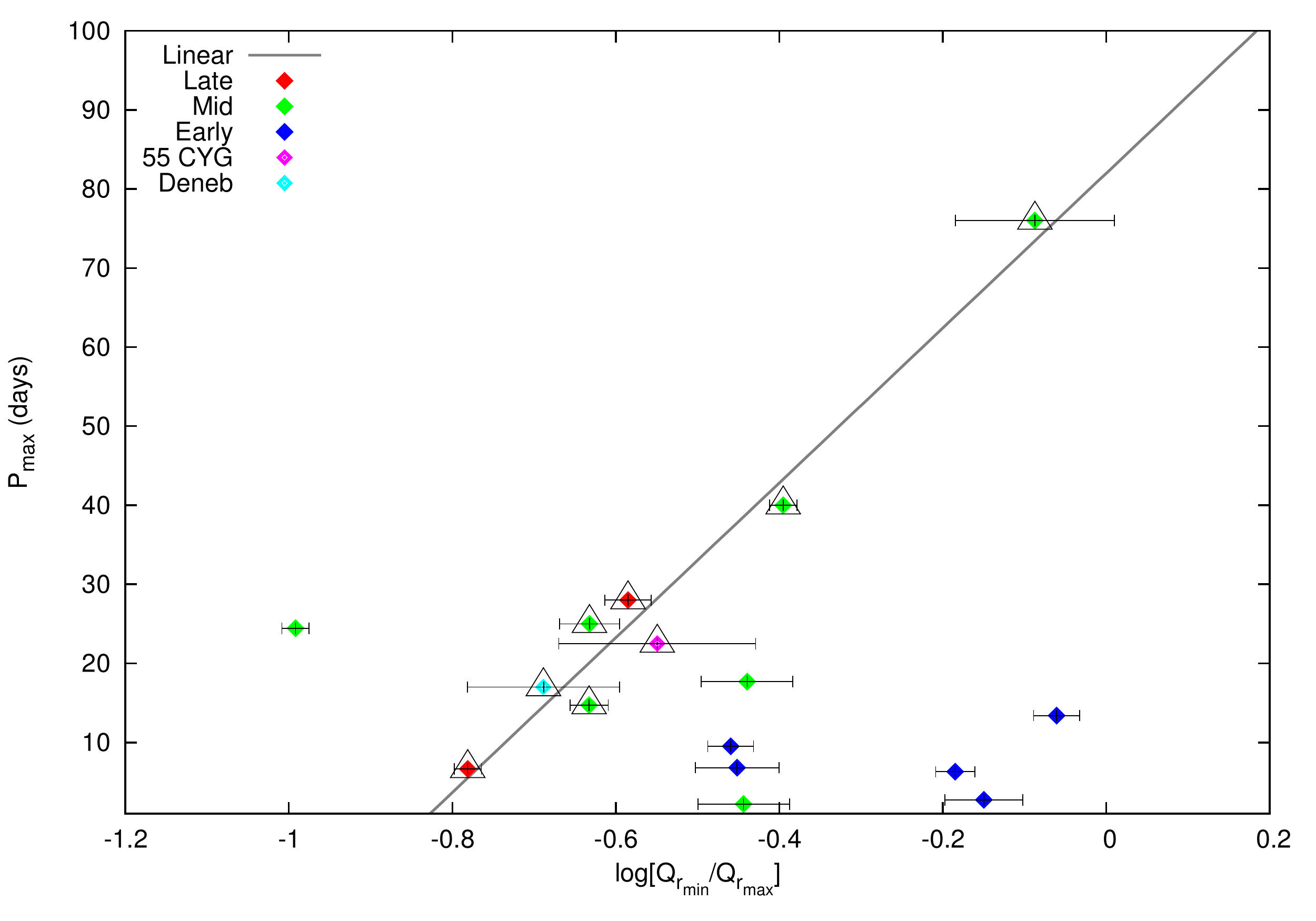}
\caption{Ratios between minimum and maximum $Q_r$ values as a function of the longest photometric/spectroscopic period, where $Q_r= \dot{M}/R_\star^{1.5}$. The linear correlation was obtained  using the up-pointing triangle.
\label{lomaximo}}
\end{figure}

According to the models of \citet{Saio2013}, the radial modes, especially in late-type stars, 
display the longest periods. Therefore, for each star we  searched for correlations with the longest known period.
As shown in Fig. \ref{lomaximo}, we found a  linear correlation  for stars mostly represented by mid/late BSGs (i.e. spectral type from B2 to B9)  with  P $>$ 6 days. The longer the period the lower 
  the amplitude of the parameter $\rm Q_r$. We  added to our sample two well-studied variable stars having measurements of their stellar and wind parameters \citep[\object{Deneb} and \object{55\,Cyg},][]{Scuderi1992, Chesneau2010,Kraus2015}. Three mid-type BSGs with measured  values of P and $\dot{M}$ were rejected from the sample: \object{HD 53138}, \object{HD\,75149}, and  \object{HD\,99953} (Fig. \ref{lomaximo}, green dots without triangles). Two of these stars \object{HD\,75149} and \object{HD\,99953} have been  poorly studied and  new periods should probably be determined. The former, in particular,  might have inaccurate  minimum values of $\dot{M}$ since two of our observed profiles are in absorption.  The star \object{HD\,111973} is not included since we have observations during an interval of time of two days which is less than the reported period of variation.


The relationship we obtained is $P_{\rm{max}} = 100.27(\pm 7.64)\, X + 83.97(\pm 4.47)$, where X = $\log\,\rm Q_{r_{\rm min}}\,/\,Q_{r_{\rm max}}$, and was derived using the data  enclosed in triangles (the used values of $Q_{r_{\rm min}}$ and $Q_{r_{\rm max}}$ are listed in bold  in Table \ref{table:A1}). The differences in the selected values are two or three times larger than their corresponding error bars, suggesting real variations in the mass-loss rates.

The group that comprises the early BSGs (\object{HD\,38771}, the binary star \object{HD\,47420}, \object{HD\,52089}, \object{HD\,64760,} and \object{HD\,148688}, blue  diamonds) does not show a clear correlation with their periods. Some of these stars show differences in their  $Q_r$ parameters of the order of the errors suggesting that  they do not show strong wind variability.

We can then conclude that as the observed line profiles can be matched using a velocity $\beta$-law, radiation pressure is the main mechanism that drives the wind. However, pulsation modes may also play a significant role in (qua\-si-)periodically lifting the material, which will then be accelerated outwards by radiation. In this way, radial strange-mode pulsation with periods $>$ 6 days might be  responsible for the wind variability in mid- and late-type BSGs.

\subsection{Location of the BSGs in the HR diagram.}

\begin{figure}
  \centering
 \includegraphics[width=0.35\textwidth, angle=270]{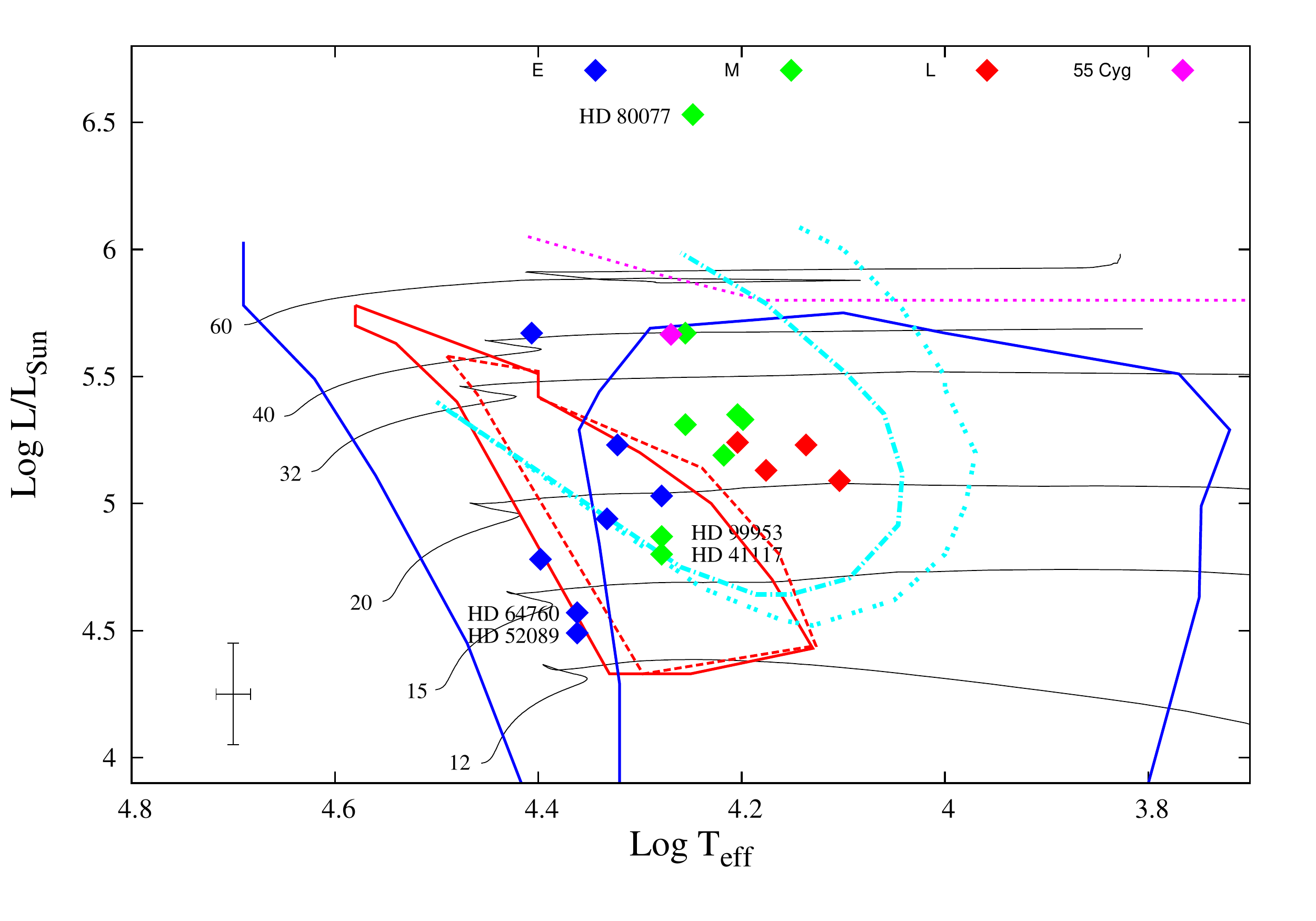}
  \caption{Position of the sample of stars in the HR
      diagram.  Evolutionary tracks \citep[taken from][black solid lines]{Ekstrom2012} correspond to non-rotating stars  with 
      $M_\star\,$= 12, 15, 20, 32, 40, and 60 M$_{\odot}$. The coloured curves  define instability boundaries of various modes \citep{Saio2011}:    a)  stability boundary for low-order radial and non-radial p modes (blue solid line); b) non-radial g modes with $l$=1 and 2 (red short-dashed and solid lines, respectively); c) oscillatory convection modes (cyan dot-dashed lines); d) monotonically unstable radial modes (above the magenta dotted-line, where the hypergiant star \object{HD\,80077} is found). The blue, green, red, and magenta diamonds  correspond to early (E), mid (M), and late (L) BSGs, and \object{55\,Cyg}, respectively. \label{HR}}

\end{figure}

Figure \ref{HR} displays the position of our sample stars in the HR diagram, using data from Table \ref{table:2},  together with the stability boundaries  for the  different pulsation modes, as  defined by \citet{Saio2013}. The stability boundary for low-order radial and non-radial $p$ modes is represented by the blue solid line. This bends and becomes horizontal due to strange-mode  instability  which occurs when $\log\,L/M > 4$. Non-radial modes (low-degree high-order g-modes) can be  excited  in the region delimited by red lines,  whilst monotonically unstable modes exist above the magenta dotted line in the most luminous part of the HR diagram (as is the case for the hypergiant star \object{HD\,80077}). On the other hand, two stars in our  sample (\object{HD\,52089} and \object{HD\,64760},  with $\log\, L/L_\sun\,  <$ 4.6, $\log\,L/M < 4$) are located inside the instability strip of the $\beta$ Cep, found  by \citet{Saio2013} and \citet{Georgy2014}, near the cool border of that region.  Particularly, \object{HD\,64740}  presents in the UV spectral region clear signs of phase bowing and ionization changes in the wind that could be associated with strong shocks originating in the interface between slow and fast wind  streams, such as in a model of co-rotating interaction regions 
 \citep{Prinja2002}, and its WISE image reveals a kind of wind-lobe structure. We propose that this star could be evolving towards the RSG stage.

In the same HR diagram  we can also see that the mid-/late-type stars are located in the region where oscillatory convection modes would be observable \citep[cyan lines,][]{Saio2011,Saio2013}, with the exception of \object{HD\,41117} and  \object {HD\,99953,} which are located together with the early B types. The star \object{55 Cyg} \citep[taken from][]{Kraus2015} was also included.

It is worth mentioning that WISE images reveal signs of previous strong wind-ISM interaction in three objects in our sample (\object{HD\,52382}, 
\object{HD\,99953,} and \object{HD\,115842}) pointing out  that these stars could be evolving towards the blue-loop after a post-RSG phase. This same  also holds true for  \object{55 Cyg} \citep{Kraus2015}.

\section{Conclusions}\label{con}

We studied a sample of 19 Galactic BSGs by fitting theoretical line profiles of H, He, and Si
to the  observed profiles. The synthetic line profiles were computed with the NLTE atmosphere code FASTWIND to obtain new estimates for their 
stellar and wind parameters. The mass-loss rates for  \object{HD\,74371},  \object{HD\,99953},  and  \object{HD\,111973} have  been derived for the  first time. In addition, some BSGs in our sample show short-term variability  in the  H$\alpha$ and UV resonance lines, indicating important changes in the wind structure.

In relation with the global properties of the stellar winds of our sample, we find two different behaviours in the WLR.  The early and mid B-type stars  have mostly $\beta < 2$ and terminal velocities greater than 500 km s$^{-1}$, while the other group is comprised mostly by  mid and  late B-type stars with $\beta \geq 2$  and terminal velocities lower than 500 km s$^{-1}$. In addition, we find that the coefficients for the linear regression of early BSGs agree with theoretical or empirical values found by \citet{Vink2000, Herrero2002, Mokiem2007}. As well, our coefficients would agree with data from \citet{Searle2008} if we separated their sample of stars into two groups: early and mid BSGs.

Furthermore, we observe that  a decrease in the ratio $V_\infty/V_{\rm{esc}}$ in the bi-stability region, and on its cool side, is accompanied by a decrease in mass-loss rate. This observation is contrary to the theoretical result obtained by \citet{Vink1999}, based on a fast wind regime. Instead, this behaviour seems to be consistent with predictions from highly ionized stellar winds  \citep[$\delta-slow$ solution,][]{Venero2016}. The possible switch between fast and $\delta-slow$ solutions should  be explored in detail.

Finally, we found an empirical correlation (period--mass-loss relationship) that associates the amplitude of the mass-loss variations with long-term
photometric and/or spectroscopic variability. This relation is expressed in terms of the $Q_r$ parameter.  The found period--mass-loss relationship could indicate that radial (or  strange) pulsation modes with periods $>$ 6 days  are related to the  wind variability. To confirm 
this trend, it is necessary to perform long-term high-resolution spectroscopic campaigns to properly measure oscillation periods and simultaneously derive  homogeneous sets of mass-loss rates.




\begin{acknowledgements}

  We thanks our anonymous referee for the  helpful comments and suggestions. We also want to  acknowledge  J. Puls  for  allowing  us  to  use the  FASTWIND  code,  for  his  help  and  advice  with  the  code, and for the  fruitful suggestions.

  This research has also made use of the SIMBAD database, operated at CDS, Strasbourg, France, and  the  International Variable Star Index (VSX)  database, operated at AAVSO, Cambridge, Massachusetts, USA.

  L.C. and M.C. acknowledge support from the project CONICYT + PAI/Atracci\'on de capital humano avanzado del extranjero (folio PAI80160057). L.C. also acknowledges financial support from CONICET (PIP 0177), La Agencia Nacional de Promoción Científica y Tecnológica (PICT 2016-1971) and the Programa de Incentivos (G11/137) of the Universidad Nacional de La Plata (UNLP), Argentina. R.V. is grateful for the  finantial support from the UNLP under programme PPID/G003. M.C., S.K., and C.A. acknowledge support from Centro de Astrof\'{\i}sica de Valpara\'{\i}so. C.A. also thanks  BECAS DE DOCTORADO NACIONAL CONICYT 2016-2017.

M.K. acknowledges support from GA\v{C}R (17-02337S). The Astronomical Institute Ond\v{r}ejov is supported by the project RVO:67985815.

This work was partly supported by the European Union European Regional Development Fund,

project ``Benefits for Estonian Society from Space Research and Application'' (KOMEET, 

2014\,-\,2020.\,4.\,01.\,16\,-\,0029), and by the institutional research funding IUT$40-1$ of the Estonian Ministry of Education and Research.

Financial support for International Cooperation of the Czech Republic and Argentina (AVCR-CONICET/14/003) is acknowledged.

\end{acknowledgements}





\bibliographystyle{aa} 

\bibliography{references} 




\begin{appendix}

  \section{Photospheric line-fitting and compiled data from the literature}

  \label{ape1}

The following figures present a comparison between observed and synthetic photospheric line profiles calculated with FASTWIND and the parameters quoted in 
Table \ref{table:2}. These parameters are also given in Table \ref{table:A1} together with a set of stellar and wind parameters gathered from the literature.  

\clearpage

\onecolumn

\begin{landscape}

  \tiny

   \tabcolsep 2.5pt

   \begin{longtable}{lcccccccccccccl}

       \caption{\label{table:A1} Stellar and wind parameters.}\\

\hline\hline

Name   & $T_{\rm{eff}}$ & $\log\,g_{\rm{eff}}$  & $\log\,g_{\rm{cor}}$ & $\beta$  & $\dot{M}$           & $V_{\infty}$ &  $V_{\rm{micro}}$ & $V_{\rm{macro}}$  & $V\,sin\,i$ & $R_{\star}$ & $\log\,(L/L_{\sun})$ & $\log\,Q$ &$\log\,Q_{\rm r}$ & References\\

          &\footnotesize{K} &  \footnotesize{dex}    &  \footnotesize{dex}    & &\footnotesize{$10^{-6}$ $\rm M_{\odot}\,\rm yr^{-1}$}    & \footnotesize{km\,s$^{-1}$}     &  \footnotesize{km\,s$^{-1}$}   & \footnotesize{km\,s$^{-1}$} &\footnotesize{km\,s$^{-1}$}  &\footnotesize{$R_{\odot}$} & \footnotesize{dex} & & &\\

\hline

\endfirsthead

\caption{Continued.}\\

\hline\hline

Name    & $T_{\rm{eff}}$& $\log\,g_{\rm{eff}}$  & $\log\,g_{\rm{cor}}$& $\beta$  & $\dot{M}$           & $V_{\infty}$ &  $V_{\rm{micro}}$ & $V_{\rm{macro}}$  & $V\,sin\,i$ & $R_{\star}$ & $\log\,L/L_{\sun}$ & $\log\,Q$ & $\log\,Q_{\rm r}$ & References\\

 &\footnotesize{K} &  \footnotesize{dex}    &  \footnotesize{dex}    & &\footnotesize{$10^{-6}$ $M_{\odot}\,yr^{-1}$}    & \footnotesize{km\,s$^{-1}$}     &  \footnotesize{km\,s$^{-1}$}   & \footnotesize{km\,s$^{-1}$} &\footnotesize{km\,s$^{-1}$}  &\footnotesize{$\rm R_{\odot}$} & \footnotesize{dex} & &  & \\

\hline

\endhead

\hline

\endfoot

\object{HD\,34085} & $12\,700$ & $1.70$   &  $1.72$   &  $2.6$   &  $0.23$ &   $155$      & $10$  &  $52$ &  $30$ & $72$   &  $5.09$ & $-12.71$& $-9.42\pm0.17$ & This work\\

    & $12\,000$ &   $\cdots$   &$\cdots$ &  $\cdots$      & $0.76$ / $0.94$  &   $300$   &   $\cdots$    &    $\cdots$    &  $36$  & $115$ &  $5.45$  & $-12.93$ / $-12.83$ & $-9.21$ / $-9.12$ &\citet{Chesneau2014}\\

    & $12\,100$ & $1.75$ &$\cdots$&  $\cdots$      &   $\cdots$           &       $\cdots$        &   $\cdots$  &   $\cdots$     &  $25$  &   $\cdots$     &   $\cdots$        &$\cdots$& $\cdots$&\citet{Firnstein2012}\\

    & $12\,500$ & $1.70$ &  $\cdots$ & $\cdots$ & $0.34$  & $230$ / $350$ &   8   &  $35$  &  $30$  & $129$ &  $5.56$  & $-13.35$ / $-13.08$ &  {\bf $-$9.63$\pm$0.17} &\citet{Markova2008}\\

    & $12\,000$ & $1.75$ & $\cdots$&  $\cdots$      &    $\cdots$ &    $\cdots$     & $\cdots$    &  $22$  &  $36$  & $109$ &  $5.34$ &$\cdots$& $\cdots$&\citet{Przybilla2006}\\

    & $13\,000$ & $1.75$ &$\cdots$&  $\cdots$      &     $\cdots$         &    $\cdots$     &  $\cdots$   &  $40$   &    $\cdots$   &   $\cdots$     &  $4.87$  &$\cdots$& $\cdots$&\citet{McErlean1999}\\

    & $13\,000$ & $1.60$ &$\cdots$&  $\cdots$      &     $\cdots$         & $400$ / $600$ &  $\cdots$    &  $40$   &    $\cdots$   &  $\cdots$      &  $\cdots$         &$\cdots$& $\cdots$&\citet{Israelian1997}\\

    &  $\cdots$   &$\cdots$&$\cdots$&  $\cdots$      & $0.86$ /  $1.4$   & $530$     &  $\cdots$   &   $\cdots$      &  $33$  & $135$ &  $\cdots$   & $-13.35$ / $-13.14$& $-9.26$ /  {\bf $-$9.05} & \citet{Nerney1980} \\

\hline            

\object{HD\,38771}         & $25\,000$ & $2.70$ &  $2.76$ & $1.5$  & $0.14$ &  $1\,500$    &  $13$   &  $60$   & $ 80$ &  $13$     & $4.78$ &   $-13.29$& {\bf $-$8.52$\pm$0.20} &This work\\

                                          &  $26\,000 $ &  $3.00 $ & $\cdots$ &   $ 1.1 $  &  $1.20 $ &   $1390 $    &  $15 $    &  $\cdots$       &  $ 91 $ &   $27.0 $   &  $5.48 $ & $-12.78 $ & {\bf $-$8.07$\pm$0.17} & \citet{Searle2008} \\

                                          & $26\,500$ & $2.70$ &  $2.90$ &   $1.5$  & $0.90$ &  $1\,525$    & $12.5$ &  $\cdots$ &  $83$ &  $22.2$  &  $5.35 $ & $-12.84 $ &  $-8.07$ & \citet{Crowther2006}\\

                                          & $27\,500$ & $3.00$ &  $\cdots$       &    $1.0 $  &  $0.27$ &   $1\,350 $    &  $10   $  &        &   $80 $ &   $13.0 $   &  $4.9 $4 & $-12.94$& $-8.24\pm0.13$ &\citet{Kudritzki1999}\\

                                          & $27\,500$ & $3.00$ &   $\cdots$      &      $\cdots$     &     $\cdots$      &        $\cdots$         &   $\cdots$      &     $\cdots$    &  $80$ &       $\cdots$      & $4.68$  & $\cdots$&  $\cdots$&\citet{McErlean1999}\\

& $26\,000$ & $3.07$ & $2.94$ &       $\cdots$    &      $\cdots$     &  $1\,870$    &     $\cdots$     &    $\cdots$     &  $81$ &  $28$     & $5.52$ &  $\cdots$&  $\cdots$&\citet{Gathier1981} \\

&    $\cdots$      &     $\cdots$  & $\cdots$&    $\cdots$     & $0.7$ /  $1.1$   & $1\,900$ / $3\,450$     &   $\cdots$   &    $\cdots$      &  $82$  & $30$ &    $\cdots$       & $-13.68$ / $-13.09$ & $-8.37$ / $-8.10$ &\citet{Nerney1980} \\

\hline

\object{HD\,41117} & $19\,000$ & $2.30$  &  $2.32$       & $2.0$  & $0.17$   &   $510$   & $10$  & $65$ &  $40$  & $23$    &  $4.84$  & $-12.78$& $-8.81\pm0.14$ &This work\\

                                 & $20\,000$ & $2.40$ &     $\cdots$    &     $\cdots$   &     $\cdots$      &     $\cdots$       &  $\cdots$     &   $\cdots$   &     $\cdots$   &    $\cdots$      &     $\cdots$      & $\cdots$ &  $\cdots$&\citet{Prugniel2011}\\ 

                                 & $19\,000$ & $2.10$ & $2.35$ & $2.0$  & $0.90$   &   $510$   & $10$  &  $\cdots$   &   $72 $  &  $61.9 $ &  $5.65 $  & $-12.79$ & $-8.73$ & \citet{Crowther2006}\\

                                 & $19\,500$ & $2.25$ &     $\cdots$     & $1.0$  & $0.85$   &   $500$   & $20$  &   $\cdots$    &  $ 40 $  &  $61.7 $ & $5.70$  & $-12.8$& {\bf $-$8.76$\pm$0.13} &\citet{Kudritzki1999}    \\

                                 & $19\,500$ & $2.20$ &    $\cdots$      &     $\cdots$    &       $\cdots$     &      $\cdots$       &    $\cdots$    & $65$  &    $\cdots$    &    $\cdots$      & 5.55  & $\cdots$ &  $\cdots$&\citet{McErlean1999}\\

& $18\,500$ &    $\cdots$     &      $\cdots$     &    $\cdots$      & $1.6$     &  $510$   &    $\cdots$   &     $\cdots$   &   $\cdots$     & $58$     &     $\cdots$     & $-12.5$&  $-8.44$ &\citet{Scuderi1998}  \\

&      $\cdots$    &   $\cdots$    & $\cdots$&      $\cdots$   & $2.2$   & $1\,000$    &   $\cdots$   &      $\cdots$    &  $36$  & $63.4$ &      $\cdots$     & $-12.86$ & {\bf $-$8.36}&\citet{Nerney1980} \\

\hline

\object{HD\,42087}         & $16\,500$ & $2.45$ &  $2.47$     & $2.0$  & $0.57$   & $700$ & $15$ & $80$  &  $80$ & $55$   & $5.31$   & $-13.12$&  $-8.85\pm0.11$ & This work\\

                                          & $18\,000$ & $2.50$ & $\cdots$ & $1.2$  & $0.50$  & $650$ & $15$ & $\cdots$ &  $71$ & $36.6$ & $5.11$   & $-12.87$ & {\bf $-$8.65$\pm$0.33}  &\citet{Searle2008}\\

                    & $19\,000$ & $2.30$ & $\cdots$&   $\cdots$    & $0.20$  &   $\cdots$   &   $\cdots$  &    $\cdots$  &   $\cdots$   &     $\cdots$  & $5.08$   & $\cdots$ &  $\cdots$&\citet{Benaglia2007}\\

                    & $17\,150$ &   $\cdots$    & $\cdots$&   $\cdots$    & $0.20$  & $735$ &   $\cdots$  &    $\cdots$  &  $71$ & $26.0$ & $4.72$   & $-13.12$& $-8.82$&\citet{Morel2004}\\

                    & $20\,500$ & $2.50$ & $\cdots$& $3.0$  & $0.11$  & $735$ & $40$ &    $\cdots$  & $60$ & $35.2$ & $5.30$  & $-13.58$& {\bf $-$9.28$\pm$0.13}&\citet{Kudritzki1999}\\

                    & $20\,500$ & $2.55$ & $\cdots$&    $\cdots$   &  $\cdots$      &   $\cdots$   &   $\cdots$  &    $\cdots$  &  $60$ &    $\cdots$   & $4.96$   & $\cdots$ &  $\cdots$&\citet{McErlean1999}\\

\hline

\object{HD\,47240}         & $19\,000$ & $2.40$                  &    $2.47$             & $1.0$ & $0.24$           &  $450$  & $10$          &  $60$  &  $95$ & $30$   & $5.03$  & $-12.82$ & {\bf $-$8.84$\pm$0.11}&This work\\

                                         & $19\,000$ &  $2.40$        & $2.48$ & $1.5$  & $0.17-0.24$  & $1\,000$  & $15$ &  $55$  &  $94$ & $27$   & $4.93$   & $-13.42$ / $-13.27$ & {\bf $-$8.92$\pm$0.25 / $-$8.77$\pm$0.25} &\citet{Lefever2007}\\

\hline

\object{HD\,52089}         & $23\,000$ & $3.00$ &  $3.00$    & $1.0$  & $0.02$  & $900$  & $8$  & $65$  & $10$  & $11$  & $4.49$ & $-13.69$ & $-9.26\pm0.2$ & This work\\

                                         & $20\,100$ & $3.05$ &     $\cdots$          &    $\cdots$    &     $\cdots$      &   $\cdots$       & $17$ & $17$  & $22$  &   $\cdots$     &     $\cdots$     &          $\cdots$       &          $\cdots$     &\citet{Fraser2010}\\

                    & $22\,000$ & $3.20$ &   $\cdots$    &   $\cdots$    &   $\cdots$    &    $\cdots$   & $15$  & $20$  & $32$      &  $\cdots$    &   $\cdots$    &  $\cdots$&  $\cdots$&\citet{Lefever2010}\\

                    & $22\,200$ & $3.22$ &    $\cdots$   &    $\cdots$   &    $\cdots$   &    $\cdots$   & $16$  &   $\cdots$   &    $\cdots$  &   $\cdots$   &     $\cdots$  &  $\cdots$&  $\cdots$&\citet{Morel2008}\\

\hline

\object{HD\,52382}        & $21\,500$ & $2.45$ &   $2.48$    & $2.2$  & $0.24$  & $1\,000$  & $10$ & $65$ & $55$  & $21$ & $4.94$ &$-13.10$ & $-8.60\pm0.13$&This work\\

                                         & $23\,000$ &     $\cdots$     & $2.71$  &   $\cdots$    &   $\cdots$     &     $\cdots$     & $20$ & $53$  & $56$  &   $\cdots$  &    $\cdots$   & $\cdots$&  $\cdots$&\citet{Lefever2010}\\

                    & $20\,800$ &    $\cdots$   &    $\cdots$    &     $\cdots$  & $0.49$  & $1\,200$  &  $\cdots$   &     $\cdots$ &     $\cdots$ &  $\cdots$   &  $\cdots$     & $\cdots$&  $\cdots$&\citet{Kri2001}\\

\hline

\object{HD\,53138} & $18\,000$ & $2.25$ &  $2.26$       & $2.0$ & $0.2$ / $0.24$    & $600$ / $450$   & $10$   & $60$     & $40$ &  $46$  & $5.31$  &$-13.36$ / $-13.09$ & $-9.19\pm0.11$ / $-9.11\pm0.11$ &This work\\

                                  & $15\,400$ & $2.15$ &       $\cdots$          &      $\cdots$  &       $\cdots$            &          $\cdots$        & $18$   & $23$     & $35$ &     $\cdots$    &     $\cdots$      &        $\cdots$                         &           $\cdots$                   &\citet{Fraser2010}\\

                                  & $16\,500$ & $2.25$ & $\cdots$ & $1.2$ &  $0.45$      & $500$   & $20$   &  $\cdots$  & $58$ &  $54.7$  & $5.30$   &    $-13.00$            & $-8.95\pm0.17$&\citet{Searle2008}\\

                                  & $17\,000$ & $2.15$ & $2.16$ & $2.5$ & $0.21$ / $0.31$   & $490$  & $15$  & $45$     &  $38$ &  $50$    & $5.27$   & $-13.20$ & $-9.23\pm0.25$ / $-9.06\pm0.25$ &\citet{Lefever2007}\\

                                  & $15\,500$ & $1.75$ & $2.05$ & $2.0$ & $0.36$           & $865$  & $20$  & $\cdots$ & $58$  &  $65$  & $5.34$    & $-13.57$ & $-9.16$&\citet{Crowther2006}\\

                                  & $18\,500$ & $2.30$ &     $\cdots$    & $2.5$ & $0.095$          & $620$  &    $\cdots$   &  40       & $60$  &  $39.6$ & $5.22$   &       $\cdots$          & {\bf $-$9.42$\pm$0.13} &\citet{Kudritzki1999}\\

                                  & $18\,500$ & $2.35$ &     $\cdots$    &    $\cdots$    &     $\cdots$                 &    $\cdots$      &     $\cdots$   &        $\cdots$        & $55$ &      $\cdots$      & $5.04$   &        $\cdots$         &  $\cdots$&\citet{McErlean1999}\\

                                  &     $\cdots$          &    $\cdots$      &    $\cdots$     &    $\cdots$     & $1.7$               & $580$ &      $\cdots$   &       $\cdots$        &      $\cdots$   &  $59.1$ &    $\cdots$          &       $\cdots$         & {\bf $-$8.43} &\citet{Nerney1980}   \\

\hline 

\object{HD\,58350} & $15\,000-16\,000$ & $2.00$  &  $2.02$    &$ 3.0$  & $0.116$-$0.15$   & $175$-$233$    &  $11$ &  $50$ / $70$  & $40$ &  $54$  & $5.13$-$5.24$   &$-12.9$ / $-12.97$& $-9.53\pm0.13$ / $-9.42\pm0.13$&This work\\

                                  & $14\,500$              & $2.10$  &       $\cdots$        &    $\cdots$     &      $\cdots$                  &         $\cdots$           &  $18$   &  $25$  & $32$ &    $\cdots$    &     $\cdots$          & $\cdots$  &  $\cdots$ &\citet{Fraser2010}\\

            & $15\,000$         & $2.13$ &  $\cdots$  & $1.0$  &  $0.70$        & $320$        &   $20$  &  $\cdots$   & $50$ &  $57.3$  & $5.18$  & $-12.55$ & {\bf $-$8.79$\pm$0.23}&\citet{Searle2008}\\

            & $13\,500$         & $1.75$ & $1.77$ & $2.5$  & $0.14$        & $250$        &   $12$  &  $40$ & $37$ &  $65$  & $5.10$        & $-13.17$ & {\bf $-$9.57$\pm$0.25} &\citet{Lefever2007}\\

& 16\,000         & 2.10 &    $\cdots$    &   $\cdots$     &        $\cdots$       &       $\cdots$       &    $\cdots$     &  40  &  $\cdots$    &      $\cdots$  & $5.36$        & $\cdots$  &  $\cdots$ &\citet{McErlean1999}\\

\hline

 \object{HD\,64760} & $23\,000$ & $2.90$ &  $3.09$    & $0.5$  & $0.42$  & $1\,500$  & $15$  &  $100$ &  $230$ &  $12$  & $4.57$  & $-12.76$ & {\bf $-$8.00$\pm$0.13} & This work \\

                                   & $26\,000$     & $3.25$ &           $\cdots$       &     $\cdots$          &       $\cdots$         &       $\cdots$            & $22$ &  $\cdots$  &  $255$ &    $\cdots$    &      $\cdots$    & $\cdots$  & $\cdots$  &\citet{Fraser2010}\\

            & $28\,000$ &   $3.38$  &  $\cdots$  & $1.0$ & $1.10$  & $1\,600$ & 15 &  $\cdots$   &  $265$ & $23.3$ & $5.48$   & $-12.82$  & $-8.01\pm0.17$ &\citet{Searle2008}\\

            & $24\,000$ &  $3.20$ & $3.27$ & $0.8$     & $0.42$  & $1\,400$ &  $14$  &  $\cdots$  &  $230$ &  $24$  & $5.23$   & $-13.16$ & {\bf $-$8.45$\pm$0.25} &\citet{Lefever2007}\\

\hline

\object{HD\,74371} & $13\,700$   & $1.80$  & $1.81$ & $2$   & $0.28$ &  $200$  &  $10$  &   $60$  &   $30$ &   $73$  & $5.23$ & $-12.8$ & $-9.35\pm0.11$ &This work\\

            & $13\,400$ & $1.90$ & $\cdots$ &   $\cdots$     &     $\cdots$   &    $\cdots$     &   $\cdots$     &   $20$  &   $30$ &   $31$  &    $\cdots$    & $\cdots$  &  $\cdots$ &\citet{Fraser2010}  \\

\hline

 \object{HD\,75149} & $16\,000$ & $2.10$   &   $2.11$       &  $2.5$  & $0.09$ / $0.25$   & $400$ /  $350$ &  $9$ / $17$      & $55$ / $60$ &  $40$    & $61$   & $5.35$  & $-13.63$ / $-13.1$ & {\bf $-$9.72$\pm$0.15 / $-$9.28$\pm$0.15} &This work\\

                                                          & $15\,900$ & $2.20$  &    $\cdots$      &    $\cdots$    &     $\cdots$               &      $\cdots$         & $20$          & $34$      & $30$     &     $\cdots$  &      $\cdots$       &  $\cdots$ &  $\cdots$ &\citet{Fraser2010}\\

                                                          & $16\,000$ & $2.05$  & $2.06$ & $2.5$   & $0.10$           & $500$        & $15$          & $60$      & $30$     & $39$  & $4.95$   & $-13.43$& $-9.39\pm0.25$ &\citet{Lefever2007}\\

\hline

\object{HD\,79186} & $15\,800$ & $2.0$ &  $2.02$     & $3.3$   & $0.40$   & $400$ &  $11$ &  $53$  & $40$  & $61$   & $5.33$   & $-12.98$& $-9.08\pm0.09$ &This work\\

                                  & $15\,100$ & $2.0$ & $\cdots$ & $\cdots$ & $\cdots$ & $\cdots$ &  $14$ & $36$  & $39$  &   $\cdots$ & $\cdots$ & $\cdots$ & $\cdots$ & \citet{Fraser2010}\\

                                  & $15\,000$ &    $\cdots$   & $\cdots$ &  $\cdots$      & $0.80$   & $435$ &   $\cdots$    &    $\cdots$    &  $\cdots$     &    $\cdots$     &      $\cdots$    &$\cdots$ & $\cdots$& \citet{Prinja2010}\\          

                                  & $13\,600$ &  $\cdots$     & $\cdots$&  $\cdots$      & $0.66$  & $450$ &   $\cdots$     &    $\cdots$    &     $\cdots$   & $62.4$ &   $\cdots$      & $-12.85$& $-8.87\pm0.09$ &\citet{Kri2001}  \\

                                  & $13\,900$ &   $\cdots$    & $\cdots$&  $\cdots$      &      $\cdots$     &   $\cdots$     &  $\cdots$      &      $\cdots$   & $45$  &  $81$  &    $\cdots$      & $\cdots$& $\cdots$&\citet{Under1984} \\

\hline 

\object{HD\,80077}  & $17\,700$ & $2.20$   & $2.20$  & $3-3.2$ & $5.4$  &      $150$ / $200$     &  $10$    &   $60$  &   $10$ &    $195$ & $6.53$ & $-11.97$ / $12.15$& {\bf $-$8.70$\pm$0.11} &This work\\

                                   & $17\,000$ & $2.00$   & $\cdots$&        $\cdots$  & $5.00$      & $ 140$    &      $\cdots$       &      $\cdots$        &     $\cdots$       &    $162$ & $6.3$   & $-11.83$& {\bf $-$8.62$\pm$0.65}&\citet{Carpay1989} \\

                                   & $18\,500$ & $2.25$   & $\cdots$&     $\cdots$     & $1.70$      &    $ 140$    &   $\cdots$           &   $\cdots$           &     $\cdots$       &      $\cdots$        & $5.4$   & $\cdots$& $\cdots$&\citet{Benaglia2007} \\ 

\hline 

\object{HD\,92964}  & $18\,000$ & $2.20$ &  $2.21$     & $2.0$    & $ 0.49$        & $370$        &   $11$  & $40$ & $45$ &  $70$   & $5.67$  & $-12.93$ & $-9.08\pm0.09$&This work\\

                                   & $15\,600$ & $2.00$ &   $\cdots$    &    $\cdots$      &        $\cdots$        &      $\cdots$         &   $22$  & $28$ & $36$ &    $\cdots$    &  $\cdots$        &$\cdots$ & $\cdots$&\citet{Fraser2010}\\

& $18\,000$ & $2.10$ & $2.10$ &  $3.0$   & $0.25$ / $0.28$        & $520$        &   $15$  & $50$ & $31$ &  $48$ &  $5.33$   & $-13.14$ / $-13.19$ & {\bf $-$9.07$\pm$0.25} & \citet{Lefever2007}\\

                                   & $17\,400$ &    $\cdots$   &   $\cdots$    &     $\cdots$     & $0.11$        & $530$ / $807$ &      $\cdots$      &     $\cdots$  &     $\cdots$   & $68$   &          & $-13.79$ / $-14.07$& {\bf $-$9.71} &\citet{Kri2001}    \\

\hline

\object{HD\,99953}  & $19\,000$ & $2.30$  & $2.33$ & $2.0$  & $0.08$ / $0.22$     & $250$ / $700$    &  $18$   &  $50$  & $50$      & $25$   & $4.87$ & $-12.79$ / $-13.02$ & {\bf $-$9.19$\pm$0.11 / $-$8.75$\pm$0.11} &This work \\

                                   & $16\,800$ & $2.15$  &$\cdots$&    $\cdots$        &         $\cdots$                &         $\cdots$             &  $22$   &  $37$  & $49$      &    $\cdots$        &      $\cdots$       & $\cdots$ & $\cdots$  &\citet{Fraser2010}\\

\hline            

\object{HD\,111973} & $16\,500$ & $2.10$  &  $2.11$ & $2.0$  & $0.14$ / $0.21$ &   $350$    & $12$ & $\sim60$  & $35$ & $48$   & $5.16$  & $-13.19$ / $-13.02$ & $-9.38\pm0.09$ / $-9.20\pm0.09$ &This work \\

                                    & $16\,000$ & $2.30$  &&      $\cdots$      &         $\cdots$            &     $\cdots$            & $19$ &  $28$ & $36$ &   $\cdots$         &         $\cdots$     & $\cdots$&$\cdots$ &\citet{Fraser2010}\\

                                    & $16\,500$ &         $\cdots$     &$\cdots$& $0.5$  &           $\cdots$           &   $520$   &     $\cdots$     &     $\cdots$      &       $\cdots$   &  $\cdots$          &     $\cdots$         & $\cdots$& $\cdots$&\citet{Prinja2010}\\ 

\hline            

\object{HD\,115842} & $25\,500$ & $2.75$ &  $2.76$    & $2.5$  & $1.8$  & $1\,700$    &  $10$  & $120$ &  $70$  & $35$   & $5.68$  & $-12.91$ & {\bf $-$8.06$\pm$0.13} &This work \\

            & $24\,800$ & $2.75$ &   $\cdots$   &  $\cdots$    &  $\cdots$    &  $\cdots$         &  $14$  &  $66$ &  $39$  &  $\cdots$    &   $\cdots$    & $\cdots$& $\cdots$&\citet{Fraser2010}\\

& $25\,500$ & $2.65$ & $2.85$ & $1.5$  & $2.0$  & $1\,180$    &  $10$ & $\cdots$    &  $84$  & $34.2$ & $5.65$  & $-12.61$ & {\bf $-$8.00}&\citet{Crowther2006}\\

\hline

\object{HD\,148688} & $21\,000$ & $2.45$  &  $2.47$    & $2.5$  &  $1.15$  & $1\,200   $ & $ 11 $ & $ 65 $ & $ 50  $ & $ 31 $ & $ 5.23   $ &  $-12.8$  & $-8.18\pm0.15$ &This work \\

                                    & $20\,700$ & $2.45$  &    $\cdots$         &   $\cdots$      &    $\cdots$         &      $\cdots$       & $16$ &  $44$  &  $48$   &  $\cdots$     &     $\cdots$       &       $\cdots$         &      $\cdots$         & \citet{Fraser2010}\\

                                    & $21\,000$ & $2.50$  & $2.5$ &  $3.0$  &   $1.10$ / $1.40$   & $1\,200$   &  $15$ &  $40$      &   $50$  &  $42$    & $5.49$   & $-12.90$ / $-13.01$ & {\bf $-8.29\pm0.25$} & \citet{Lefever2007}\\

                                    & $22\,000$ & $2.40$  &   $2.60$ &  $2.0$  &   $1.75$         & $725$     &  $15$ &     $\cdots$     &  $72$            & $36.7$  & $5.45$   & $-12.39$   & {\bf $-$8.10} & \citet{Crowther2006}\\


   \end{longtable}

   \let\thefootnote\relax\footnote{Highlighted values 
were used to make Fig.\,\ref{lomaximo}.}   

\end{landscape}

\begin{figure*}[h]

  \includegraphics[width=0.9\textwidth, angle=0]{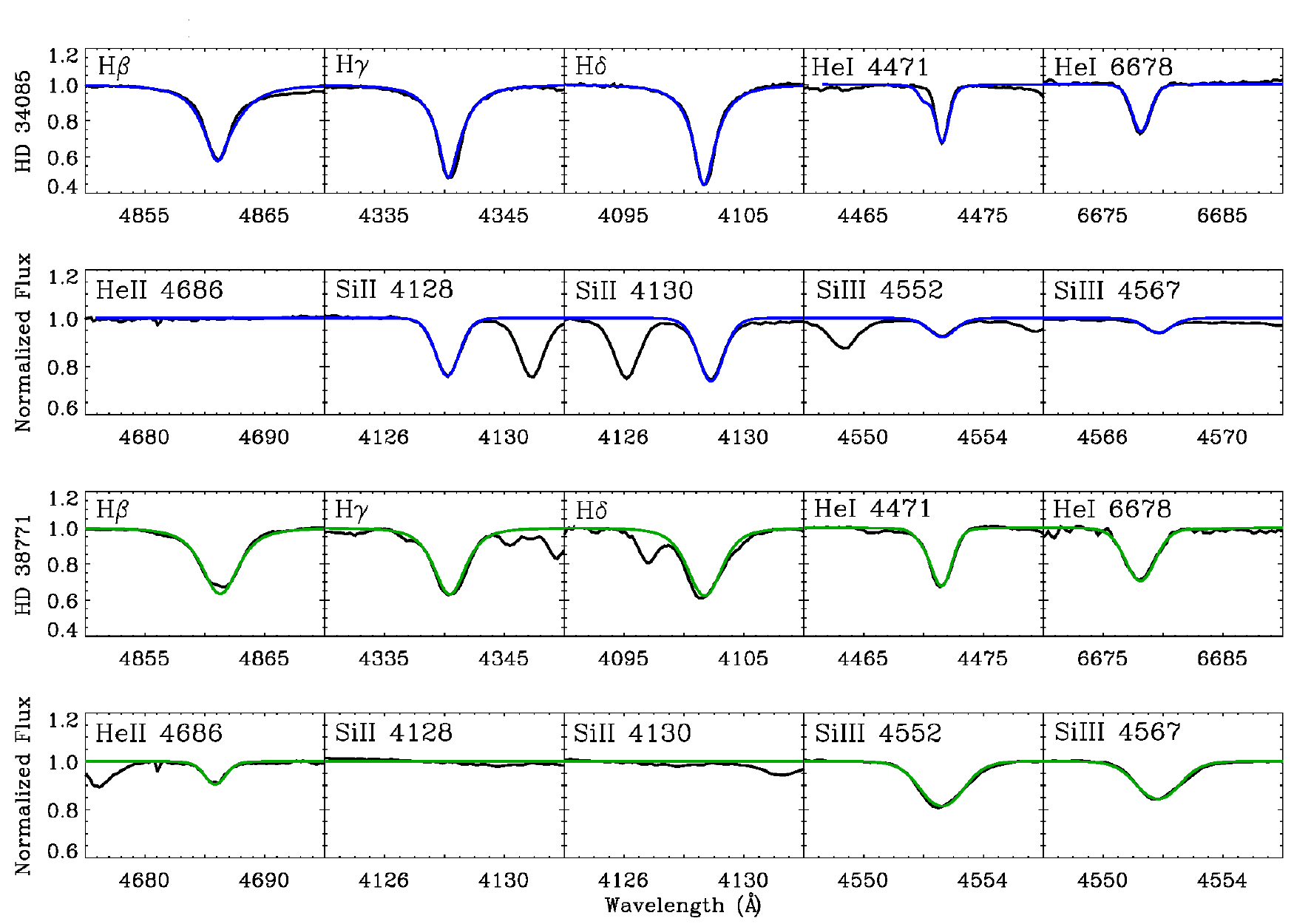}

\includegraphics[width=0.9\textwidth, angle=0]{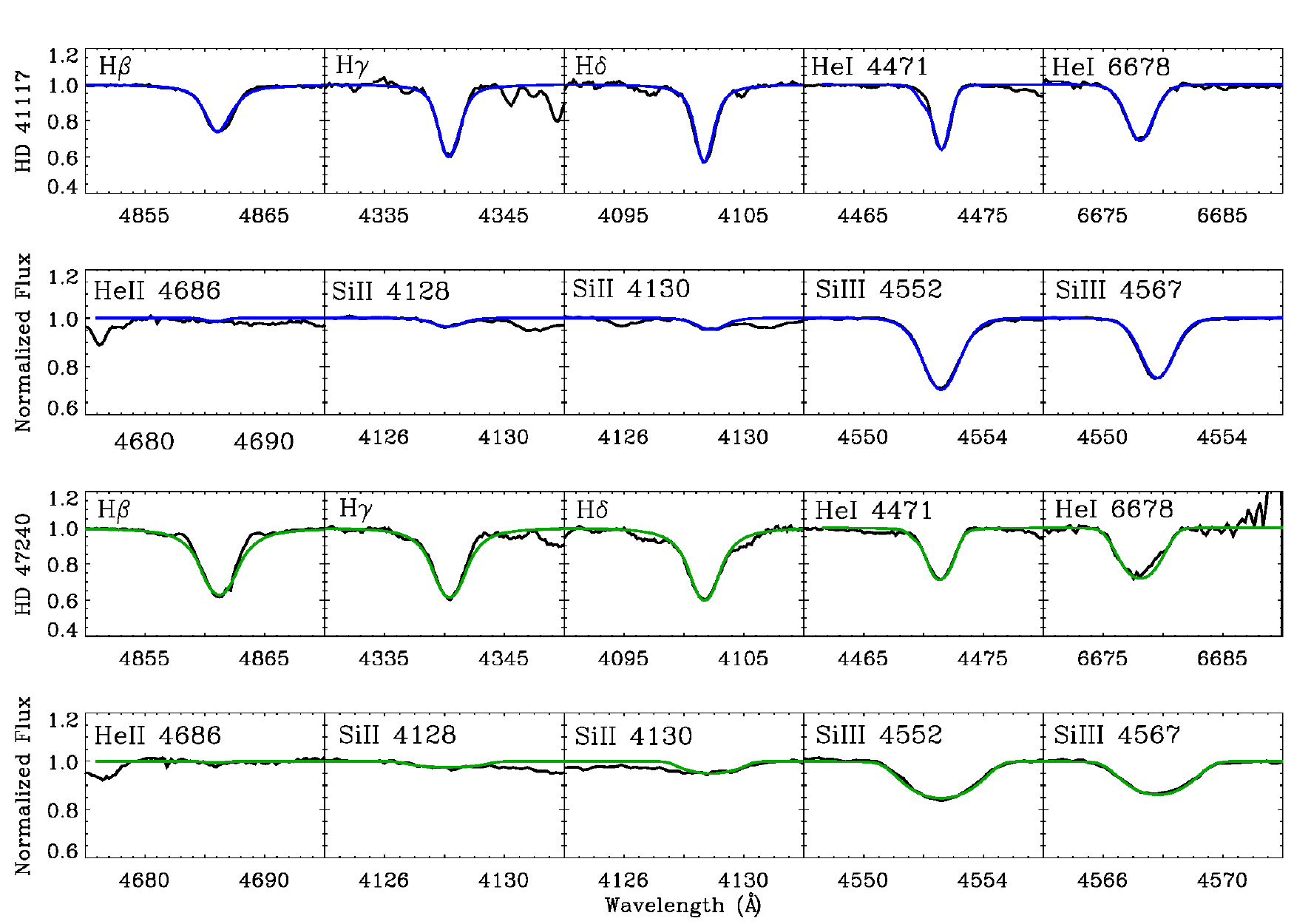}

\caption{\object{HD\,34085}, \object{HD\,38771}, \object{HD\,41117}, and  \object{HD\,47240}: Line model fittings to observations.\label{fig:7}}

\end{figure*}

\begin{figure*}[h]

\includegraphics[width=0.9\textwidth, angle=0]{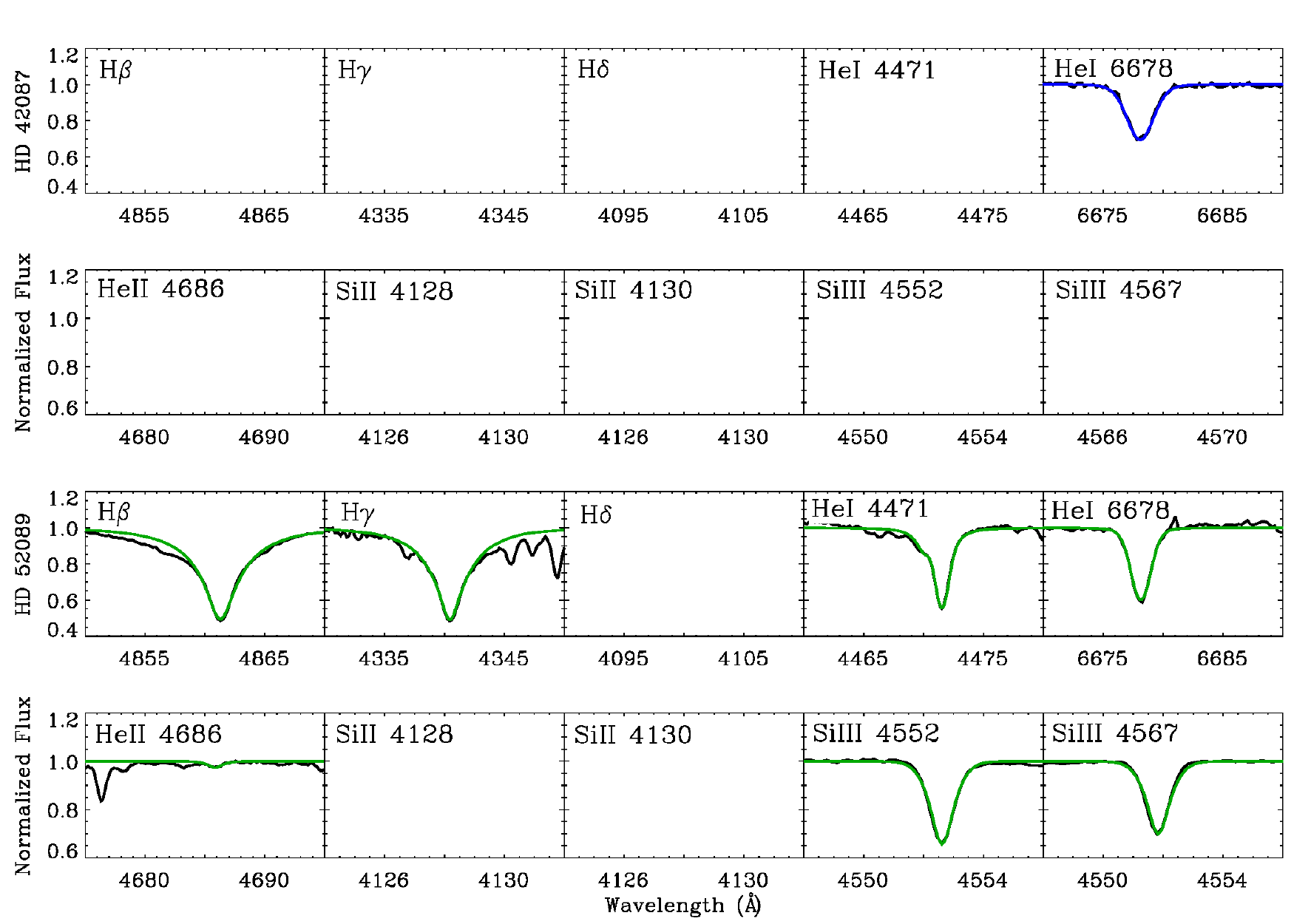}

\includegraphics[width=0.9\textwidth, angle=0]{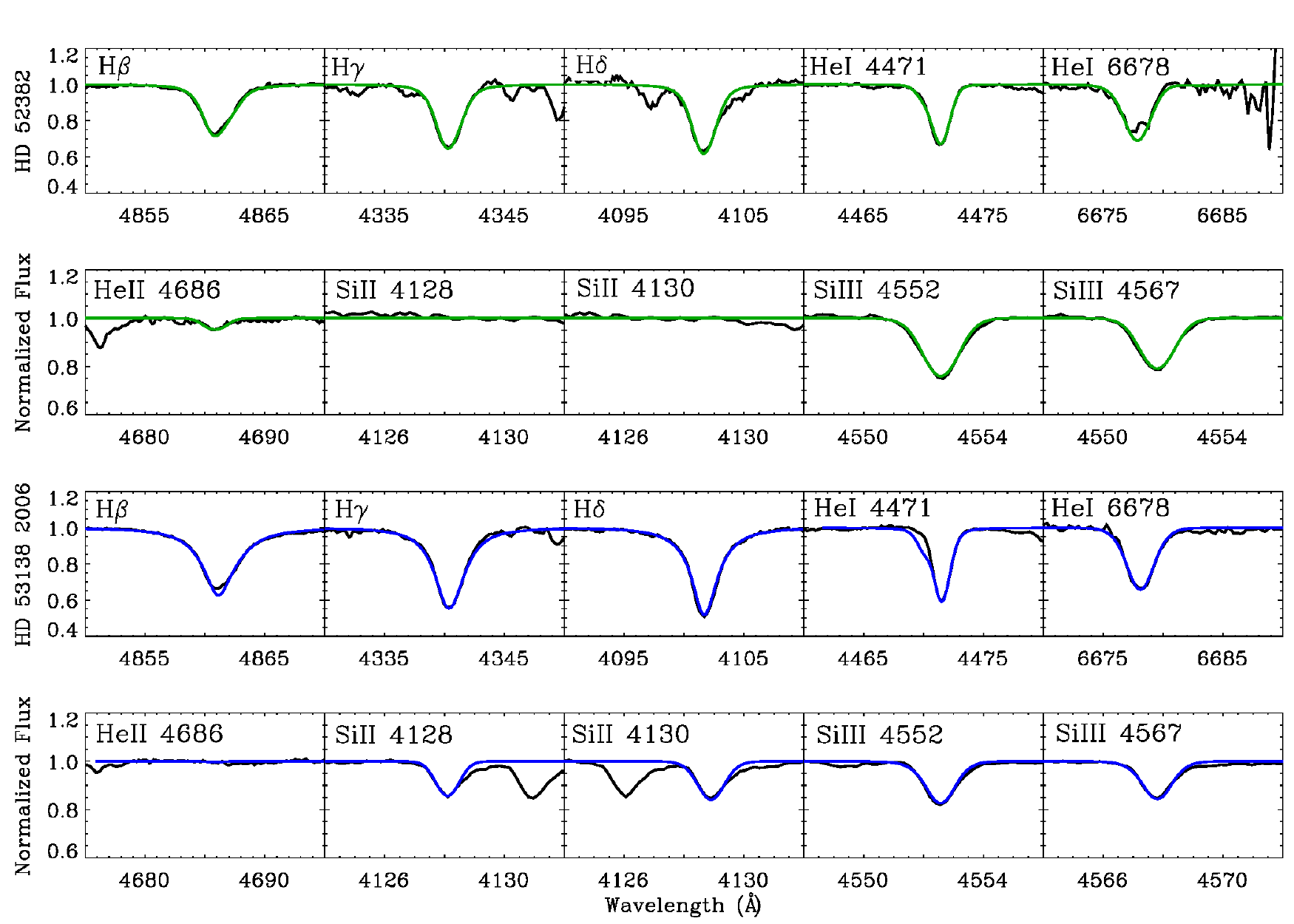}

\caption{ \object{HD\,42087},  \object{HD\,52089},  \object{HD\,52382}, and  \object{HD\,53138}: Line model fittings to observations.}

\end{figure*}

\begin{figure*}[h]

\includegraphics[width=0.9\textwidth, angle=0]{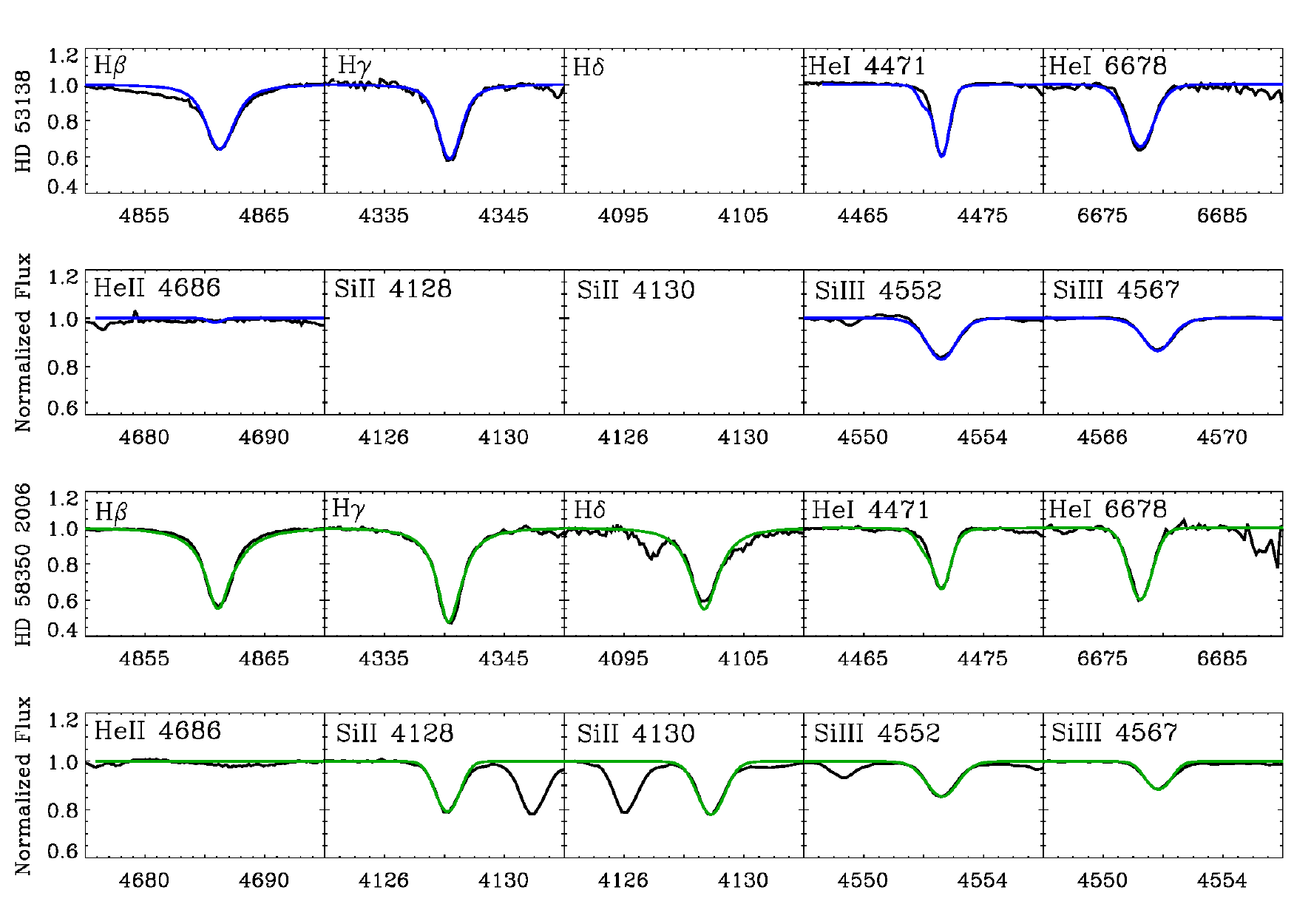}

\includegraphics[width=0.9\textwidth, angle=0]{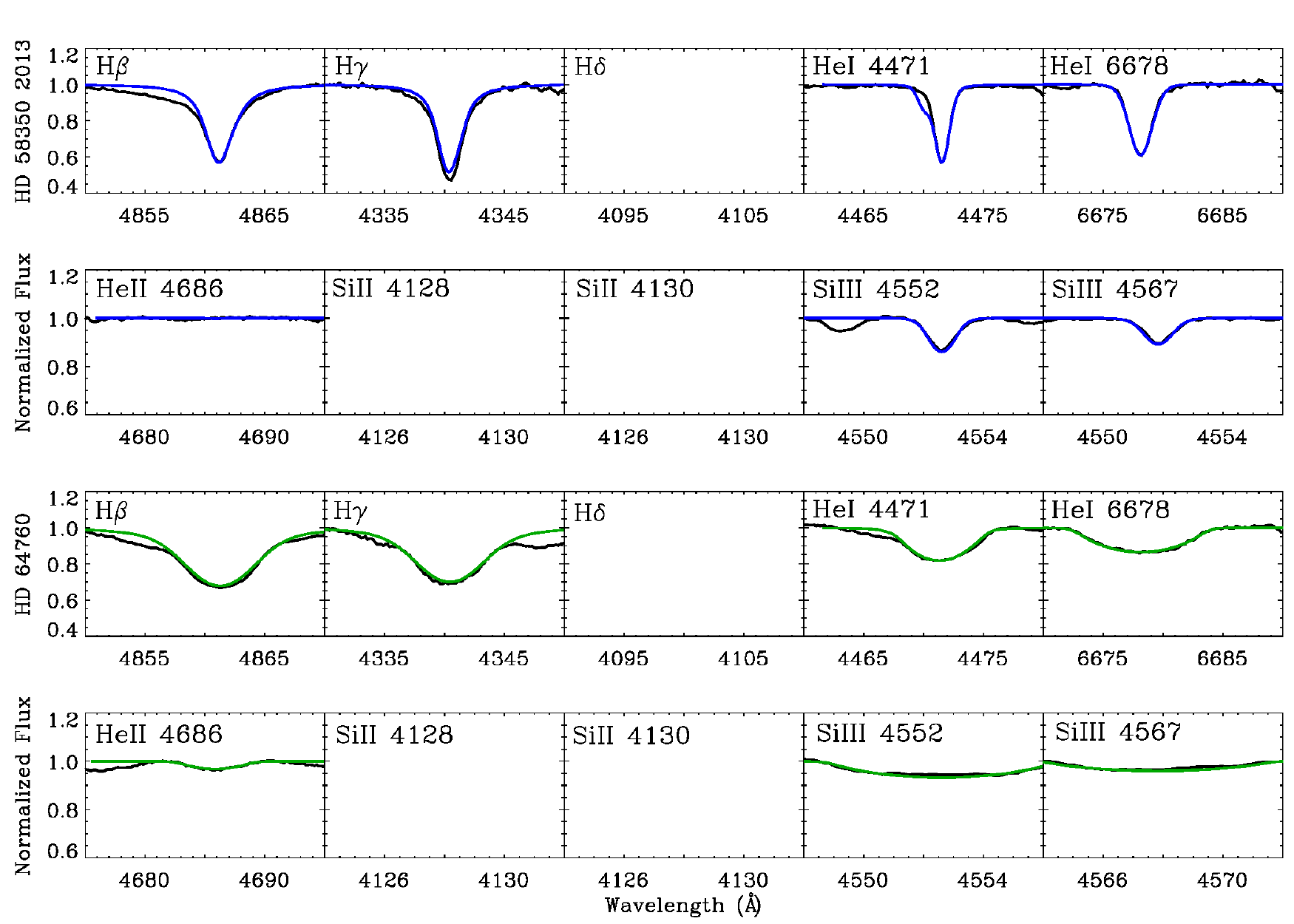}

\caption {\object{HD\,53138},  \object{HD\,58350}, and  \object{HD\,64760:} Line model fittings to observations.}.

\end{figure*}

\begin{figure*}[h]

\includegraphics[width=0.9\textwidth, angle=0]{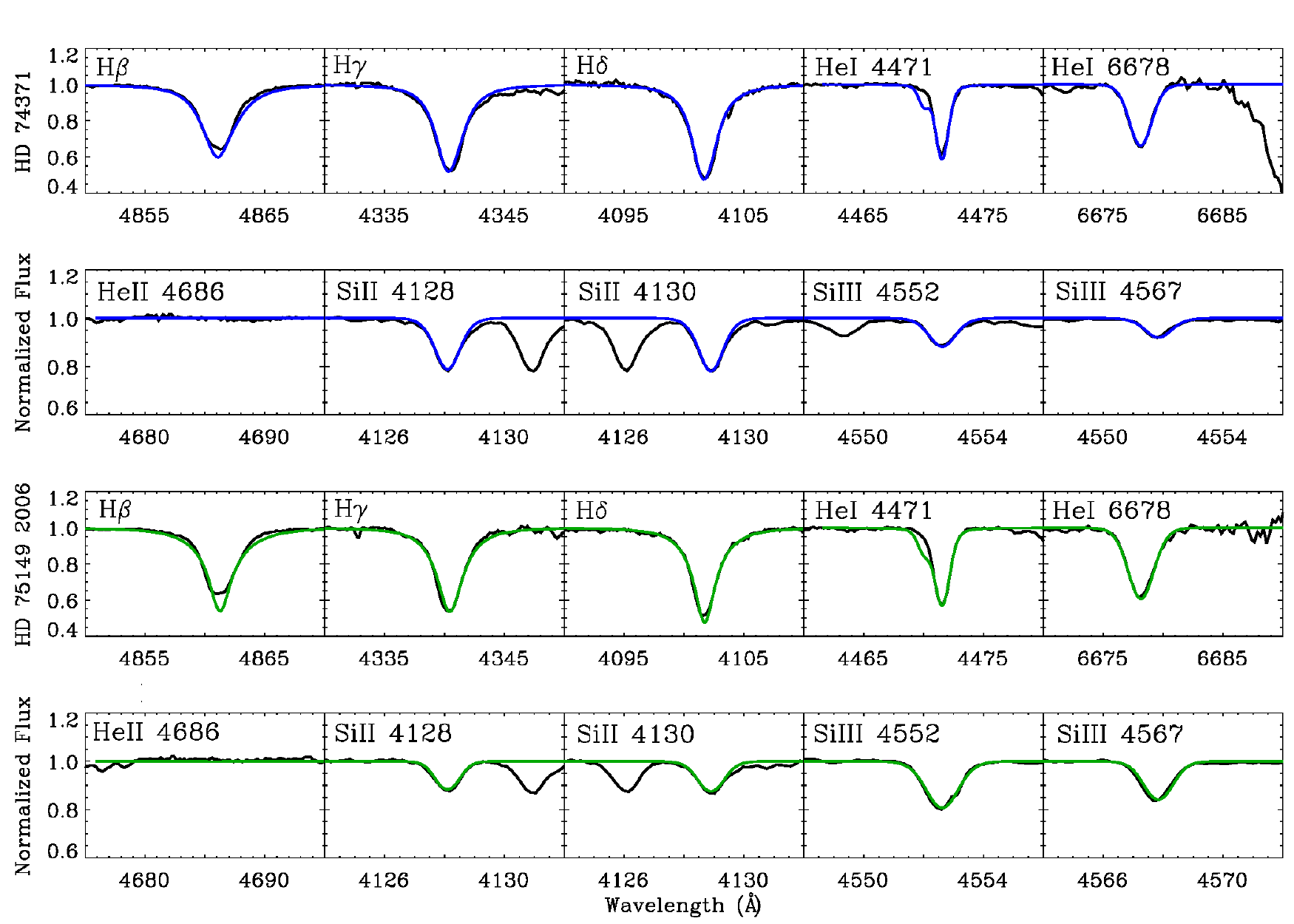}

\includegraphics[width=0.9\textwidth, angle=0]{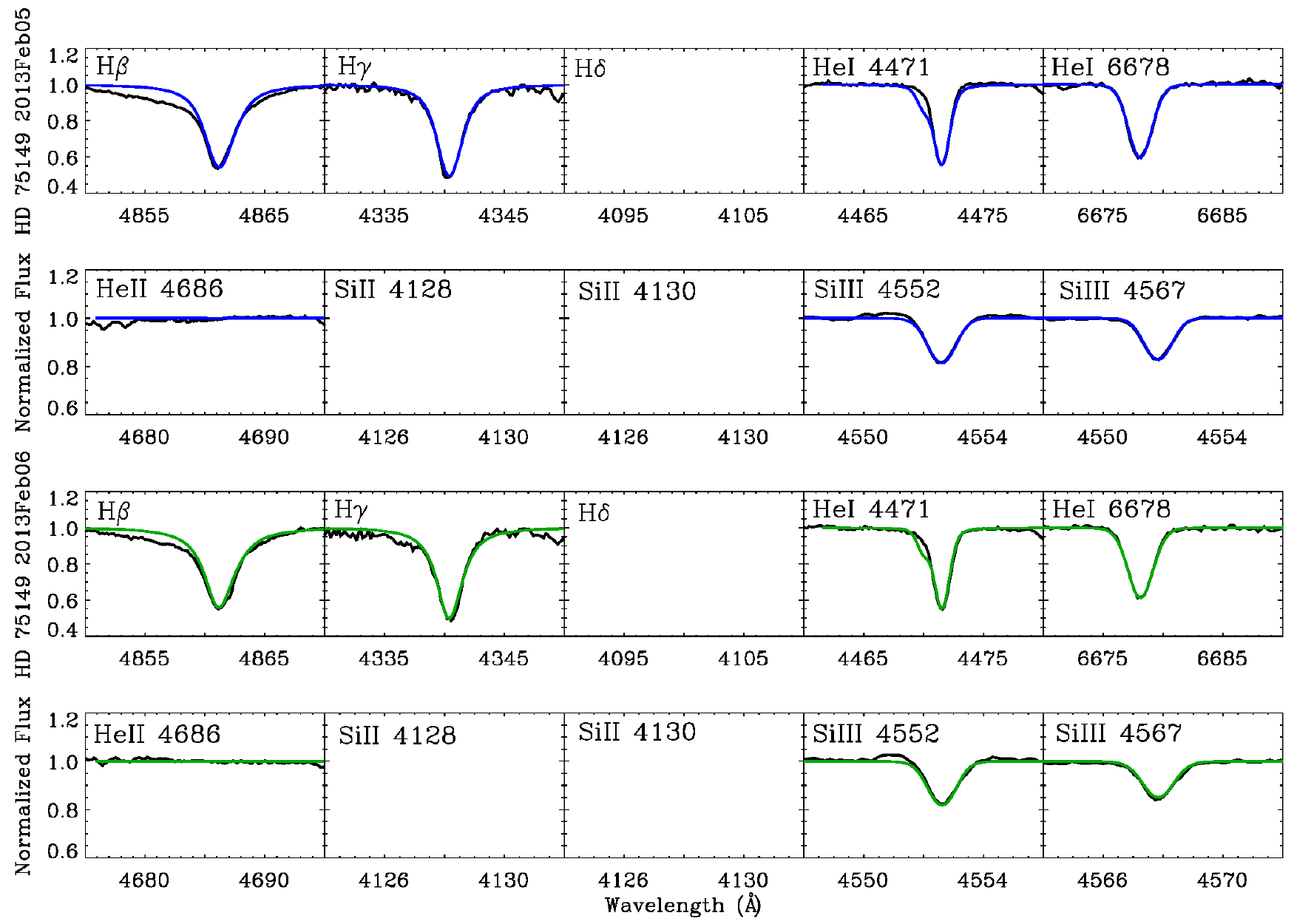}

\caption{\object{HD\,74371} and \object{HD\,75149}: Line model fittings to observations.\label{fig:10}}

\end{figure*}

\begin{figure*}[h]

\includegraphics[width=0.9\textwidth, angle=0]{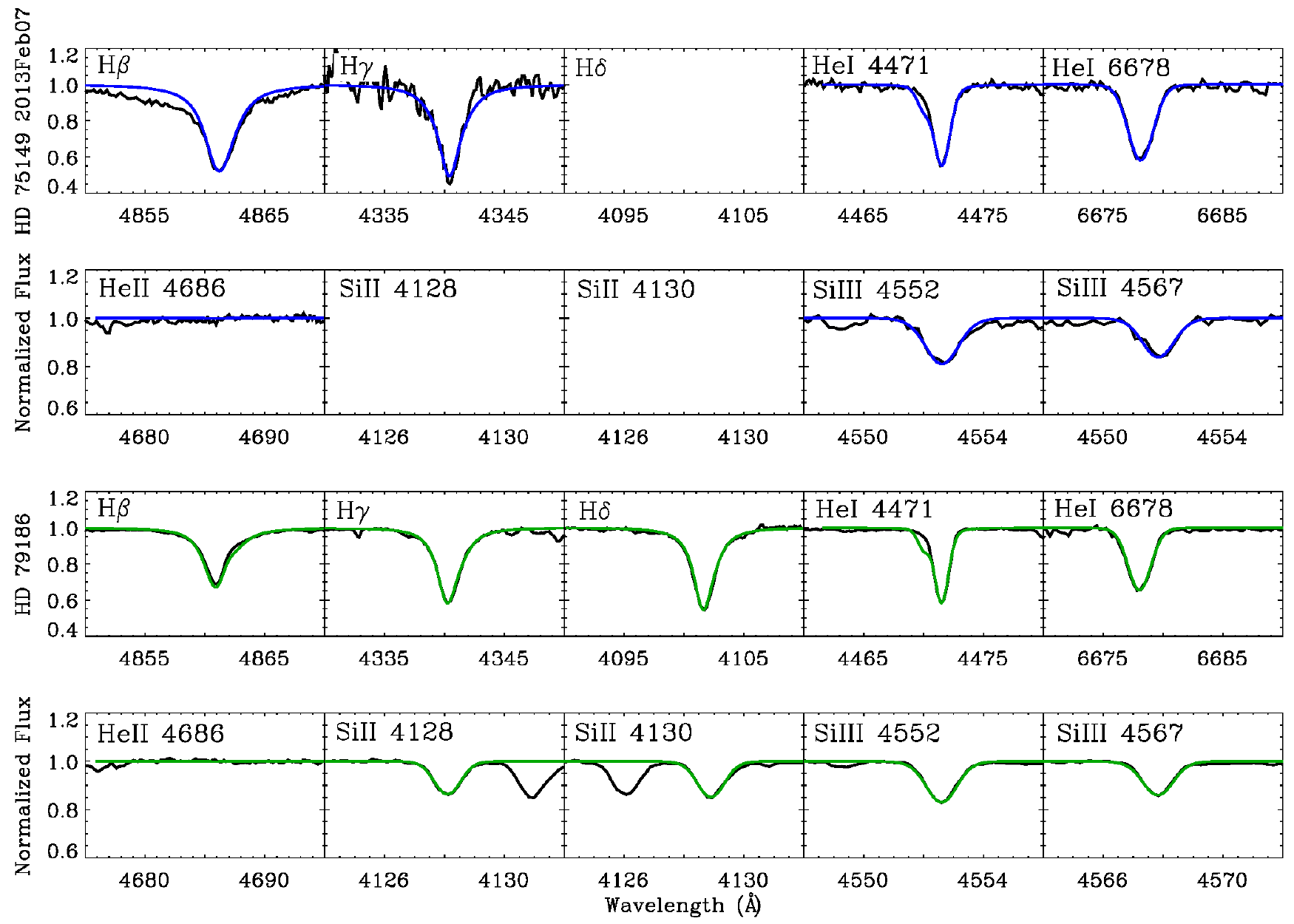}

\includegraphics[width=0.9\textwidth, angle=0]{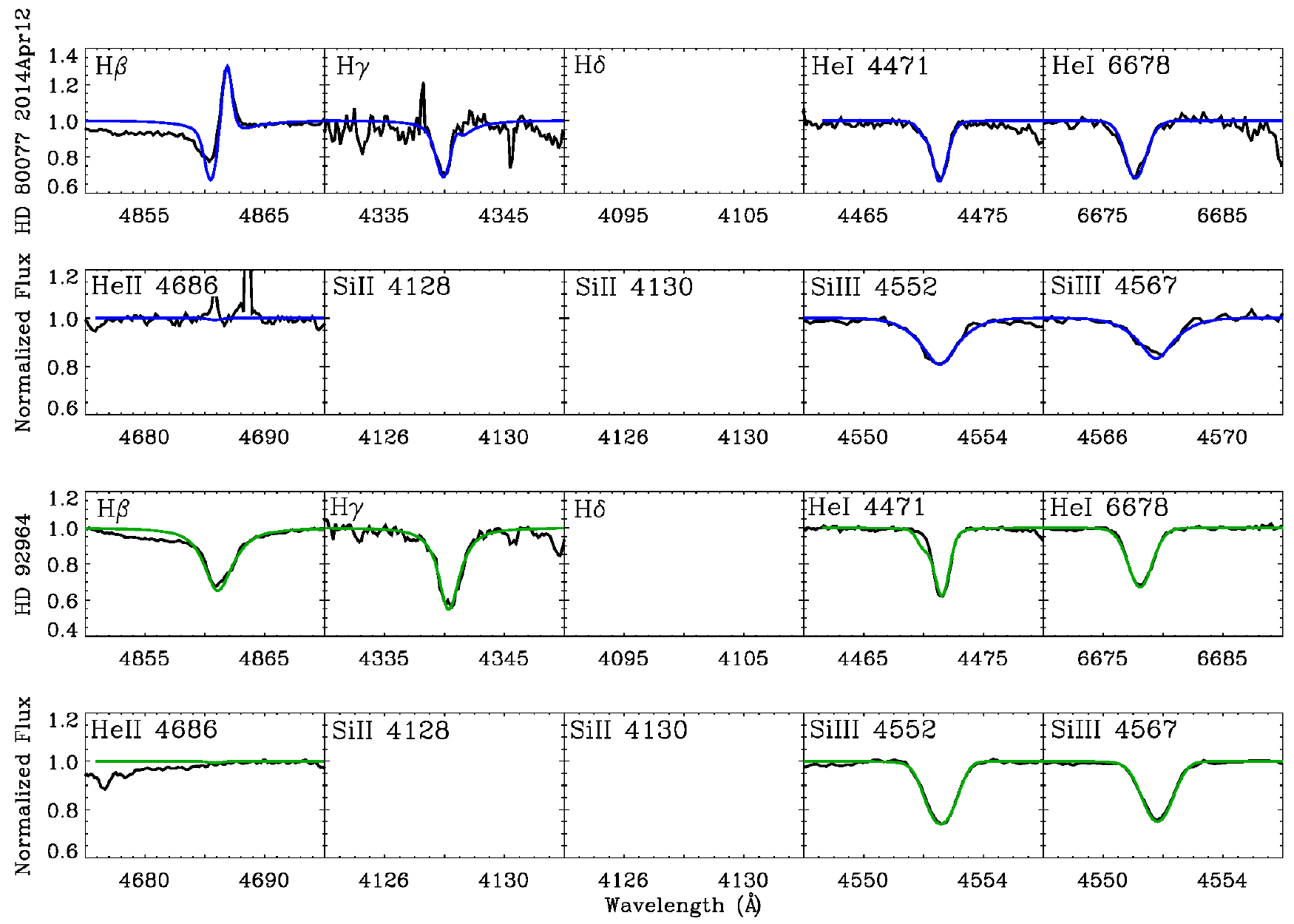}

\caption{\object{HD\,75149}, \object{HD\,79186}, \object{HD\,92964,} and \object{HD\,80077}: Line model fittings to observations.}.

\end{figure*}

\begin{figure*}[h]

\includegraphics[width=0.9\textwidth, angle=0]{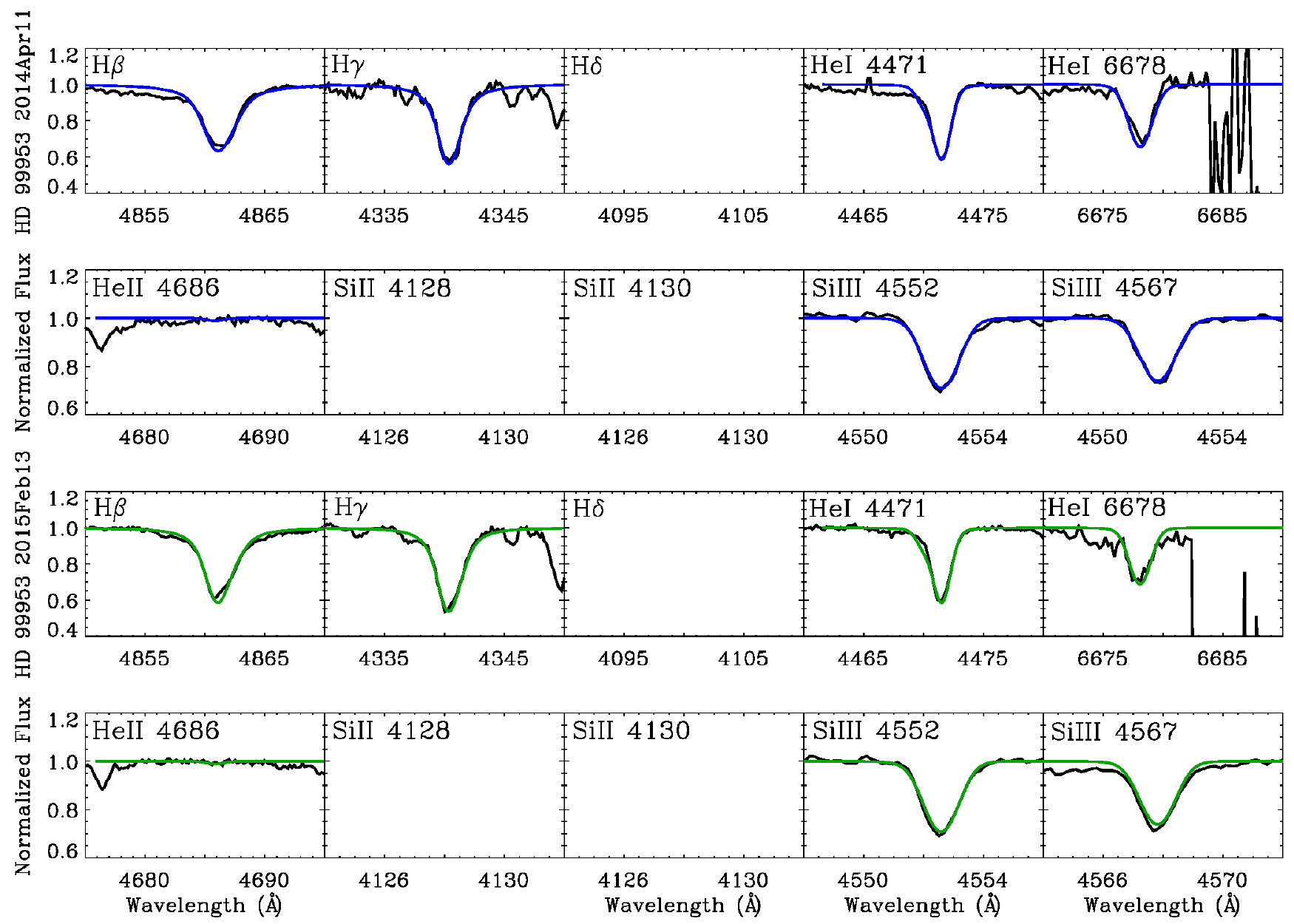}

\includegraphics[width=0.9\textwidth, angle=0]{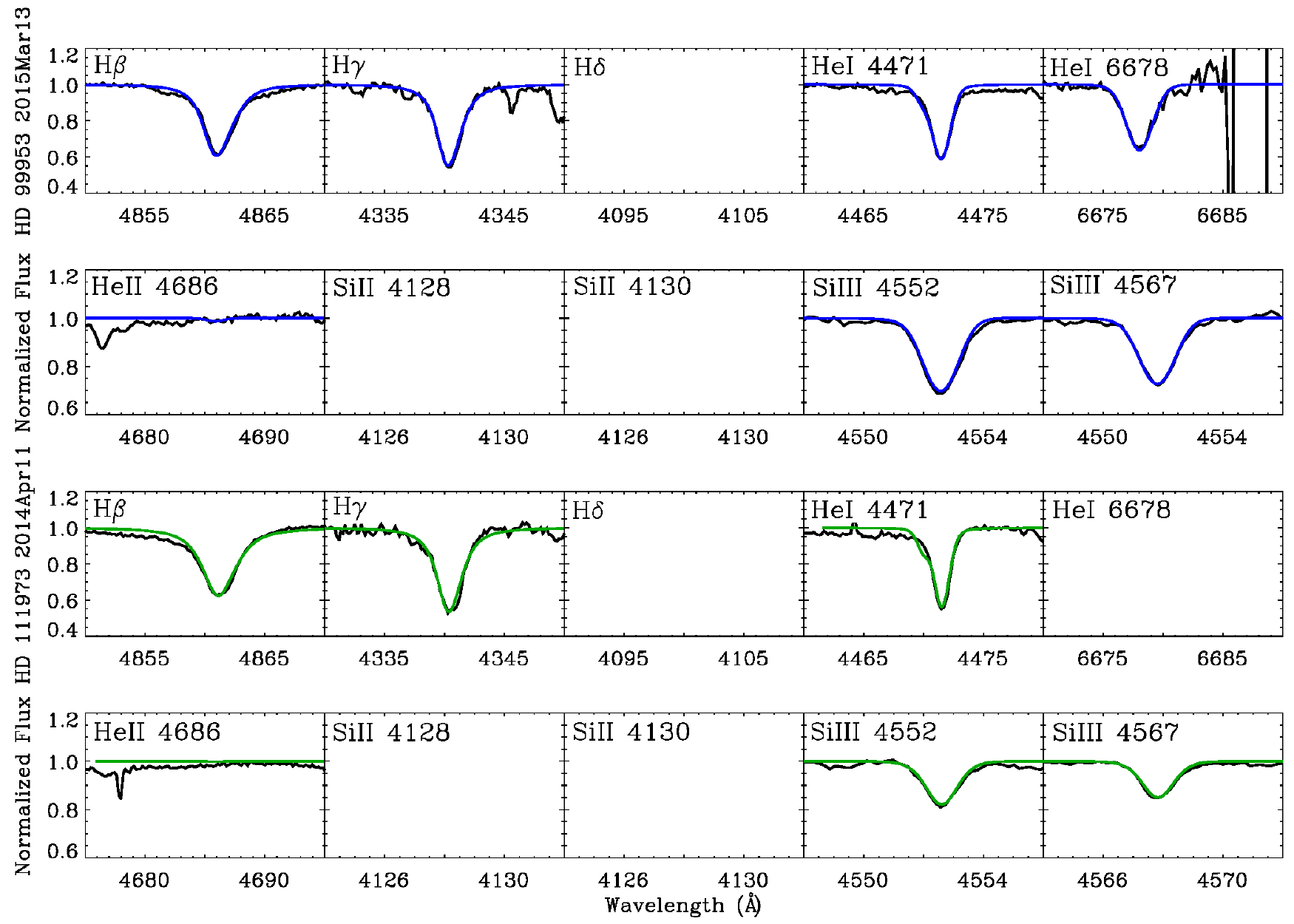}

\caption{\object{HD\,99953} and \object{HD\,111973}: Line model fittings to observations.}

\end{figure*}

\begin{figure*}[h]

\includegraphics[width=0.9\textwidth, angle=0]{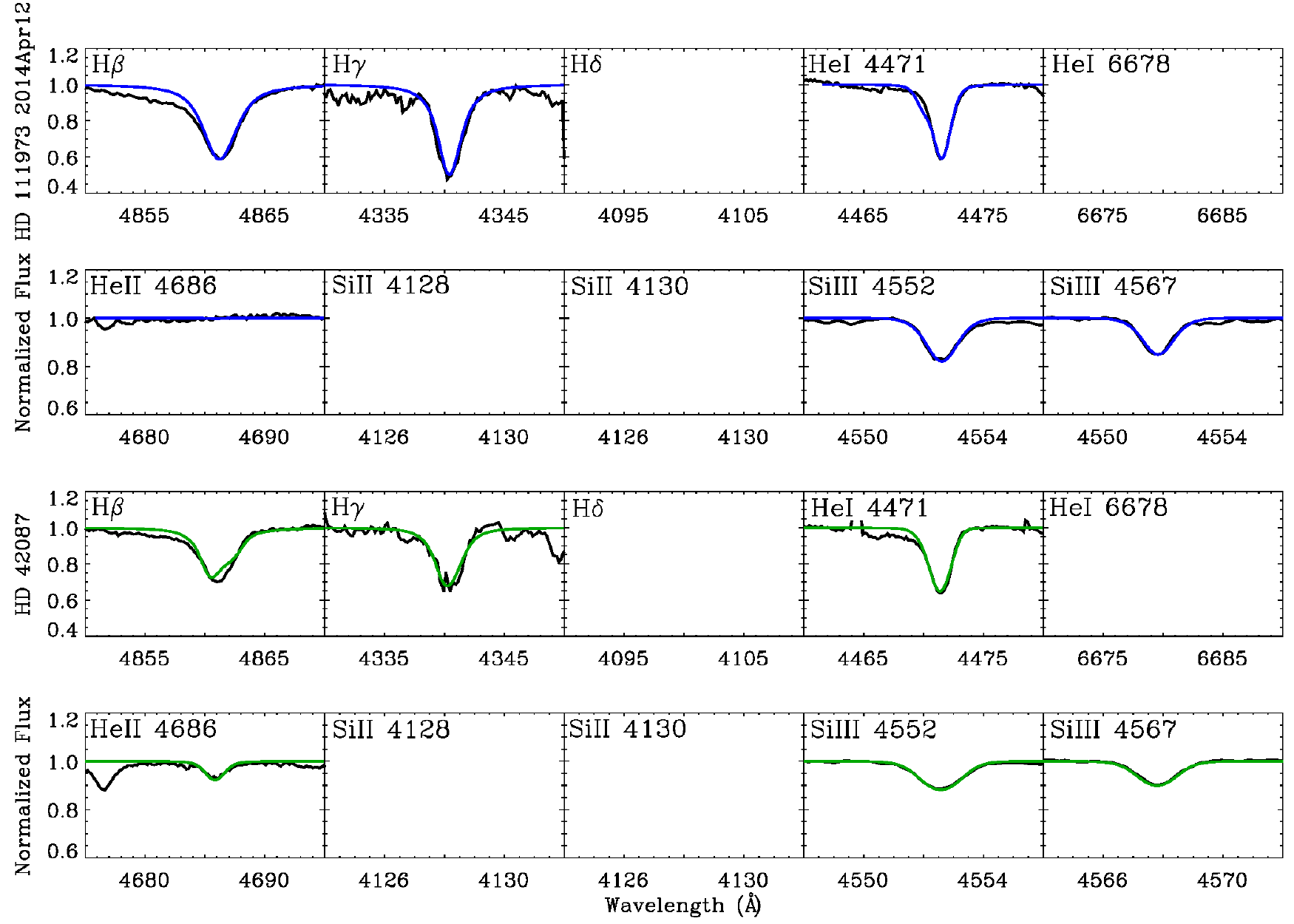}

\includegraphics[width=0.9\textwidth, angle=0]{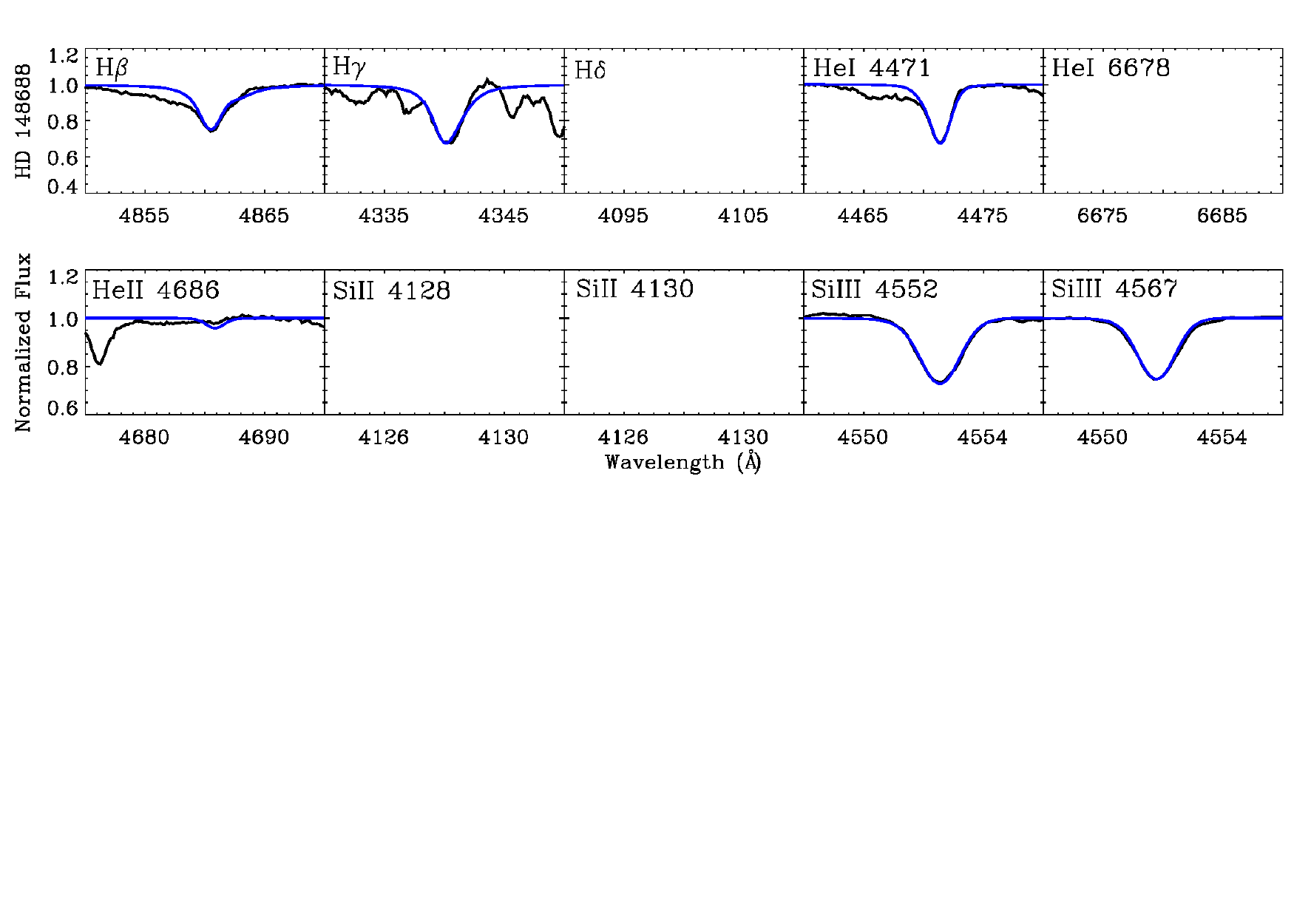}

\caption{\object{HD\,111973}, \object{HD\,115842,} and \object{HD\,148688}: Line model fittings to observations. \label{fig:13}}

\end{figure*}

\end{appendix}

\end{document}